\documentclass[12pt]{msuthesis}
\usepackage[dvips]{graphics}
\usepackage{epsfig}
\def\a{\alpha}

\def\d{\delta}
\def\e{\epsilon}
\def\p{\pi}
\def\m{\mu}
\def\n{\nu}
\def\ln{\mathrm{ln}}

\def\H{{\rm\scriptscriptstyle H}}

\def\ph{p_\H}
\def\mh{M_\H}
\def\frac#1#2{{#1 \over #2}}
\def\slash#1{\rlap/#1}

\def\la{\langle}
\def\ra{\rangle}
\def\l{\left}
\def\r{\right}

\def\o{\over}

\def\lp{\left ( }
\def\rp{\right ) }

\def\beq{\begin{equation}}
\def\beqa{\begin{eqnarray}}
\def\eeq{\end{equation}}
\def\eeqa{\end{eqnarray}}

\def\no{\nonumber}

\def\qq{\qquad}

\def\s{\hat{s}}
\def\t{\hat{t}}

\def\su{S_{u}}
\def\st{S_{t}}
\department{Physics and Astronomy}
\degree{DOCTOR OF PHILOSOPHY}
\author{Christopher James Glosser}
\title{Higgs Boson Production in Hadron-Hadron Colliders}
\gyear{2001}

\begin{document}
%
%
%
%
%
%
\message{FEYNMAN:  For generating Feynman Diagrams in LaTex}
\message{Mark 1.0 Last Altered by MJSL 2/89}
\setlength{\unitlength}{0.01pt}
\gdef\Feynmanlength{\setlength{\unitlength}{0.01pt}}  
\gdef\unlock{\catcode`\@=11}
\gdef\lock{\catcode`\@=12}
\global\newcount\LINETYPE
\global\newcount\LINEDIRECTION
\global\newcount\LINECONFIGURATION
\newcommand{\LTYPE}{\LINETYPE}
\newcommand{\LDIR}{\LINEDIRECTION}
\newcommand{\LCONFIG}{\LINECONFIGURATION}
\global\LINETYPE=1  \global\LINEDIRECTION=0  \global\LINECONFIGURATION=0
\global\newcount\fermion    \fermion=1
\global\newcount\Scalar     \Scalar=2
\global\newcount\photon     \photon=3
\global\newcount\gluon      \gluon=4
\global\newcount\SPECIAL    \SPECIAL=5
\gdef\N{0}  \gdef\NE{1}  \gdef\E{2}   \gdef\SE{3}
\gdef\S{4}  \gdef\SW{5}  \gdef\W{6}   \gdef\NW{7}
\global\newcount\REG            \global\REG=0
\global\newcount\FLIPPED        \global\FLIPPED=1
\global\newcount\CURLY          \global\CURLY=2
\global\newcount\FLIPPEDCURLY   \global\FLIPPEDCURLY=3
\global\newcount\FLAT           \global\FLAT=4
\global\newcount\FLIPPEDFLAT    \global\FLIPPEDFLAT=5
\global\newcount\CENTRAL        \global\CENTRAL=6
\global\newcount\FLIPPEDCENTRAL \global\FLIPPEDCENTRAL=7
\gdef\LONGPHOTON{6}             \gdef\FLIPPEDLONG{7}
\global\newcount\SQUASHEDGLUON  \global\SQUASHEDGLUON=8
\gdef\SQUASHED{\SQUASHEDGLUON}
%
\newcount\adjx \adjx=0
\newcount\adjy \adjy=0
\global\newdimen\BIGPHOTONS     \BIGPHOTONS=0pt  
\gdef\bigphotons{\global\BIGPHOTONS=12pt}
\global\newdimen\THICKPHOTONS     \THICKPHOTONS=0pt  
\global\newdimen\THICKPHOTONSWITCH    \THICKPHOTONSWITCH=0pt
\gdef\THICKPHOTONTEST{
\THICKPHOTONSWITCH=0pt
\ifdim\THICKPHOTONS=0pt \relax
  \else \ifnum\LTYPE=3
           \ifnum\LDIR=2 \THICKPHOTONSWITCH=1pt \fi 
           \ifnum\LDIR=6 \THICKPHOTONSWITCH=1pt \fi 
        \fi
\fi
}  
\gdef\THICKLINES{\thicklines  \THICKPHOTONS=1pt}
\gdef\THINLINES{\thinlines  \THICKPHOTONS=0pt}
\global\newcount\phantomswitch   \global\phantomswitch=0
\global\newcount\stemlength   \global\stemlength=275   
\global\newcount\absstemlength        
\global\newcount\stemlengthx          
\global\newcount\stemlengthy          
\newdimen\FRONTSTEM  \FRONTSTEM=0pt   
\newdimen\BACKSTEM   \BACKSTEM=0pt    
\newdimen\EITHERSTEM \EITHERSTEM=0pt  
\gdef\frontstemmed{\FRONTSTEM=1pt}            
\gdef\backstemmed{\BACKSTEM=1pt}              
\gdef\stemmed{\FRONTSTEM=1pt  \BACKSTEM=1pt}    
\global\newcount\arrowlength                
\global\newdimen\ATTIP   \global\ATTIP=0pt  
\global\newdimen\ATBASE  \global\ATBASE=1pt 
\global\newcount\unitboxnumber  
\global\newcount\unitboxnumberpo  
\global\newcount\particlelengthx  
\gdef\plengthx{\particlelengthx}
\global\newcount\particlelengthy  
\gdef\plengthy{\particlelengthy}
\global\newcount\boxlengthx  
\global\newcount\boxlengthy  
\global\newcount\particleadjustx  
\global\newcount\particleadjusty  
\global\newcount\particlelength   
\global\newcount\particlefrontx
\gdef\pfrontx{\particlefrontx}
\global\newcount\PFRONTx
\global\newcount\particlefronty
\gdef\pfronty{\particlefronty}
\global\newcount\PFRONTy
\global\newcount\particlebackx
\gdef\pbackx{\particlebackx}
\global\newcount\particlebacky
\gdef\pbacky{\particlebacky}
\global\newcount\particlemidx
\gdef\pmidx{\particlemidx}
\global\newcount\particlemidy
\gdef\pmidy{\particlemidy}
\global\newcount\seglength  \global\newcount\gaplength
\global\gaplength=850  
\global\seglength=1416  
\global\newcount\Xone    \global\newcount\Yone    
\global\newcount\Xtwo    \global\newcount\Ytwo    
\global\newcount\Xthree  \global\newcount\Ythree  
\global\newcount\Xfour   \global\newcount\Yfour   
\global\newcount\Xfive   \global\newcount\Yfive   
\global\newcount\Xsix    \global\newcount\Ysix    
\global\newcount\Xseven  \global\newcount\Yseven  
\global\newcount\Xeight  \global\newcount\Yeight  
%
%
\newsavebox{\lastline}  
\global\newcount\numlineparts   
\global\newcount\upperlineadjx  \upperlineadjx=0  
\global\newcount\upperlineadjy  \upperlineadjy=0  
\global\newcount\lowerlineadjx  \lowerlineadjx=0  
\global\newcount\lowerlineadjy  \lowerlineadjy=0  
\global\newcount\thirdlineadjx  \thirdlineadjx=0  
\global\newcount\thirdlineadjy  \thirdlineadjy=0  
\global\newcount\fourthlineadjx \fourthlineadjx=0  
\global\newcount\fourthlineadjy \fourthlineadjy=0  
\global\newcount\unitboxwidth   \unitboxwidth=1000
\global\newcount\unitboxheight  \unitboxheight=0  
\global\newcount\numupperunits  \numupperunits=8  
\global\newcount\numlowerunits  \numlowerunits=8  
\global\newcount\numthirdunits  \numthirdunits=8  
\global\newcount\numfourthunits \numfourthunits=8  
\global\newcount\fermioncount   \global\fermioncount=0
\global\newcount\Scalarcount    \global\Scalarcount=0
\global\newcount\photoncount    \global\photoncount=0
\global\newcount\gluoncount     \global\gluoncount=0
\global\newcount\SPECIALcount   \global\SPECIALcount=0
\global\newcount\vertexcount    \global\vertexcount=-1
%
\global\newcount\XDIR
\global\newcount\YDIR
\gdef\SETDIR{  
\ifcase\LDIR
     \global\XDIR=0  \global\YDIR=1   
\or  \global\XDIR=1  \global\YDIR=1   
\or  \global\XDIR=1  \global\YDIR=0   
\or  \global\XDIR=1  \global\YDIR=-1  
\or  \global\XDIR=0  \global\YDIR=-1  
\or  \global\XDIR=-1 \global\YDIR=-1  
\or  \global\XDIR=-1 \global\YDIR=0   
\or  \global\XDIR=-1 \global\YDIR=1   
\else\DIRECTERROR
\fi}  
\gdef\moduloeight#1{
\ifnum#1>7 \global\advance #1 by -8
\relax
\moduloeight#1
\relax
\else \relax
\fi}
\gdef\multroothalf#1{\global\multiply #1 by 7071 \global\divide #1 by 10000}
\gdef\negate#1{\global\multiply #1 by -1}
\gdef\double#1{\global\multiply #1 by 2}
\gdef\slanttest(#1,#2){
\ifodd\LDIR
\multiply #1 by 7071  \divide #1 by 10000
\multiply #2 by 7071  \divide #2 by 10000
\fi
}
\gdef\gslanttest(#1,#2){
\ifodd\LDIR
\multroothalf#1
\multroothalf#2
\fi
}
%
%
\gdef\setplength{ 
\global\particlelengthx=\unitboxwidth
\global\particlelengthy=\unitboxheight
\global\multiply \particlelengthx by \unitboxnumber
\global\multiply \particlelengthy by \unitboxnumber
\global\advance \particlelengthx by \particleadjustx
\global\advance \particlelengthy by \particleadjusty
}
\gdef\boxlengthdefault{  
\global\boxlengthx=\plengthx
\global\boxlengthy=\plengthy
\ifnum\plengthx<0 \global\multiply\boxlengthx by -1 \fi
\ifnum\plengthy<0 \global\multiply\boxlengthy by -1 \fi
}
\gdef\rearcoords{  
\global\particlebacky=\particlefronty
\global\particlebackx=\particlefrontx
\global\advance \particlebackx by \particlelengthx
\global\advance \particlebacky by \particlelengthy
}
\gdef\midcoords{  
\global\particlemidy=\particlefronty
\global\particlemidx=\particlefrontx
\global\stemlengthx=\particlelengthx  
\global\stemlengthy=\particlelengthy
\global\divide\stemlengthx by 2
\global\divide\stemlengthy by 2
\global\advance \particlemidx by \stemlengthx
\global\advance \particlemidy by \stemlengthy
}
\gdef\setparticle{\setplength\rearcoords\midcoords\boxlengthdefault}  
%
\gdef\setcoords(#1,#2,#3)(#4,#5,#6)[#7,#8]{
\global\upperlineadjx=#1
\global\lowerlineadjx=#2
\global\thirdlineadjx=#3
\global\upperlineadjy=#4
\global\lowerlineadjy=#5
\global\thirdlineadjy=#6
\global\unitboxwidth=#7
\global\unitboxheight=#8
}
%
%
%
\gdef\drawoldpic#1(#2,#3){  
\global\particlefrontx=#2
\global\particlefronty=#3
\rearcoords
\midcoords
\put(#2,#3){\usebox{#1}}
}
\gdef\drawsavedline`#1' as #2[#3#4](#5,#6)[#7]{
\global\LINETYPE=#2
\global\LINEDIRECTION=#3
\global\LINECONFIGURATION=#4
\global\particlefrontx=#5
\global\particlefronty=#6
\global\unitboxnumber=#7
\selectcase
\rearcoords
\midcoords
\ifnum\phantomswitch=0 \drawas{#1}\fi
}

\gdef\startphantom{\phantomswitch=1} 
\gdef\stopphantom{\phantomswitch=0}  

\gdef\drawas#1{
\global\savebox{#1}(\boxlengthx,\boxlengthy){
\setlength{\unitlength}{0.01pt}
\begin{picture}(\boxlengthx,\boxlengthy)
\multiput(\upperlineadjx,\upperlineadjy)(\unitboxwidth,\unitboxheight)
{\numupperunits}{\upperunitbox}
\ifnum\numlineparts > 1  
\multiput(\lowerlineadjx,\lowerlineadjy)(\unitboxwidth,\unitboxheight)
{\numlowerunits}{\lowerunitbox}
\fi
\ifnum\numlineparts > 2  
\multiput(\thirdlineadjx,\thirdlineadjy)(\unitboxwidth,\unitboxheight)
{\numthirdunits}{\thirdunitbox}
\fi
\ifnum\numlineparts > 3  
\multiput(\fourthlineadjx,\fourthlineadjy)(\unitboxwidth,\unitboxheight)
{\numfourthunits}{\lowerunitbox}
\fi
\end{picture} }
\global\PFRONTx=\pfrontx  \global\PFRONTy=\pfronty   
\SETFRONTSTEM
\THICKPHOTONTEST
\ifdim\THICKPHOTONSWITCH=1pt\global\advance\PFRONTy by 20  \fi
\put(\PFRONTx,\PFRONTy) {\usebox{#1}}   
\ifdim\THICKPHOTONSWITCH=1pt
\global\advance\PFRONTy by -40
\put(\PFRONTx,\PFRONTy) {\usebox{#1}}   
\global\advance \PFRONTy by 20  
\fi  
\SETBACKSTEM
\seglength=1416   \gaplength=850   
}
%
%

\gdef\drawandsaveline`#1' as #2[#3#4](#5,#6)[#7]{
\global\newsavebox{#1}
\drawsavedline`#1' as #2[#3#4](#5,#6)[#7]
}

\gdef\drawline#1[#2#3](#4,#5)[#6]{   
\drawsavedline`\lastline' as #1[#2#3](#4,#5)[#6]}

\gdef\saveas#1{  
\global\newsavebox#1
\drawas#1}
%
%
%
\gdef\TYPEERROR{\message{*** ERROR IN PARTICLE TYPE SELECTION ***}
\message{+++ Try with line type \fermion,\Scalar,\photon,\gluon
(see manual) +++}\SETERR}
\gdef\DIRECTERROR{\SETERR\message{*** ERROR IN PARTICLE DIRECTION SELECTION
***}
\message{+++ Try again with direction N, NE, E, SE  etc. or see manual +++}}
\gdef\UNIMPERROR{\message{*** ERROR IN PARTICLE OPTIONS SELECTION ***}
\message{
+++ The requested options combination has not yet been implemented +++}\SETERR}
\gdef\SETERR{\gdef\upperunitbox{{\tiny Error}}  
\gdef\lowerunitbox{\relax}
\gdef\thirdunitbox{\relax}
}
\gdef\neglengthcheck{\ifnum\unitboxnumber < 1
\message{   *** ERROR:  PARTICLE OF NEGATIVE OR ZERO LENGTH REQUESTED. ***   }
\message{   ***         TAKING ABSOLUTE VALUE. ***   }\negate\unitboxnumber
\fi}
\gdef\selectcase{
\neglengthcheck   
\SETDIR
\ifcase\LINETYPE
\TYPEERROR  
\or \selectfermion  
\or \selectScalar   
\or \selectphoton   
\or \selectgluon    
\or \selectspecial  
\else \TYPEERROR \fi  }
\gdef\selectfermion{
\ifnum\fermioncount=0 
\global\newcount\fermionlength  
\global\newcount\fermionlengthx
\global\newcount\fermionlengthy
\global\newcount\fermionfrontx  
\global\newcount\fermionfronty  
\global\newcount\fermionbackx
\global\newcount\fermionbacky
\gdef\ALLfermion{  
\global\fermionfrontx=\particlefrontx \global\fermionfronty=\particlefronty
\ifnum\unitboxnumber > 50000
\message{   *** WARNING *** Fermion of length
\the\unitboxnumber\space requested ***   }
\ifnum\unitboxnumber > 80000
\message{   *** Reducing fermion length to 30000 (max 80000) ***   }
\global\unitboxnumber=30000 \fi \fi  
\global\fermionlength=\unitboxnumber 
\global\particleadjustx=0   \global\particleadjusty=0 
\global\numlineparts = 1    \global\numupperunits=1
\global\upperlineadjx=-200  \global\upperlineadjy=0
\global\fermionlengthx=\fermionlength    \global\fermionlengthy=\fermionlength
\gslanttest(\fermionlengthx,\fermionlengthy)  
\global\multiply\fermionlengthx by \XDIR  
\global\multiply\fermionlengthy by \YDIR  
\global\unitboxheight=\fermionlengthy   \global\unitboxwidth=\fermionlengthx
\global\advance \fermionlengthx by \particleadjustx
\global\advance \fermionlengthy by \particleadjusty
\global\particlelengthx=\fermionlengthx
\global\particlelengthy=\fermionlengthy
\boxlengthdefault    \rearcoords    \midcoords
\global\fermionbackx=\particlebackx     \global\fermionbacky=\particlebacky
\ifcase\LINECONFIGURATION  
\ifnum\XDIR=0
\gdef\upperunitbox{\line(\XDIR,\YDIR){\boxlengthy}} 
\else
\gdef\upperunitbox{\line(\XDIR,\YDIR){\boxlengthx}}
\fi
\else \UNIMPERROR
\fi
}
 \fi
\global\advance\fermioncount by 1  
\ALLfermion
}
\gdef\selectScalar{
\ifnum\Scalarcount=0 
\newcount\Scalarlength
\newcount\Scalarlengthx
\newcount\Scalarlengthy
\newcount\Scalarfrontx  
\newcount\Scalarfronty  
\newcount\Scalarbackx
\newcount\Scalarbacky
\gdef\ALLScalar{
\global\Scalarfrontx=\particlefrontx   
\global\Scalarfronty=\particlefronty   
\numlineparts = 1      \numupperunits=\unitboxnumber
\ifcase\LINECONFIGURATION
\global\upperlineadjx=-200     \global\upperlineadjy=0
\slanttest(\seglength,\gaplength)   
\gdef\upperunitbox{\line(\XDIR,\YDIR){\seglength}}
\else \UNIMPERROR 
\fi
\global\unitboxwidth=\seglength  \global\advance\unitboxwidth by \gaplength
\global\multiply \unitboxwidth by \XDIR
\global\unitboxheight=\seglength  \global\advance\unitboxheight by \gaplength
\global\multiply \unitboxheight by \YDIR
\global\particleadjustx=\gaplength \global\multiply\particleadjustx by \XDIR
\global\particleadjusty=\gaplength \global\multiply\particleadjusty by \YDIR
\negate\particleadjustx   \negate\particleadjusty   
\setparticle  
\global\Scalarlengthx=\particlelengthx  
\global\Scalarlengthy=\particlelengthy  
\ifnum\boxlengthx > 50000
\message{   *** WARNING *** Scalar of length in excess of 50000cp
requested!}\fi
\ifnum\boxlengthy > 50000
\message{   *** WARNING *** Scalar of length in excess of 50000cp
requested!}\fi
\global\Scalarbackx=\pbackx      \global\Scalarbacky=\pbacky   
}
 \fi
\global\advance\Scalarcount by 1  
\ALLScalar
}
\gdef\selectphoton{   
\ifnum\photoncount=0 
\newcount\numwiggles    \newcount\numwigglespo
\global\newcount\photonlengthx
\global\newcount\photonlengthy
\global\newcount\photonfrontx  
\global\newcount\photonfronty  
\global\newcount\photonbackx
\global\newcount\photonbacky
\newcount\halfwigglelength
\global\font\Twelverom=cmr12
\global\font\Tenrom=cmr10
\gdef\Lbr{{\Twelverom(}}   \gdef\Rbr{{\Twelverom)}}
\gdef\SLbr{{\Tenrom(}}     \gdef\SRbr{{\Tenrom)}}
\gdef\Smile{{\large$\smile$}}  
\gdef\Frown{{\large$\frown$}}  
\ifdim\BIGPHOTONS>0pt  \gdef\Smile{$\smile$} \gdef\Frown{$\frown$} \fi
%
\gdef\selectphoton{   
\global\advance\photoncount by 1  
\global\photonfrontx=\particlefrontx   
\global\photonfronty=\particlefronty   
\ifnum\unitboxnumber > 50
\message{   *** WARNING *** Photon with
\the\unitboxnumber\space half-wiggles requested ***   }
\ifnum\unitboxnumber > 150
\message{   *** Reducing photon length to 10 half-wiggles (max 150) ***   }
\ifnum\unitboxnumber > 1000
\message{   *** Probable Cause:  Photon selected instead of Fermion ***   }
\fi \global\unitboxnumber=10 \fi \fi  
\numwiggles=\unitboxnumber
\divide\numwiggles by 2
\global\unitboxnumberpo=\numwiggles 
\global\multiply \unitboxnumberpo by -1
\numwigglespo=\unitboxnumber
\advance\numwigglespo by \unitboxnumberpo 
\global\numlineparts = 2  
\global\numupperunits=\numwigglespo  
\global\numlowerunits=\numwiggles  
\particleadjustx=0  
\particleadjusty=0  
\ifcase\LINEDIRECTION
     \Nphoton    
\or  \NEphoton   
\or  \Ephoton    
\or  \SEphoton   
\or  \Sphoton    
\or  \SWphoton   
\or  \Wphoton    
\or  \NWphoton   
\else\DIRECTERROR \fi
\setplength
\global\divide\plengthx by 2  \global\divide\plengthy by 2
\rearcoords  \boxlengthdefault   \midcoords
\global\photonbackx=\pbackx  
\global\photonbacky=\pbacky  
\global\photonlengthx=\plengthx  
\global\photonlengthy=\plengthy  
}
\gdef\SETUNITBOX(#1)[#2][#3]{ 
\gdef\upperunitbox{\oval(#1,#1)[#2]}
\gdef\lowerunitbox{\oval(#1,#1)[#3]}
}
\gdef\Nphoton{  
\ifcase\LINECONFIGURATION  
\setcoords(-490,-250,0)(260,1250,0)[0,2000]
\gdef\upperunitbox{\SLbr}   \gdef\lowerunitbox{\SRbr}
\particleadjusty=10
\or 
\setcoords(-271,-501,0)(250,1250,0)[0,2000]
\gdef\upperunitbox{\SRbr}   \gdef\lowerunitbox{\SLbr}
\or 
\particleadjusty=0
\setcoords(-501,-351,0)(300,1400,0)[0,2200]
\gdef\upperunitbox{\Lbr}   \gdef\lowerunitbox{\Rbr}
\or 
\setcoords(-353,-499,0)(300,1400,0)[0,2200]
\gdef\upperunitbox{\Rbr}   \gdef\lowerunitbox{\Lbr}
\or 
\setcoords(-481,-371,0)(280,1300,0)[0,2000]
\gdef\upperunitbox{\Lbr}   \gdef\lowerunitbox{\Rbr}
\particleadjusty=150
\ifnum\numwiggles=\number\numwigglespo \particleadjustx=-50 \fi
\or 
\setcoords(-321,-391,0)(280,1300,0)[0,2000]
\gdef\upperunitbox{\Rbr}   \gdef\lowerunitbox{\Lbr}
\particleadjusty=150
\ifnum\numwiggles=\number\numwigglespo \particleadjustx=80 \fi
\or 
\setcoords(-490,-260,0)(300,1500,0)[0,2400]
\gdef\upperunitbox{\Lbr}   \gdef\lowerunitbox{\Rbr}
\or 
\setcoords(-301,-531,0)(300,1500,0)[0,2400]
\gdef\upperunitbox{\Rbr}   \gdef\lowerunitbox{\Lbr}
\else \UNIMPERROR
\fi
}
\gdef\NEphoton{    
\ifcase\LINECONFIGURATION  
\setcoords(425,425,0)(1250,0,0)[1250,1250]       \SETUNITBOX(1250)[br][tl]
\ifnum\numwigglespo > \number \numwiggles \particleadjustx=15 \fi
\or 
\setcoords(1050,-200,0)(625,625,0)[1250,1250]    \SETUNITBOX(1250)[tl][br]
\ifnum\numwigglespo > \number \numwiggles \particleadjustx=25 \fi
\or 
\setcoords(500,500,0)(1400,0,0)[1400,1400]       \SETUNITBOX(1400)[br][tl]
\or 
\setcoords(1200,-200,0)(700,700,0)[1400,1400]    \SETUNITBOX(1400)[tl][br]
\or 
\setcoords(400,400,0)(1200,0,0)[1200,1200]       \SETUNITBOX(1200)[br][tl]
\or 
\setcoords(1000,-200,0)(600,600,0)[1200,1200]    \SETUNITBOX(1200)[tl][br]
\else \UNIMPERROR
\fi
\numupperunits=\numwiggles   \numlowerunits=\numwigglespo
}
\gdef\Ephoton{    
\ifcase\LINECONFIGURATION  
\setcoords(-285,715,0)(-150,-400,0)[2005,0]
\gdef\upperunitbox{\Frown}   \gdef\lowerunitbox{\Smile}
\or  
\setcoords(-285,715,0)(-420,-170,0)[2005,0]
\gdef\upperunitbox{\Smile}   \gdef\lowerunitbox{\Frown}
\else \UNIMPERROR
\fi
\particleadjustx=-15 
}
\gdef\SEphoton{   
\ifcase\LINECONFIGURATION  
\setcoords(-200,1050,0)(-625,-625,0)[1250,-1250] \SETUNITBOX(1250)[tr][bl]
\ifnum\numwigglespo > \number \numwiggles \particleadjustx=25 \fi
\or 
\setcoords(425,425,0)(0,-1250,0)[1250,-1250]     \SETUNITBOX(1250)[bl][tr]
\ifnum\numwigglespo > \number \numwiggles \particleadjustx=15 \fi
\or 
\setcoords(-200,1200,0)(-700,-700,0)[1400,-1400] \SETUNITBOX(1400)[tr][bl]
\or 
\setcoords(500,500,0)(0,-1400,0)[1400,-1400]     \SETUNITBOX(1400)[bl][tr]
\or 
\setcoords(-200,1000,0)(-600,-600,0)[1200,-1200] \SETUNITBOX(1200)[tr][bl]
\particleadjustx=-20
\or 
\setcoords(420,420,0)(0,-1200,0)[1200,-1200]     \SETUNITBOX(1200)[bl][tr]
\particleadjustx=40
\else \UNIMPERROR
\fi
}
\gdef\Sphoton{  
\ifcase\LINECONFIGURATION  
\setcoords(-252,-490,0)(-740,-1740,0)[0,-2000]
\gdef\upperunitbox{\SRbr}   \gdef\lowerunitbox{\SLbr}
\or 
\setcoords(-490,-260,0)(-740,-1740,0)[0,-2002]
\gdef\upperunitbox{\SLbr}   \gdef\lowerunitbox{\SRbr}
\or 
\setcoords(-299,-449,0)(-870,-1970,0)[0,-2200]
\gdef\upperunitbox{\Rbr}    \gdef\lowerunitbox{\Lbr}
\particleadjusty=-95
\or 
\setcoords(-517,-371,0)(-900,-2000,0)[0,-2200]
\gdef\upperunitbox{\Lbr}    \gdef\lowerunitbox{\Rbr}
\particleadjusty=-165
\or 
\setcoords(-299,-409,0)(-885,-1905,0)[0,-2000]
\gdef\upperunitbox{\Rbr}   \gdef\lowerunitbox{\Lbr}
\particleadjustx=50     \particleadjusty=-380
\ifodd\unitboxnumber\relax\else\particleadjustx=250 \particleadjusty=-400 \fi
\or 
\setcoords(-519,-449,0)(-900,-1920,0)[0,-2000]
\gdef\upperunitbox{\Lbr}   \gdef\lowerunitbox{\Rbr}
\particleadjusty=-370
\ifodd\unitboxnumber\relax\else\particleadjustx=-240 \particleadjusty=-400 \fi
\or 
\gdef\upperunitbox{\Rbr}   \gdef\lowerunitbox{\Lbr}
\setcoords(-325,-555,0)(-900,-2100,0)[0,-2400]
\particleadjusty=-40
\or 
\setcoords(-505,-275,0)(-900,-2100,0)[0,-2400]
\gdef\upperunitbox{\Lbr}   \gdef\lowerunitbox{\Rbr}
\particleadjusty=-30  
\else \UNIMPERROR
\fi
}
\gdef\SWphoton{  
\ifcase\LINECONFIGURATION  
\setcoords(-825,-825,0)(0,-1250,0)[-1250,-1250]     \SETUNITBOX(1250)[br][tl]
\or 
\setcoords(-175,-1425,0)(-625,-625,0)[-1250,-1250]  \SETUNITBOX(1250)[tl][br]
\or 
\setcoords(-900,-900,0)(0,-1410,0)[-1400,-1400]     \SETUNITBOX(1400)[br][tl]
\or 
\setcoords(-200,-1600,0)(-700,-700,0)[-1400,-1400]  \SETUNITBOX(1400)[tl][br]
\or 
\setcoords(-800,-800,0)(0,-1200,0)[-1200,-1200]     \SETUNITBOX(1200)[br][tl]
\or 
\setcoords(-200,-1400,0)(-600,-600,0)[-1200,-1200]  \SETUNITBOX(1200)[tl][br]
\else \UNIMPERROR
\fi
}
\gdef\Wphoton{
\ifcase\LINECONFIGURATION 
\setcoords(-2245,-1245,0)(-150,-400,0)[-2005,0]
\gdef\upperunitbox{\Frown}   \gdef\lowerunitbox{\Smile}
\or 
\setcoords(-2245,-1245,0)(-400,-150,0)[-2005,0]
\gdef\upperunitbox{\Smile}   \gdef\lowerunitbox{\Frown}
\else \UNIMPERROR
\fi
\particleadjustx=57 
\ifnum\numwigglespo=\number\numwiggles \particleadjustx=0  \fi
\numlowerunits=\numwigglespo   \numupperunits=\numwiggles
}
\gdef\NWphoton{  
\ifcase\LINECONFIGURATION  
\setcoords(-200,-1425,0)(625,625,0)[-1250,1250]   \SETUNITBOX(1250)[bl][tr]
\or 
\setcoords(-825,-825,0)(0,1250,0)[-1250,1250]     \SETUNITBOX(1250)[tr][bl]
\ifnum\numwigglespo > \number \numwiggles \particleadjusty=-15 \fi
\or 
\setcoords(-200,-1600,0)(700,700,0)[-1400,1400]   \SETUNITBOX(1400)[bl][tr]
\or 
\setcoords(-900,-900,0)(0,1400,0)[-1400,1400]     \SETUNITBOX(1400)[tr][bl]
\or 
\setcoords(-200,-1400,0)(600,600,0)[-1200,1200]   \SETUNITBOX(1200)[bl][tr]
\or 
\setcoords(-800,-800,0)(0,1200,0)[-1200,1200]     \SETUNITBOX(1200)[tr][bl]
\else \UNIMPERROR
\fi
}
  \fi
\selectphoton
}
\gdef\selectgluon{   
\ifnum\gluoncount=0 
\global\newcount\gluonlength
\global\newcount\gluonlengthx
\global\newcount\gluonlengthy
\global\newcount\gluonfrontx  
\global\newcount\gluonfronty  
\global\newcount\gluonbackx
\global\newcount\gluonbacky
%
\gdef\setunitbox(#1)[#2][#3](#4)[#5]{
\gdef\upperunitbox{\oval(#1,#1)[#2]}
\gdef\lowerunitbox{\oval(401,401)[#3]}
\gdef\thirdunitbox{\oval(#4,#4)[#5]}
}
\gdef\selectgluon{  
\global\advance\gluoncount by 1  
\global\gluonfrontx=\particlefrontx   
\global\gluonfronty=\particlefronty   
\global\particleadjustx=0     \global\particleadjusty=0
\ifnum\unitboxnumber > 40
\message{   *** WARNING *** Gluon with
\the\unitboxnumber\space loops requested ***   }
\ifnum\unitboxnumber > 85
\message{   *** Reducing gluon length to 6 loops (max 85) ***   }
\ifnum\unitboxnumber > 1000
\message{   *** Probable Cause:  Gluon selected instead of Fermion ***   }
\fi \global\unitboxnumber=6 \fi \fi  
\global\unitboxnumberpo=\unitboxnumber  
\global\advance\unitboxnumberpo by 1 
\global\numlineparts = 3
\global\numupperunits=\unitboxnumber
\global\numlowerunits=\unitboxnumber
\global\numthirdunits=\unitboxnumber
\ifcase\LINEDIRECTION
\Ngluon    
\or  \NEgluon  
\or  \Egluon   
\or  \SEgluon
\or  \Sgluon
\or  \SWgluon
\or  \Wgluon
\or  \NWgluon
\else\DIRECTERROR \fi
\setparticle
\global\gluonlengthx=\particlelengthx  \global\gluonlengthy=\particlelengthy
\global\gluonbackx=\particlebackx      \global\gluonbacky=\particlebacky
}
\gdef\Ngluon{   
\ifcase\LINECONFIGURATION   
\setcoords(600,540,600)(20,620,1220)[0,1050]
\setunitbox(1600)[tl][r](1600)[bl]
\particleadjusty=195
\or 
\setcoords(-990,-930,-990)(12,615,1215)[0,1050]
\setunitbox(1600)[tr][l](1600)[br]
\particleadjusty=195
\or 
\setcoords(440,390,440)(-10,415,840)[0,850]
\setunitbox(1250)[tl][r](1250)[bl]
\particleadjustx=0
\particleadjusty=-10
\or 
\setcoords(-820,-770,-820)(-25,400,825)[0,850]  
\particleadjusty=-10  
\setunitbox(1250)[tr][l](1250)[br]
\or \UNIMPERROR  
\or \UNIMPERROR  
\or 
\numupperunits=\unitboxnumberpo
\numlowerunits=\unitboxnumber
\numthirdunits=\unitboxnumberpo
\setcoords(-200,-200,-200)(616,1041,616)[0,850]
\setunitbox(1250)[tl][r](1250)[bl]
\particleadjusty=1238
\particleadjusty=1233
\or 
\numupperunits=\unitboxnumberpo
\numlowerunits=\unitboxnumber
\numthirdunits=\unitboxnumberpo
\setcoords(-200,-200,-200)(620,1045,620)[0,850]
\setunitbox(1250)[tr][l](1250)[br]
\particleadjusty=1245
\else \UNIMPERROR 
\fi
}
\gdef\NEgluon{
\numupperunits=\unitboxnumberpo
\numlowerunits=\unitboxnumber
\numthirdunits=\unitboxnumber
\ifcase\LINECONFIGURATION
\setcoords(900,900,900)(0,900,900)[900,900]
\setunitbox(2200)[tl][tr](401)[b]
\particleadjustx=1100     \particleadjusty=1100
\or 
\setcoords(-180,720,720)(1090,1091,1091)[900,900]
\setunitbox(2200)[br][tr](401)[l]
\particleadjustx=1110     \particleadjusty=1050
\else \UNIMPERROR 
\fi
}
\gdef\Egluon{     
\ifcase\LINECONFIGURATION
\setcoords(-210,390,990)(-800,-745,-800)[1050,0]  
\setunitbox(1600)[tr][b](1600)[tl]
\particleadjustx=130  
\or 
\setcoords(-210,390,990)(800,745,800)[1050,0]  
\setunitbox(1600)[br][t](1600)[bl]
\particleadjustx=130
\or 
\setcoords(-200,225,650)(-625,-575,-625)[850,0]
\setunitbox(1250)[tr][b](1250)[tl]
\or 
\setcoords(-200,225,650)(625,575,625)[850,0]
\setunitbox(1250)[br][t](1250)[bl]
\or 
\setcoords(-200,430,1060)(-830,-780,-830)[1260,0]
\setunitbox(1660)[tr][b](1660)[tl]
\or 
\setcoords(-200,430,1060)(830,780,830)[1260,0]
\setunitbox(1660)[br][t](1660)[bl]
\or 
\numupperunits=\unitboxnumberpo
\numlowerunits=\unitboxnumber
\numthirdunits=\unitboxnumberpo
\setcoords(440,865,440)(0,50,0)[850,0]
\setunitbox(1250)[tr][b](1250)[tl]
\particleadjustx=1260
\or 
\numupperunits=\unitboxnumberpo
\numlowerunits=\unitboxnumber
\numthirdunits=\unitboxnumberpo
\setcoords(430,855,430)(0,-50,0)[850,0]
\setunitbox(1250)[br][t](1250)[bl]
\particleadjustx=1250
\or 
\setcoords(-160,440,1040)(-600,-550,-600)[1200,0]
\gdef\upperunitbox{\oval(1600,1200)[tr]}
\gdef\thirdunitbox{\oval(1600,1200)[tl]}
\gdef\lowerunitbox{\oval(401,401)[b]}
\else \UNIMPERROR
\fi
}
\gdef\SEgluon{
\numupperunits=\unitboxnumberpo
\numlowerunits=\unitboxnumber
\numthirdunits=\unitboxnumber
\ifcase\LINECONFIGURATION
\setcoords(-200,700,700)(-1100,-1100,-1100)[900,-900]
\setunitbox(2200)[tr][br](401)[l]
\particleadjustx=1100     \particleadjusty=-1100
\or 
\setcoords(890,890,890)(0,-900,-900)[900,-900]
\setunitbox(2200)[bl][br](401)[t]
\particleadjustx=1050     \particleadjusty=-1100
\else \UNIMPERROR 
\fi
}
\gdef\Sgluon{   
\ifcase\LINECONFIGURATION  
\setcoords(-1000,-940,-1000)(0,-595,-1195)[0,-1050]
\setunitbox(1600)[br][l](1600)[tr]
\particleadjusty=-150
\or 
\setcoords(605,545,605)(-20,-615,-1215)[0,-1050]
\setunitbox(1600)[bl][r](1600)[tl]
\particleadjusty=-150
\or 
\setcoords(-820,-770,-820)(0,-425,-850)[0,-850]
\setunitbox(1250)[br][l](1250)[tr]
\or 
\setcoords(440,390,440)(0,-425,-850)[0,-850]
\setunitbox(1250)[bl][r](1250)[tl]
\or \UNIMPERROR 
\or \UNIMPERROR
\or 
\numupperunits=\unitboxnumberpo
\numlowerunits=\unitboxnumber
\numthirdunits=\unitboxnumberpo
\setcoords(-180,-180,-180)(-635,-1060,-635)[0,-850]
\setunitbox(1250)[br][l](1250)[tr]
\particleadjusty=-1290
\or 
\numupperunits=\unitboxnumberpo
\numlowerunits=\unitboxnumber
\numthirdunits=\unitboxnumberpo
\setcoords(-180,-180,-180)(-635,-1060,-635)[0,-850]
\setunitbox(1250)[bl][r](1250)[tl]
\particleadjusty=-1290
\else \UNIMPERROR 
\fi
}
\gdef\SWgluon{
\numupperunits=\unitboxnumberpo
\numlowerunits=\unitboxnumber
\numthirdunits=\unitboxnumber
\ifcase\LINECONFIGURATION
\setcoords(-1300,-1300,-1300)(0,-900,-900)[-900,-900]
\setunitbox(2200)[br][bl](401)[t]
\particleadjustx=-1100     \particleadjusty=-1100
\or 
\setcoords(-215,-1115,-1115)(-1107,-1107,-1107)[-900,-900]
\setunitbox(2200)[tl][bl](401)[r]
\particleadjustx=-1120     \particleadjusty=-1120
\else \UNIMPERROR 
\fi
}
\gdef\Wgluon{   
\ifcase\LINECONFIGURATION
\setcoords(-190,-790,-1390)(800,745,800)[-1050,0]
\setunitbox(1600)[bl][t](1600)[br]
\particleadjustx=-150  
\or 
\setcoords(-190,-790,-1390)(-800,-745,-800)[-1050,0]
\setunitbox(1600)[tl][b](1600)[tr]
\particleadjustx=-150  
\or 
\setcoords(-200,-625,-1050)(625,575,625)[-850,0]
\setunitbox(1250)[bl][t](1250)[br]
\or 
\setcoords(-200,-625,-1050)(-625,-575,-625)[-850,0]
\setunitbox(1250)[tl][b](1250)[tr]
\or 
\setcoords(-230,-860,-1490)(830,780,830)[-1260,0]
\setunitbox(1660)[bl][t](1660)[br]
\or 
\setcoords(-230,-860,-1490)(-830,-780,-830)[-1260,0]
\setunitbox(1660)[tl][b](1660)[tr]
\or 
\numupperunits=\unitboxnumberpo
\numlowerunits=\unitboxnumber
\numthirdunits=\unitboxnumberpo
\setcoords(-825,-1250,-825)(0,-50,0)[-850,0]
\setunitbox(1250)[bl][t](1250)[br]
\particleadjustx=-1250
\or  
\numupperunits=\unitboxnumberpo
\numlowerunits=\unitboxnumber
\numthirdunits=\unitboxnumberpo
\setcoords(-825,-1250,-825)(0,50,0)[-850,0]
\setunitbox(1250)[tl][b](1250)[tr]
\particleadjustx=-1250
\else \UNIMPERROR 
\fi
}
\gdef\NWgluon{
\numupperunits=\unitboxnumberpo
\numlowerunits=\unitboxnumber
\numthirdunits=\unitboxnumber
\ifcase\LINECONFIGURATION
\setcoords(-200,-1100,-1100)(1100,1100,1100)[-900,900]
\setunitbox(2200)[bl][tl](401)[r]
\particleadjustx=-1110   \particleadjusty=1100
\or  
\setcoords(-1309,-1309,-1309)(-15,885,885)[-900,900]
\setunitbox(2200)[tr][tl](401)[b]
\particleadjustx=-1120   \particleadjusty=1065
\else \UNIMPERROR 
\fi
}
%
%
%
\gdef\gluonlink{    
%
%
%
%
%
%
\gdef\gluonlink{
\global\stemlengthx=401 \SETDIR \setadjxy \divide\adjx by -2 \divide\adjy by -2
\ifcase\LDIR
     \linksetupB \ifodd\LCONFIG\LINKPUT[l]  \else\LINKPUT[r] \fi 
\or  \ifodd\LCONFIG \linksetupBx \LINKPUT[tr]\LINKPUT[l]         
     \else          \linksetupBy \LINKPUT[tr]\LINKPUT[b]     \fi
\or  \linksetupB \ifodd\LCONFIG\LINKPUT[t]  \else\LINKPUT[b] \fi 
\or  \ifodd\LCONFIG \linksetupBy \LINKPUT[br]\LINKPUT[t]         
     \else          \linksetupBx \LINKPUT[br]\LINKPUT[l]     \fi
\or  \linksetupB \ifodd\LCONFIG\LINKPUT[r]  \else\LINKPUT[l] \fi 
\or  \ifodd\LCONFIG \linksetupBx \LINKPUT[bl]\LINKPUT[r]         
     \else          \linksetupBy \LINKPUT[bl]\LINKPUT[t]     \fi
\or  \linksetupB \ifodd\LCONFIG\LINKPUT[b]  \else\LINKPUT[t] \fi 
\or  \ifodd\LCONFIG \linksetupBy \LINKPUT[tl]\LINKPUT[b]         
     \else          \linksetupBx \LINKPUT[tl]\LINKPUT[r]     \fi
\else \UNIMPERROR
\fi
\ifodd\LDIR\relax
\else \global\advance\pbackx by \adjx   \global\advance\pbacky by \adjy \fi
\linksetupC
}
%
%
%
\gdef\gluoncap{  
\global\stemlengthx=0
\ifodd\LDIR\message{NOTE:  Diagonal Gluons are not Capped}\relax
\else
\ifcase\LCONFIG \global\stemlengthx=1000  
\or \global\stemlengthx=1000  
\or \global\stemlengthx=825   
\or \global\stemlengthx=825   
\or \global\stemlengthx=1030  
\or \global\stemlengthx=1030  
\or \global\stemlengthx=0     
\or \global\stemlengthx=0     
\or \global\stemlengthx=800   
\else\UNIMPERROR
\fi 
\ifnum\stemlengthx>400
\global\advance\LDIR by 2 \moduloeight\LDIR   \SETDIR
\global\advance\LDIR by 6 \moduloeight\LDIR   \setadjxy
\ifodd\LCONFIG \multiply \adjx by -1   \multiply \adjy by -1 \fi
\divide\adjx by 2  \divide\adjy by 2 \linksetupB
\ifcase\LDIR
     \ifodd\LCONFIG\LINKPUT[tr]  \else\LINKPUT[tl]   \fi   
\or  \relax 
\or  \ifodd\LCONFIG\LINKPUT[br]  \else\LINKPUT[tr]   \fi   
\or  \relax 
\or  \ifodd\LCONFIG\LINKPUT[bl]  \else\LINKPUT[br]   \fi   
\or  \relax 
\or  \ifodd\LCONFIG\LINKPUT[tl]  \else\LINKPUT[bl]   \fi   
\or  \relax 
\else \UNIMPERROR
\fi
\linksetupD
\else\message{* NOTE:  Attempt to use gluoncap of less that 401 centipoints *}
\fi 
\fi 
}
%
%
\gdef\setadjxy{
\adjx=\stemlengthx   \adjy=\stemlengthx 
\multiply \adjx by \XDIR%
\multiply \adjy by \YDIR%
}
\gdef\advancegluonlength{
\global\advance\particlelengthx by \adjx \global\advance\particlelengthy
by\adjy
\global\advance\gluonlengthx by \adjx  \global\advance\gluonlengthy by \adjy
}
\gdef\LINKPUT[#1]{\ifnum\phantomswitch=0\put(\gluonbackx,\gluonbacky)
{\oval(\stemlengthx,\stemlengthx)[#1]}\fi}
\gdef\linksetupBx{\global\advance\gluonbackx by \adjx}
\gdef\linksetupBy{\global\advance\gluonbacky by \adjy}
\gdef\linksetupB{\linksetupBx\linksetupBy}
\gdef\linksetupC{
\global\advance\pbackx by \adjx  \global\advance\pbacky by \adjy
\global\gluonbackx=\pbackx   \global\gluonbacky=\pbacky
\ifnum\plengthx<0 \multiply\adjx by -1 \fi
\ifnum\plengthy<0 \multiply\adjy by -1 \fi
\advancegluonlength
}
\gdef\linksetupD{
\linksetupC
\SETDIR
\setadjxy
\divide\adjx by 2  \divide\adjy by 2 \linksetupB
\global\pbackx=\gluonbackx   \global\pbacky=\gluonbacky
\advancegluonlength
\ifnum\phantomswitch=0\put(\pbackx,\pbacky){\line(\XDIR,\YDIR){\stemlength}}\fi
\stemlengthx=\stemlength  
\setadjxy
\global\advance\pbackx by \adjx  \global\advance\pbacky by \adjy
\global\gluonbackx=\pbackx   \global\gluonbacky=\pbacky
\advancegluonlength
}
\gluonlink}  
\gdef\gluoncap{    
\gluoncap}  
  \fi
\selectgluon
}
\gdef\selectspecial{\UNIMPERROR}
%
%
\gdef\checkvertex{ 
\ifnum\vertexcount=-1   
%
\global\advance\vertexcount by 1 
\newsavebox{\vertexbox}
\global\newcount\LDIRcount      
\global\newcount\VERTEXNUMBER   
\global\newdimen\VERTEXLINKONE  
\global\newdimen\VERTEXLINKTWO
\global\newdimen\VERTEXLINKTHREE
\global\newdimen\VERTEXLINKFOUR 
\global\newdimen\VERTEXCAPONE   
\global\newdimen\VERTEXCAPTWO
\global\newdimen\VERTEXCAPTHREE
\global\newdimen\VERTEXCAPFOUR  
\global\newdimen\STEMVERTEXONE
\global\newdimen\STEMVERTEXTWO
\global\newdimen\STEMVERTEXTHREE
\global\newdimen\STEMVERTEXFOUR 
\global\newcount\stemlengthcopy
\global\newcount\vertexonex    \global\newcount\vertexoney
\global\newcount\vertextwox    \global\newcount\vertextwoy
\global\newcount\vertexthreex  \global\newcount\vertexthreey
\global\newcount\vertexfourx   \global\newcount\vertexfoury
\global\newcount\vertexmidx    \global\newcount\vertexmidy
\global\newcount\VERTEXLINE
\global\newcount\FLIPVERTEX    \global\FLIPVERTEX=0
\gdef\flipvertex{\global\FLIPVERTEX=1}  
\newcount\vertadj              \newcount\negvertadj
\gdef\drawvertex#1[#2#3](#4,#5)[#6]{
\global\advance\vertexcount by 1 
\global\LINETYPE=#1           
\global\LINEDIRECTION=#2
\global\VERTEXNUMBER=#3  
\global\vertexonex=#4
\global\vertexoney=#5
\global\unitboxnumber=#6
\global\stemlengthcopy=\stemlength   
%
%
\ifnum\LINETYPE<3 \LINEERROR \fi
\ifnum\LINETYPE=3   
\ifnum\gluoncount=0 \def\gluonlink{\relax} \def\gluoncap{\relax} \fi
  \ifnum\VERTEXNUMBER<3 \UNIMPERROR \fi
  \ifnum\VERTEXNUMBER=3 \THREEPHOTON\fi 
  \ifnum\VERTEXNUMBER=4 \FOURPHOTON \fi 
  \ifnum\VERTEXNUMBER>4 \UNIMPERROR \fi
\fi
\ifnum\LINETYPE=4   
  \ifnum\VERTEXNUMBER<3 \UNIMPERROR \fi
  \ifnum\VERTEXNUMBER=3 \THREEGLUON \fi 
  \ifnum\VERTEXNUMBER=4 \FOURGLUON  \fi 
  \ifnum\VERTEXNUMBER>4 \UNIMPERROR \fi
\fi
\ifnum\LINETYPE>4 \LINEERROR \fi
\clearvertex   
}  
%
\gdef\advtwomodeight#1{
\global\advance\LDIRcount by2\moduloeight\LDIRcount
\diagFOURVERT#1[\LDIRcount]}
%
\gdef\THREEPHOTON{
\ifcase\LDIR  
\setvertexA[\S\REG]
\setvertexB[\NW\CURLY](0,0)[2]  \setvertexB[\NE\FLIPPEDCURLY](0,0)[3]
\or  
\setvertexA[\SW\CURLY]
\setvertexB[\N\REG](70,-100)[2] \setvertexB[\E\FLIPPED](0,0)[3]
\or  
\setvertexA[\W0]
\setvertexB[\NE\CURLY](20,0)[2] \setvertexB[\SE\FLIPPEDCURLY](20,0)[3]
\or  
\setvertexA[\NW\CURLY]
\setvertexB[\E\REG](0,0)[2] \setvertexB[\S1](20,100)[3]
\or  
\setvertexA[\N\REG]
\setvertexB[\SE\CURLY](0,0)[2]  \setvertexB[\SW\FLIPPEDCURLY](0,0)[3]
\or  
\setvertexA[\NE\CURLY]
\setvertexB[\S\REG](-20,100)[2]
\setvertexB[\W\FLIPPED](0,0)[3]
\or  
\setvertexA[\E\REG]
\setvertexB[\SW\CURLY](0,0)[2]  \setvertexB[\NW\FLIPPEDCURLY](0,0)[3]
\or  
\setvertexA[\SE\CURLY]
\setvertexB[\W\REG](0,0)[2]
\setvertexB[\N\FLIPPED](-40,-100)[3]
\else \DIRECTERROR
\fi
} 
%
%
\gdef\FOURPHOTON{
\ifnum\LDIR>-1
\global\LDIRcount=\LDIR  \global\advance\LDIRcount by 4 \moduloeight\LDIRcount
\setvertexA[\LDIRcount\REG] \advtwomodeight2 \advtwomodeight3 \advtwomodeight4
\else \UNIMPERROR \fi
\global\FLIPVERTEX=0
}
%
%
\gdef\THREEGLUON{
\vertadj=0   \adjvert  
\ifcase\LDIR  
\setvertexA[\S\CENTRAL]
\setvertexB[\NW\REG](0,0)[2]  \setvertexB[\NE\FLIPPED](0,0)[3]
\or  
\setvertexA[\SW\REG]
\setvertexB[\N\CURLY](-170,442)[2] \setvertexB[\E 3](420,-183)[3]
\setvertexC(442,442)[bl]
\or  
\setvertexA[\W6]
\setvertexB[\NE\REG](0,0)[2] \setvertexB[\SE\FLIPPED](0,0)[3]
\or  
\setvertexA[\NW\REG]
\setvertexB[\E\CURLY](420,183)[2]  \setvertexB[\S3](-183,-442)[3]
\setvertexC(442,-442)[tl]
\or  
\setvertexA[\N\CENTRAL]
\setvertexB[\SE\REG](0,0)[2]  \setvertexB[\SW\FLIPPED](0,0)[3]
\or  
\setvertexA[\NE\REG]
\setvertexB[\S\CURLY](170,-442)[2]
\setvertexB[\W\FLIPPEDCURLY](-420,183)[3]
\setvertexC(-442,-442)[tr]
\or  
\setvertexA[\E\CENTRAL]
\setvertexB[\SW\REG](0,0)[2]  \setvertexB[\NW\FLIPPED](0,0)[3]
\or  
\setvertexA[\SE\REG]
\setvertexB[\W\CURLY](-420,-183)[2]
\setvertexB[\N\FLIPPEDCURLY](170,442)[3]
\setvertexC(-442,442)[br]
\else \DIRECTERROR
\fi
} 
%
%
%
%
\gdef\FOURGLUON{
\ifodd\LDIR\vertadj=0 \else  \vertadj=412 \fi    \adjvert
\ifcase\LDIR  
\setvertexA[\S\CURLY]         \MIDADJUST(\negvertadj,\vertadj)
\WFOURVERT2    \NFOURVERT3    \EFOURVERT4
\or \FOURPHOTON  
\or  
\setvertexA[\W\CURLY]         \MIDADJUST(\vertadj,\vertadj)
\NFOURVERT2  \EFOURVERT3  \SFOURVERT4
\or \FOURPHOTON    
\or  
\setvertexA[\N\CURLY]         \MIDADJUST(\vertadj,\negvertadj)
\EFOURVERT2   \SFOURVERT3   \WFOURVERT4
\or \FOURPHOTON    
\or  
\setvertexA[\E\CURLY]         \MIDADJUST(\negvertadj,\negvertadj)
\SFOURVERT2  \WFOURVERT3  \NFOURVERT4
\or \FOURPHOTON    
\else \DIRECTERROR
\fi
\global\FLIPVERTEX=0   
} 
%
%
\gdef\MIDADJUST(#1,#2){
\ifnum\FLIPVERTEX=0
\global\advance\vertexmidx by #1   \global\advance\vertexmidy by #2
\setvertexC(0,\vertadj)[bl]        \setvertexC(0,\negvertadj)[tr]
\setvertexC(\vertadj,0)[tl]        \setvertexC(\negvertadj,0)[br]
\else  
\global\particleadjustx=#1  \global\particleadjusty=#2
\ifnum\LDIR=0 \global\multiply\particleadjustx by -1 \fi
\ifnum\LDIR=2 \global\multiply\particleadjusty by -1 \fi
\ifnum\LDIR=4 \global\multiply\particleadjustx by -1 \fi
\ifnum\LDIR=6 \global\multiply\particleadjusty by -1 \fi
\global\advance\vertexmidx by \particleadjustx
\global\advance\vertexmidy by \particleadjusty
\setvertexC(0,\negvertadj)[tl]        \setvertexC(0,\vertadj)[br]
\setvertexC(\negvertadj,0)[tr]        \setvertexC(\vertadj,0)[bl]
\fi
}
\gdef\NFOURVERT#1{
  \ifnum\FLIPVERTEX=0 \setvertexB[\N\CURLY](\negvertadj,\vertadj)[#1]
  \else \setvertexB[\N\FLIPPEDCURLY](\vertadj,\vertadj)[#1] \fi}
\gdef\SFOURVERT#1{
  \ifnum\FLIPVERTEX=0 \setvertexB[\S\CURLY](\vertadj,\negvertadj)[#1]
  \else \setvertexB[\S\FLIPPEDCURLY](\negvertadj,\negvertadj)[#1] \fi}
\gdef\EFOURVERT#1{
  \ifnum\FLIPVERTEX=0 \setvertexB[\E\CURLY](\vertadj,\vertadj)[#1]
  \else \setvertexB[\E\FLIPPEDCURLY](\vertadj,\negvertadj)[#1] \fi}
\gdef\WFOURVERT#1{
  \ifnum\FLIPVERTEX=0 \setvertexB[\W\CURLY](\negvertadj,\negvertadj)[#1]
  \else \setvertexB[\W\FLIPPEDCURLY](\negvertadj,\vertadj)[#1] \fi}
\gdef\diagFOURVERT#1[#2]{\setvertexB[#2\FLIPVERTEX](0,0)[#1]}
%
%
%
\gdef\setvertexA[#1#2]{
\global\savebox\vertexbox(0,0){
\begin{picture}(0,0)
\global\adjx=#2
\ifnum\FLIPVERTEX=1 \global\advance\adjx by 1
      \ifnum\VERTEXNUMBER=3 \global\FLIPVERTEX=0 \fi
  \fi  
\ifdim\STEMVERTEXONE=1pt\backstemmed \fi
\drawline\LINETYPE[#1\adjx](0,0)[\unitboxnumber]
\global\stemlength=\stemlengthcopy  
\ifdim\VERTEXLINKONE=1pt\gluonlink \fi 
\ifdim\VERTEXCAPONE=1pt\gluoncap \fi 
\global\vertexmidx=\particlebackx  \global\vertexmidy=\particlebacky
\end{picture}
}
\global\multiply\vertexmidx by -1  \global\multiply\vertexmidy by -1
\global\advance\vertexmidx by \vertexonex
\global\advance\vertexmidy by \vertexoney
\ifdim\STEMVERTEXONE=1pt\backstemmed \fi
\drawline\LINETYPE[#1\adjx](\vertexmidx,\vertexmidy)[\unitboxnumber]
\global\stemlength=\stemlengthcopy  
\ifdim\VERTEXLINKONE=1pt\gluonlink \fi 
\ifdim\VERTEXCAPONE=1pt\gluoncap \fi 
} 
\gdef\setvertexB[#1#2](#3,#4)[#5]{
\global\adjx=\vertexmidx   \global\adjy=\vertexmidy
\global\advance\adjx by #3   \global\advance\adjy by #4
\VERTEXLINE=#5
\ifcase\VERTEXLINE\UNIMPERROR  
\or \UNIMPERROR  
\or \ifdim\STEMVERTEXTWO=1pt\backstemmed \fi  
\or \ifdim\STEMVERTEXTHREE=1pt\backstemmed\fi 
\or \ifdim\STEMVERTEXFOUR=1pt\backstemmed\fi  
\else \UNIMPERROR                             
\fi
\drawline\LINETYPE[#1#2](\adjx,\adjy)[\unitboxnumber]
\global\stemlength=\stemlengthcopy  
\ifcase\VERTEXLINE\UNIMPERROR  
\or \UNIMPERROR  
\or \ifdim\VERTEXLINKTWO=1pt\gluonlink \fi 
    \ifdim\VERTEXCAPTWO=1pt\gluoncap \fi 
    \global\vertextwox=\particlebackx  \global\vertextwoy=\particlebacky
\or \ifdim\VERTEXLINKTHREE=1pt\gluonlink\fi 
    \ifdim\VERTEXCAPTHREE=1pt\gluoncap\fi 
    \global\vertexthreex=\particlebackx  \global\vertexthreey=\particlebacky
\or \ifdim\VERTEXLINKFOUR=1pt\gluonlink\fi 
    \ifdim\VERTEXCAPFOUR=1pt\gluoncap\fi 
    \global\vertexfourx=\particlebackx  \global\vertexfoury=\particlebacky
\else \UNIMPERROR
\fi
}
\gdef\setvertexC(#1,#2)[#3#4]{
\global\adjx=\vertexmidx   \global\adjy=\vertexmidy
\global\advance\adjx by #1   \global\advance\adjy by #2
\absstemlength=1250  
\ifnum \VERTEXNUMBER=4 \absstemlength=\vertadj\double\absstemlength \fi
\ifnum\phantomswitch=0
   \put(\adjx,\adjy) {\oval(\absstemlength,\absstemlength)[#3#4]}\fi
}
\gdef\adjvert{\negvertadj=\vertadj  \multiply\negvertadj by -1}
%
%
\gdef\vertexlink#1{
\global\VERTEXLINE=#1
\ifcase\VERTEXLINE\UNIMPERROR
\or \global\VERTEXLINKONE=1pt    \or \global\VERTEXLINKTWO=1pt
\or \global\VERTEXLINKTHREE=1pt  \or \global\VERTEXLINKFOUR=1pt
\else\UNIMPERROR\fi}
\gdef\vertexlinks{
\global\VERTEXLINKONE=1pt     \global\VERTEXLINKTWO=1pt
\global\VERTEXLINKTHREE=1pt  \global\VERTEXLINKFOUR=1pt  }
%
%
%
\gdef\vertexcap#1{
\global\VERTEXLINE=#1
\ifcase\VERTEXLINE\UNIMPERROR
\or \global\VERTEXCAPONE=1pt    \or \global\VERTEXCAPTWO=1pt
\or \global\VERTEXCAPTHREE=1pt  \or \global\VERTEXCAPFOUR=1pt
\else\UNIMPERROR\fi}
\gdef\vertexcaps{
\global\VERTEXCAPONE=1pt     \global\VERTEXCAPTWO=1pt
\global\VERTEXCAPTHREE=1pt  \global\VERTEXCAPFOUR=1pt  }
%
%
%
\gdef\stemvertex#1{
\global\VERTEXLINE=#1
\ifcase\VERTEXLINE\UNIMPERROR
\or \global\STEMVERTEXONE=1pt    \or \global\STEMVERTEXTWO=1pt
\or \global\STEMVERTEXTHREE=1pt  \or \global\STEMVERTEXFOUR=1pt
\else\UNIMPERROR\fi}
\gdef\stemvertices{
\global\STEMVERTEXONE=1pt     \global\STEMVERTEXTWO=1pt
\global\STEMVERTEXTHREE=1pt  \global\STEMVERTEXFOUR=1pt  }
\gdef\clearvertex{
\global\STEMVERTEXONE=0pt    \global\STEMVERTEXTWO=0pt
\global\STEMVERTEXTHREE=0pt  \global\STEMVERTEXFOUR=0pt
\global\VERTEXLINKONE=0pt    \global\VERTEXLINKTWO=0pt
\global\VERTEXLINKTHREE=0pt  \global\VERTEXLINKFOUR=0pt
\global\VERTEXCAPONE=0pt    \global\VERTEXCAPTWO=0pt
\global\VERTEXCAPTHREE=0pt  \global\VERTEXCAPFOUR=0pt
\global\stemlength=275}  
\global\stemlengthcopy=\stemlength
\clearvertex  
\global\stemlength=\stemlengthcopy
  \fi}
\gdef\drawvertex#1[#2#3](#4,#5)[#6]{\checkvertex\drawvertex#1[#2#3](#4,#5)[#6]}
\gdef\vertexcap#1{\checkvertex\vertexcap#1}
\gdef\vertexcaps{\checkvertex\vertexcaps}
\gdef\vertexlink#1{\checkvertex\vertexlink#1}
\gdef\vertexlinks{\checkvertex\vertexlinks}
\gdef\stemvertex#1{\checkvertex\stemvertex#1}
\gdef\stemvertices{\checkvertex\stemvertices}
\gdef\flipvertex{\checkvertex\flipvertex}
%
%
\global\arrowlength=349  
\gdef\drawarrow[#1#2](#3,#4){
\global\LDIR=#1
\SETDIR
\global\boxlengthx=#3  
\global\boxlengthy=#4  
\ifdim#2=1pt  
\adjx=\arrowlength      \adjy=\arrowlength
\multiply\adjx by \XDIR \multiply\adjy by \YDIR  
\slanttest(\adjx,\adjy)
\global\advance\boxlengthx by \adjx    \global\advance\boxlengthy by \adjy
\fi
\ifnum\phantomswitch=0\put(\boxlengthx,\boxlengthy){\vector(\XDIR,\YDIR){0}}\fi
}  
%
%
\gdef\SETFRONTSTEM{
\EITHERSTEM=\FRONTSTEM   \advance\EITHERSTEM by \BACKSTEM
\ifdim\EITHERSTEM>0pt
\global\stemlengthx=\stemlength   \global\stemlengthy=\stemlength
\global\absstemlength=\stemlength
\SETDIR
\gslanttest(\stemlengthx,\stemlengthy)
\gslanttest(\absstemlength,\REG)  
\ifnum\XDIR=0 \stemlengthx=0 \fi
\ifnum\YDIR=0 \stemlengthy=0 \fi
\global\multiply\stemlengthx by \XDIR
\global\multiply\stemlengthy by \YDIR
\ifdim\FRONTSTEM=1pt
\ifnum\phantomswitch=0
          \put(\pfrontx,\pfronty){\line(\XDIR,\YDIR){\absstemlength}}\fi
\global\advance\plengthx by \stemlengthx
\global\advance\plengthy by \stemlengthy
\global\advance\PFRONTx by \stemlengthx
\global\advance\PFRONTy by \stemlengthy
\global\advance\pmidx by \stemlengthx
\global\advance\pmidy by \stemlengthy
\global\advance\pbackx by \stemlengthx
\global\advance\pbacky by \stemlengthy
\ifnum\LTYPE=3
\global\photonfrontx=\PFRONTx  \global\photonfronty=\PFRONTy
\global\photonbackx=\pbackx    \global\photonbacky=\pbacky
\fi  
\ifnum\LTYPE=4
\global\gluonfrontx=\PFRONTx  \global\gluonfronty=\PFRONTy
\global\gluonbackx=\pbackx    \global\gluonbacky=\pbacky
\fi  
\fi  
\fi  
}    
\gdef\SETBACKSTEM{
\ifdim\BACKSTEM=1pt
\ifnum\phantomswitch=0
       \put(\pbackx,\pbacky){\line(\XDIR,\YDIR){\absstemlength}}\fi
\global\advance\plengthx by \stemlengthx
\global\advance\plengthy by \stemlengthy
\global\advance\pbackx by \stemlengthx
\global\advance\pbacky by \stemlengthy
\fi  
\global\stemlength=275  \FRONTSTEM=0pt  \BACKSTEM=0pt 
}    
\gdef\drawloop#1[#2#3](#4,#5){  
\global\newcount\loopfrontx    \global\newcount\loopfronty
\global\newcount\loopbackx    \global\newcount\loopbacky
\global\newcount\loopmidx    \global\newcount\loopmidy
\global\newdimen\CENTRALLOOP
\gdef\drawloop#1[#2#3](#4,#5){
\global\CENTRALLOOP=0pt  
\global\LINETYPE=#1
\ifnum\LTYPE=\gluon\relax\else\UNIMPERROR\LTYPE=1\message{Reverting to Gluons}
\fi
\global\LINEDIRECTION=#2  
\global\fourthlineadjx=#3 
\ifnum\fourthlineadjx=0 
  \global\CENTRALLOOP=1pt  
  \global\fourthlineadjx=8
  \global\LDIR=0
\fi
\global\fourthlineadjy=\fourthlineadjx  
\global\advance\fourthlineadjy by -4
\global\loopfrontx=#4   \global\loopfronty=#5
\ifdim\CENTRALLOOP=1pt
  \global\advance\loopfrontx by -2413  \global\advance\loopfronty by -425
\fi                          
\global\unitboxnumber=1  
\ifnum\LINETYPE=\photon \unitboxnumber=2 \fi
\checkdir
\drawline\LINETYPE[\LDIR\LCONFIG](\loopfrontx,\loopfronty)[\unitboxnumber]
\DRAWLOOP
\ifnum\fourthlineadjy>-1 
\global\loopmidx=\loopfrontx   \global\loopmidy=\loopfronty
\global\advance\loopmidx by \loopbackx  \global\advance\loopmidy by \loopbacky
\divide\loopmidx  by 2 \divide\loopmidy by 2  
\ifdim\CENTRALLOOP=1pt
  \global\advance\loopfrontx by 200    \global\advance\loopfronty by 425
  \global\advance\loopbackx by -200    \global\advance\loopbacky by -425
\fi
\fi 
}
\gdef\DRAWLOOP{
\global\advance\fourthlineadjx by -1
\ifnum\fourthlineadjx=0\relax  
\else
\ifnum\fourthlineadjx=\fourthlineadjy 
   \global\loopbackx=\pbackx   \global\loopbacky=\pbacky
\fi
\global\advance\LDIR by 1
\moduloeight\LDIR
\checkdir
\drawline\LINETYPE[\LDIR\LCONFIG](\pbackx,\pbacky)[\unitboxnumber]
\fi 
\ifnum\fourthlineadjx>1 \DRAWLOOP  \fi  
}
\gdef\checkdir{
\ifnum\LTYPE=\gluon
\ifodd\LDIR \global\LCONFIG=0 \else \global\LCONFIG=2 \fi
\fi 
}

\drawloop#1[#2#3](#4,#5)}
\Feynmanlength  

\newcount\Scalarlength
\newcount\Scalarlengthx
\newcount\Scalarlengthy
\newcount\Scalarfrontx  
\newcount\Scalarfronty  
\newcount\Scalarbackx
\newcount\Scalarbacky
\gdef\ALLScalar{
\global\Scalarfrontx=\particlefrontx   
\global\Scalarfronty=\particlefronty   
\numlineparts = 1      \numupperunits=\unitboxnumber
\ifcase\LINECONFIGURATION
\global\upperlineadjx=-200     \global\upperlineadjy=0
\slanttest(\seglength,\gaplength)   
\gdef\upperunitbox{\line(\XDIR,\YDIR){\seglength}}
\else \UNIMPERROR 
\fi
\global\unitboxwidth=\seglength  \global\advance\unitboxwidth by \gaplength
\global\multiply \unitboxwidth by \XDIR
\global\unitboxheight=\seglength  \global\advance\unitboxheight by \gaplength
\global\multiply \unitboxheight by \YDIR
\global\particleadjustx=\gaplength \global\multiply\particleadjustx by \XDIR
\global\particleadjusty=\gaplength \global\multiply\particleadjusty by \YDIR
\negate\particleadjustx   \negate\particleadjusty   
\setparticle  
\global\Scalarlengthx=\particlelengthx  
\global\Scalarlengthy=\particlelengthy  
\ifnum\boxlengthx > 50000
\message{   *** WARNING *** Scalar of length in excess of 50000cp
requested!}\fi
\ifnum\boxlengthy > 50000
\message{   *** WARNING *** Scalar of length in excess of 50000cp
requested!}\fi
\global\Scalarbackx=\pbackx      \global\Scalarbacky=\pbacky   
}

\doublespacing


\begin{preliminary}

\maketitle

\begin{abstract}
The Higgs boson is the last remaining component of the Standard Model
of Particle Physics to be discovered.
Within the context of the Standard Model, the
Higgs boson is responsible for the breaking of the $\mathrm{SU_L(2)
\times U(1)}$ gauge symmetry and provides a mechanism for the
generation of the masses of the corresponding gauges bosons, the $W^\pm$
and the $Z$.  In addition, this same mechanism provides masses for the
leptons and quarks via Yukawa couplings.  Therefore, the discovery and
subsequent study of the Higgs boson and its properties is of the
highest priority in particle physics today.

The dominant production mechanism for Higgs bosons in hadron-hadron
colliders is via gluon-gluon fusion, in which gluons fuse via a
virtual top quark to produce a Higgs. At the Large hadron Collider,
which has a center of mass of approximately 14 TeV, one expects
4000-40000 of these particles to be produced via this process in one
year of running. At leading order in Quantum
Chromodynamics (QCD), the process $gg \to H$ produces a Higgs with no momentum
component transverse to the beam axis. Higgs bosons of non-trivial
transverse momentum may be produced if the the Higgs recoils against
one or more emitted partons.

In this work we seek to compute precisely the rate at which these
Higgs particles are produced as a function of the transverse momentum
and the rapidity at the Large Hadron Collider.  This calculation is
carried out to next leading order in the QCD strong coupling
$\alpha_s$.   In particular, this calculation includes the dominant
gluon-initiated mechanism, which is responsible for the majority of
the cross section.  We work in the infinite top mass approximation,
and argue that this is very precise in the domain of applicability.

We find an increase in the cross section by a factor of  1.9 - 2.0 in
the intermediate transverse momentum range, where this theory applies.
We also find that the renormalization scale dependence is much reduced
in our calculation as compared to the leading order prediction.
\end{abstract}

\begin{ded}
I would like to dedicate this work my parents, Natalie Miller and Donn 
Miller, and to my friend Kelly R. Carmichael, for their unending support
and encouragement.

I would like to dedicate this work to the memory of James W. Glosser,
who instilled a love and passion for science in his son, and hence
made this work possible.
\end{ded}

\begin{ack}
I would like to acknowledge the following people.  Carl R. Schmidt, for 
his useful advice and guidance, Tim Tait for his interesting topical
discussions, Tom Rockwell and Jim Amundson for their endless patience
with me in computer matters, Wu-Ki Tung, Chien-Peng Yuan, Csaba
Balazs, Pavel Nadolsky, Hung-Jian He for their physics input, and
finally, Jim Linnemann for his mentorship.
\end{ack}

\tableofcontents
\listoftables
\addcontentsline{toc}{chapter}{\listtablename}
\listoffigures
\addcontentsline{toc}{chapter}{\listfigurename}

\end{preliminary}


\chapter{Introduction:  The Standard Model of Particle Physics}

The Standard Model (SM) of Particle Physics is the most
stringently tested theoretical description of the world in which we
live.  It has succeeded in describing all subatomic phenomenology seen 
to date, to an incredible degree of accuracy \cite{RPP}.  However, even
with all of the successes of the SM, a number of theoretical questions
remain.  Of these the most  crucial is the mechanism
in which the electroweak symmetry is broken.  Over the past twenty
years or so, a number of different methods have been proposed to
generate this symmetry breaking.  All models using a perturbative
approach  to this symmetry breaking 
require at least one additional scalar electroweak isospin doublet.
The phenomenological manifestation of this scalar is generically
dubbed a ``Higgs boson.'' 

In order to understand why such a field is necessary, a pedagogical
review of the Standard Model and its gauge theory is beneficial.
From this discussion, we will come to understand the role of the Higgs 
field in maintaining gauge invariance, as well as understanding some
of the theoretical problems that a lone Higgs doublet generates.

It is  crucial to understand the experimental signatures of
the Higgs with the next generation of high energy experiments.  As the 
current phenomenological understanding of the Higgs is plagued with
large theoretical uncertainties, new calculations are required to
improve the present predictions.   With the discovery of the Higgs 
and a detailed theoretical and experimental description of its
properties, we will gain our first insight into the question of
electroweak symmetry breaking.

\section{Yang-Mills Gauge Theory}

The entire dynamics of the SM is the result of a particular
manifestation of{ \em Yang Mills gauge theory} \cite{YaMi}, which was
inspired by General Relativity.  In Yang-Mills gauge theory, we note 
that measurable quantities are proportional to the absolute square of 
a complex field, and thus the general theory should be invariant under 
a general phase transformation.  That is, we may demand that the
fermion field  $\Psi(x_{\mu})$ be invariant under the transformation
\beq
\Psi(x_{\mu}) \rightarrow \exp(i \a) \Psi(x_{\mu}).
\eeq
Since from quantum mechanics, we know that all observables are
computed from expectation values of some differential operator, it is
quite trivial to see that these observables will remain unchanged
under this phase rotation.  This invariance is known as a {\em global
gauge symmetry}.

Now, we may require that the theory be invariant under a {\em local gauge
symmetry}, in which the phase parameter is a function of the space
time co-ordinates $x_\m$, by making the replacement
\beq
\a \rightarrow \a(x_{\m}).
\eeq
A typical fermion Lagrangian is of the form
\begin{equation}
L_{F}=\bar{\Psi}(x_{\mu})(i/  \hspace{-2.1 mm} \partial-m) \Psi(x_{\mu}).
\end{equation}
It is quite trivial to see that the Lagrangian is not invariant under
such a transformation, since the derivative in the kinetic energy may
act directly on the $\a(x_\m)$.  However, if we replace the
derivative with
 \begin{equation}
D_{\mu} = \partial_{\mu}+ig  A_{\mu}(x)
\label{eqn:abel}
\end{equation}
and demand that the field $A_\m$ transform as
\beq
 A_{\mu}(x) \rightarrow  A_{\mu}(x) +\frac{1}{g}\partial_{\m}(\a(x_\m)),
\eeq
we see that the latter term cancels the anomalous term resulting from
the differentiation of the phase.  The phase parameter $\alpha(x)$ is,
in this case, a scalar function of the space-time variables. This is
known as an {\em abelian gauge symmetry}.

Now, we wish to know what will happen if we give the fermion field
some gauge index, $i$, and demand that it transform under a local
gauge symmetry  
\begin{equation}
\Psi_{i}(x) \to \left[ \exp(i \alpha^{a}(x_{\mu}) T^{a} ) \right]_{ij} \Psi_{j}(x),
\label{eqn:phase}
\end{equation}
where we imagine that  $\Psi_{i}(x)$ 
transforms  under some gauge group $G$.  The matrices $ T^{a}$
are the gauge group generators of some representation of the gauge
group $G$.  
This is an example what is commonly called a 
{\em non-abelian gauge transformation}. 
Again, it is trivial 
to see that kinetic terms in the Lagrangian will be spoiled by this 
transformation. 
 One then notes
that the kinetic terms may be rescued, provided that the conventional
kinetic term  $i\partial_{\mu}$ is replaced with the so-called {\em
covariant derivative},  
\begin{equation}
[D_{\mu}]_{ij} = \partial_{\mu}\delta_{ij}+ig T^{a}_{ij} A^{a}_{\mu}(x).
\end{equation}
This expression is completely analogous to equation (\ref{eqn:abel}). The {\em gauge field} $ A^{a}_{\mu}(x)$ is required to transform
under the gauge transformation as
\begin{equation}
T^{a}_{ij} A^{a}_{\mu}(x)
 	\to 	[e^{i \alpha^{b}(x_{\mu}) T^{b} }]_{ik}
	\left\{
	T^{a}_{kl} A^{a}_{\mu}(x)-\frac{1}{ig}\partial_{\mu}\delta_{kl}
	\right\} 
	[e^{i \alpha^{c}(x_{\mu}) T^{c} }]^{\dagger}_{lj}
\end{equation}
in order to cancel the term coming from the differentiation of the
phase in the kinetic term.
After making the replacement $\partial \to D$, we find that the covariant
fermion Lagrangian for a non-abelian gauge theory,
\begin{equation}
L_{GF}=\bar{\Psi}(x_{\mu})(i \hspace{1 mm} / \hspace{-3 mm} D-m)
\Psi(x_{\mu}),
\end{equation}
is completely invariant under the transformation (\ref{eqn:phase}).

In order to have nontrivial dynamics, kinetic terms for the gauge
fields must be introduced.   These kinetic terms must satisfy the
following properties:
\begin{itemize}
\item They must be gauge invariant.
\item The resultant quantum theory must be renormalizable.
\footnote{
It is possible to construct models with ``higher dimension operators'' 
which are not renormalizable in the conventional sense.  The divergent 
loop terms are then cut off at some arbitrary scale at which new
physics comes into play.  These theories are beyond the scope of the
discussion, however, as they do not play a role in conventional
Standard Model dynamics.
	 }	
\item There must be one term which is quadratic in $\partial_{\mu}$.
\end{itemize}
All of these constraints are satisfied by the tensor $F^{a}_{\mu \nu}$,
where
\begin{eqnarray}
T^{a} F^{a}_{\mu \nu}&=&\frac{1}{i g}[D_{\mu}, D_{\nu}] \\
 &=& T^{a} \{ \partial_{\mu} A^{a}_{\nu}-\partial_{\nu} A^{a}_{\mu}-
	g f^{abc} A^{b}_{\nu}A^{c}_{\mu} \},
\end{eqnarray}
which allows us to write down a kinetic term for the gauge fields,
\begin{equation}
L_{GF}=-\frac{1}{4}Tr[F_{\mu \nu}F^{\mu \nu}].
\end{equation}
Note that a general gauge field is self-interacting because of the
self-coupling term proportional to $f_{abc}$ in the tensor $F^{a}_{\mu
\nu}$. \footnote{ It should also be noted that a CP violating term
$-\frac{1}{4}Tr[\epsilon_{\mu \nu \rho \sigma} F_{\rho \sigma} F_{\mu
\nu}]$ 
is also consistent with the listed conditions.  Again, this is beyond
the scope of the discussion}

So far, everything is completely consistent, although we have been
considering only the classical limit of the theory.  If we wish to
consider a Quantum Field Theory, in which the fields themselves
consist of discrete numbers of excitations, we must be more careful in 
order to ensure that the unitarity of the S-matrix is not violated.
There are two different procedures for quantizing a field theory, the
{\em canonical} approach and the {\em path integral}
\footnote{For an introduction to the path integral, a field theory
text such as reference \cite{Peskin} is useful.}
approach.  After
naive quantization of the theory, one finds that quantum
corrections apparently spoil the unitarity of the S-matrix, for
particular gauge choices.  In order to
compensate, one then adds ghost fields (scalar fields that obey fermion 
statistics) to cancel the non-unitary terms in the Lagrangian.  

While this may seem ad-hoc, the ghost terms may be derived using an
explicit gauge fixing constraint in the path integral in what is known 
as Fadeev-Popov gauge fixing, in which the ghost terms are generated
spontaneously  when the gauge in the path integral is fixed \cite{FadPop}.
Consider the path integral for a non-abelian field $A^a_{\m}$,
\beq
\int {\cal D}A \hspace{1mm} \exp \left[ i S[A] \right]=
\int {\cal D}A \hspace{1mm} \exp \left[ -\frac{i}{4} \int d^4 x \left( (G^a_{\m \n})^2 \right) \right].
\eeq
This path integral infinitely overcounts the field configurations,
since it sums over the gauge ``copies'' that represent the same
physics.
To remove this overcounting, we must establish a gauge choice by 
constraining the field $A$ through some functional relationship
$F[A]=0.$  This is known as fixing the gauge.  As an example, the
familiar Lorentz gauge is fixed through the constraint
$F[A]=\partial_{\mu} A^{a\,\mu}=0$.

To fix our choice of gauge, we begin by inserting an expression for 1
into the path integral,
\beq
1=\int {\cal D}\a \hspace{1mm} \d(F[A_\a]) Det \left[ \frac{\partial F[A_\a]}{\partial \a} \right],
\eeq
where
\beq
A^{a\mu}_{\alpha} = A^{a\mu}+\frac{1}{g} D_{\mu} \alpha^a.
\eeq
We may freely change variables 
in the path integral from 
$A \to A_{\alpha}$.
Since the
integration measure ${\cal D}A$ and the
action $S[A]$ are gauge invariant,  everything in
the path integral now is a function of  $A_{\alpha}$ and therefore
we may, without loss of generality, remove the label 
$\alpha$.
The resulting expression is
\beq
\int {\cal D}A \hspace{1mm}  \exp \left[ i S[A] \right]=
(\int {\cal D} \a ) \int {\cal D}A \hspace{1mm} \exp \left[ i S[A] \right]
\d(F[A]) Det \left[ \frac{\partial F[A]}{\partial \a} \right],
\eeq
where the term $\int {\cal D} \a$ contributes an infinite overall
constant to the normalization.  

The factor $Det \left[ \partial
F[A]/\partial \a \right]$ can be evaluated by introducing the ghost
field.
It is instructive to work this out explicitly for the Lorentz gauge.
Since the gauge field varies infinitesimally as
$(A^\a)^a_\m-(A^0)^a_\m \approx \frac{1}{g} D_\m \a^a$, the 
determinant may be rewritten;
\beqa
 Det \left[ \frac{\partial F[A_\a]}{\partial \a} \right]&=& 
 Det \left[\frac{1}{g} \partial^\m D_\m \right] \\ 
&=& \int{\cal D} c \int{\cal D} \bar{c} \hspace{1mm} \exp{ \left[ i \int d^4 x
\bar{c} \left( - \partial^\m D_\m \right) c \right]},
\eeqa
where the $c$ is an anti-commuting scalar field\footnote{Anticommuting
fields become {\it Grassman variables} in the path integral
representation.}, and therefore a
ghost.
We have thus generated the Fadeev-Popov ghost term  as a 
result of rigorously fixing the gauge.  The $c$ fields are indeed
ghosts, and have no asymptotic states.  They do, however, contribute
as closed loops in diagrams. Moreover, the infinite gauge repetitions
have been systematically eliminated in the path integral, and the
gauge has been fixed.

\section{The Standard Model}

The Standard Model is, in essence, a Yang-Mills gauge theory,
consisting of three ``copies'' of fermions, called flavors,
interacting with a set of 
gauge bosons dictated by the direct product of groups $\mathbf{G}=\mathbf{
SU}_{C}(3)  \times \mathbf{ SU}_{I}(2) \times \mathbf{ U}_{Y}(1).$  As 
was seen before, for each of these symmetries we may associate a
dynamical gauge field.
Therefore, there is a gluon field  $G^{a}_{\mu}$
that interacts with the SU(3) ``color'' quantum number, A Weak Boson field
$W^{i}_{\mu}$ that interacts with the isospin quantum number ($I$), and
an abelian field $B_{\mu}$ that interacts with the Hypercharge ($Y$).

\subsection{Spin 1 Gauge Bosons}

Since the Electroweak symmetry is broken, we will observe admixtures
of its fields, $W^{i}_{\mu}$ and $B^{\mu}$, in nature.  The 4 fields combine
into two electric charge eigenstates, $W_{\pm}$, a massive chargeless
boson, the $Z$, and a massless chargeless boson, the photon $\gamma$.
A list of the general properties of these particles may be found in
Table \ref{Bosons}

\begin{table}
\begin{tabular}{llllll}
{\bf Boson} & {\bf Symbol} & {\bf Mass(GeV)} & {\bf Charge} & {\bf
Color}  & {\bf Isospin} \\ 
&&&&&\\
gluon & $g$ &0 & 0 & Adjoint & No\\
photon & $\gamma$ &0 & 0 & No & No\\
W & $W_{\pm}$ & 80.41 & $\pm e$ & No & Yes\\
Z & $Z$ & 91.187 & 0 & No & Yes
\end{tabular}
\caption[Gauge Boson Properties]{Gauge Bosons}
\label{Bosons}
\end{table}

\subsection{Spin $1/2$ Fermions}

There are three different generations of fermions.  The first
generation consists of the electron, the electron neutrino, the up
quark and the down quark.  The second consists of the muon, the
muon neutrino, the charm quark, and the strange quark.   The third
generation consists of the tau, the tau neutrino, the top quark and
the bottom quark.

The particles of different generations may be grouped into subsets of
particles 
which are indistinguishable, save their masses and Yukawa couplings
(with the Higgs).   These are known as {\it families}.
For instance, an electron, muon and tau have exactly the same isospin
and charge characteristics, and hence belong to the same family.  
Table \ref{Fermions} organizes the various known fermions by their
family\footnote{The masses in Table 2 were obtained from reference
\cite{RPP}.  The quark masses are the so-called ``current-quark masses.''}. 

\begin{table}
\begin{tabular}{lclcc}
{\bf Fermion} & {\bf Symbol} & {\bf Mass(GeV)} & {\bf Charge} & {\bf
Color} \\
&&&&\\
electron& $\mathrm{e}$ & .000511 & $-1 e$ & Singlet \\
muon& $\mu$ & .106  & $-1 e$ & Singlet  \\
tau& $\tau$ & 1.78  & $-1 e$ & Singlet  \\
&&&&\\
electron neutrino& $\nu_\mathrm{e}$ & $<15 \times 10^{-9}$  & 0 & Singlet  \\
muon neutrino& $\nu_\mu$ & $<.00017$ & 0 & Singlet  \\
tau neutrino& $\nu_\tau$ & $<.0182$ & 0 & Singlet   \\
&&&&\\
up& $\mathrm{u}$ & .001 to .005  & $\frac{2}{3} e$ & Triplet  \\
charm& $\mathrm{c}$ & 1.15 to 1.35  & $\frac{2}{3} e$ & Triplet  \\
top& $\mathrm{t}$ & 175  & $\frac{2}{3} e$ & Triplet  \\
&&&&\\
down& $\mathrm{d}$ & .003 to .009 & $-\frac{1}{3} e$ & Triplet \\
strange& $\mathrm{s}$ & .075 to .170  & $-\frac{1}{3} e$ & Triplet \\
bottom& $\mathrm{b}$ & 4.0 tp 4.4 & $-\frac{1}{3} e$ & Triplet \\
\end{tabular}
\caption[Fermion Properties]{Standard Model Fermions}
\label{Fermions}
\end{table}

Since the left and right handed components couple differently to the
electroweak sector of the theory, it is beneficial to list the
separate components and their transformation properties in a separate
table.  The lepton doublets contain a left handed, -1 charge piece,
which is ``isospin down'' (this refers to the eigenvalues of the
$\sigma_3$ matrix), and a charge zero component that has ``isospin
up'', to be identified with the $\mathrm{e}$,$\mu$,$\tau$ (former) and their
respective neutrinos (latter).  We then arrange the components into isospin
doublets,
\beq
L_{L}=\left( \begin{array}{l}
 (\nu_\mathrm{e})_L
\\ \mathrm{e}_L 
\end{array} \right),
\eeq
with identical expressions for the remaining two families. 
The Hypercharge ($Y$) may be computed directly from the charges and
the isospin eigenvalue, using the formula $Q=I+\frac{Y}{2}$ (the
charge here is in units of $e$).  We find
that the left-handed
$\mathrm{e}$,$\mu$,$\tau$,$\nu_\mathrm{e}$,$\nu_\mu$,and $\nu_\tau$
have a hypercharge of -1, while the right-handed
$\mathrm{e}$,$\mu$,and $\tau$ have a hypercharge of -2. 


 We also arrange the left-handed quarks into an isospin doublet
\beq
Q_{L}=\left( \begin{array}{l} \mathrm{u}_L \\ \mathrm{d}_L
\end{array} \right).
\eeq
The left-handed quarks transform just as the left-handed leptons do
under  $\mathbf{SU}_I (2)$, and have a hypercharge $Y=\frac{1}{3}$.
This, of course, implies that both components have nontrivial
interactions with the electromagnetic field.  The right-hand
components of the up and down type components have hypercharges of
$\frac{4}{3}$ and $-\frac{2}{3}$ respectively.  Again, these may be
computed directly from the charge formula,  $Q=I+\frac{Y}{2}$.

All quark species transform as triplets 
in the fundamental representation of the $\mathbf{SU}_C (3)$ color
force, while all leptons transform as singlets, and thus do not
interact strongly. That is, each of the quark components may be
written as a three component spinor,
\beq
Q =\left( \begin{array}{l} Q_\mathrm{r} \\ Q_\mathrm{g} \\ Q_\mathrm{b} 
\end{array} \right),
\eeq where $Q$ may be any of the quark fields, $Q_{l}$, $u_{r}$ or $d_{r}$.
The leptons do not couple to the color sector and hence are all
singlets under $\mathbf{SU}_C (3)$.  The quantum numbers of the
standard model particles are conveniently summarized in Table \ref{FermTab2}.

One may notice that, in dividing terms up into left and right-handed
states,  and by demanding that these states transform differently
under the various gauge groups, that terms which mix the left and
right handed terms are expressedly forbidden.  Since mass terms do
exactly this, they are forbidden in (Unbroken) Electroweak gauge
theory. Thus we must break the Electroweak gauge symmetry somehow in
order to have a realistic model of the Standard Model Particles.

\begin{table}
\begin{center}
\begin{tabular}{lcccc}
{\bf Fermion} & {\bf Q} & {\bf I} & {\bf Y} & {\bf
Color} \\
&&&&\\
$L_{L}$& (-1,0)  & Doublet & -1 & Singlet \\
$Q_{L}$& (+2/3,-1/3)  & Doublet & 1/3 & Triplet \\
$u_{R}$& +2/3  & Singlet &4/3 & Triplet \\
$d_{R}$& -1/3  & Singlet &-2/3 & Triplet \\
$l_{R}$& -1  & Singlet &-2 & Singlet 
\end{tabular}
\end{center}
\caption[Fermion Quantum Numbers]{Fermion Quantum Numbers}
\label{FermTab2}
\end{table}

\section{Electroweak Symmetry Breaking}

The most glaring problem with the current particle physics
phenomenology is the existence of massive particles.  Electroweak gauge 
invariance forbids hard mass terms of the form
 $\mu^{2} B^{2}$, which is obviously not invariant under a gauge
transformation $B \to B +\partial A$.
Also, the left and
right-handed  fermions, which transform differently under $SU_{I}(2)$ 
and $U_{Y}(1)$, seem to preclude the existence of any mass terms for
the fermions, since mass terms mix the right and left-handed
components, once again spoiling gauge invariance.

We wish to introduce these terms in a gauge invariant way.  The first
clue is that we must somehow include three additional degrees of
freedom, one for each additional spin state of the three massive
bosons.  This seems to indicate that another field is necessary to
compensate for the additional degrees of freedom. 

The general way to break the electroweak symmetry is through {\em spontaneous
symmetry breaking}.  To spontaneously break a symmetry, one introduces 
a field with a vacuum state which is not invariant under the symmetry
in question. One then expands the field around this vacuum.  Since the
kinetic energy terms in the Lagrangian of this field will, in general,
contain covariant derivatives (which in turn contain couplings to the
fields W,Z), expanding the field around its {\em vacuum expectation
value} (VEV), will introduce mass terms for the gauge bosons.

If we also allow the field to interact with fermions via a
fermion-antifermion-scalar interaction known as a {\em Yukawa
coupling}, we find that we have also spontaneously generated masses for 
the fermions.  The mass of this field which we have introduced
can be arbitrary (at least at tree level) in the SM.  We therefore
have a theoretical answer to all inconsistencies in the Standard
Model.

A scalar field which has these properties is generically dubbed a {\em
Higgs field}.  In the SM, the Higgs is a complex isospin doublet, so
that it transforms under both the hypercharge and isospin rotations.
Since this introduces four degrees of freedom, there is one left over
degree of freedom after the Electroweak symmetry is broken.  The
phenomenological consequence of this is the existence of a massive,
chargeless scalar called the {\em Higgs Boson}.

One final caveat concerns the quark
masses.  The quark masses are not as well defined as those of
the other fermions, and so they must be treated with some care. Masses
in a quantum field theory are basically due to the self coupling of
the field. These, like other couplings in the theory are affected by
the renormalization process.  Since most 
masses in the SM are measurable parameters (that is, not predefined
by the theory), particle masses are generally empirically defined as
being the 
masses in the absence of other forces.  The problem is of course that
the quarks are confined inside hadrons and mesons, and thus are
never ``free particles''  in the classic sense.  
 For a quark, this limit  never exists, so the
idea of a ``rest mass'' is poorly defined.  Quarks that have large
Yukawa couplings to the Higgs may still have a rest mass defined, as long as
it is larger than the strong coupling scale, $\Lambda_{\mathrm QCD}$.

\subsection{The Higgs Boson}

To  begin our discussion of the Higgs field, we write down the
conventional Lagrangian for a standard 
model Higgs boson.  It is
\begin{equation}
L_{H}=(D_{\mu} \Phi)^{\dagger}(D^{\mu}\Phi)-\lambda (|\Phi|^2-\frac{v^2}{2})^2,
\end{equation}
where $\Phi$ is complex scalar isospin doublet, and thus has hypercharge 
+1.  The covariant derivative in this case is
\begin{equation}
[D_{\mu}]_{ij} = \partial_{\mu}\delta_{ij}
+\frac{ig_{2}}{2} \sigma^{i}_{ij} W^{i}_{\mu}(x) 
+\frac{ig_{1}}{2} \delta_{ij}B^{a}_{\mu}(x),
\end{equation}
where we include both the isospin field $W^{i}_{\mu}$ as well as the
hypercharge field $B_{\mu}$.
Now, we parameterize the isoscalar doublet as
\begin{equation}
\Phi=\frac{1}{\sqrt{2}} 
	\left( \begin{array}{c} \phi_1+i\phi_2\\
	v+\phi_3+i\phi_4 \end{array} \right);
\end{equation}
or, writing this as a gauge rotation,
\begin{equation}
\Phi=\exp(-i \frac{\sigma_{i}}{2}\theta_{i}(x))
	\left( \begin{array}{c} 0 \\
	\frac{v+H(x)}{\sqrt{2}} \end{array} \right).
\end{equation}
The dynamical field $H(x)$ is 
the Higgs boson field.
We see that for a particular gauge choice (commonly called the {\em
unitary gauge}), we may eliminate three degrees of freedom.  These
degrees of freedom will show up as the necessary third degree of
freedom for each of the gauge fields $W^{i}_{\mu}$, $B_{\mu}$, leading
to the rather  quaint
analogy that the gauge fields have ``eaten'' three of the components
of the Higgs doublet, leaving the Higgs boson field as an undiscovered
dynamical field.

\subsection{The Masses of the W and Z Bosons}

As a consequence of choosing the field $\Phi$ to have a nontrivial
vacuum, 
numerous phenomenological characteristics of the SM  emerge
from the various couplings of the weak gauge fields with the
Higgs field.  
Moreover, gauge invariant couplings with the quark and lepton fields
 generate mass terms for these particles as well,
without breaking the overall gauge invariance of the theory.
We begin by writing
the {\em most general} parameterization for the model. The mass
terms for the gauge bosons arise from the following term in the Lagrangian,
\begin{equation}
\frac{1}{4}\Phi^{\dagger}
\left( g_{2} \sigma^{i}_{ij}
W^{i}_{\mu}(x) +g_{1} \delta_{ij}B^{a}_{\mu}(x) \right)
\left( g_{2} \sigma^{i}_{jk}
W^{i}_{\mu}(x) +g_{1} \delta_{jk}B^{a}_{\mu}(x) \right)
\Phi.
\end{equation}
If we redefine the fields as
\begin{eqnarray} 
W_{\mu}^{\pm}&=&\frac{ W_{\mu}^{1} \pm i W_{\mu}^{2} }{\sqrt{2}},\\
g_{0} Z_{\mu}&=&g_{2} W_{\mu}^{3} - g_{1} B_{\mu},\\
 g_{0} A_{\mu}&=&g_{1} W_{\mu}^{3} + g_{2} B_{\mu},
\label{weakbosons}
\end{eqnarray}
where $g_{0} =\sqrt{g_{1}^2 + g_{2}^2}$,
we find that the following mass term arises:
\begin{equation}
M_{W}^{2} \left( W_{\mu}^{+} W_{\mu}^{-} \right)+
\frac{1}{2}M_{Z}^{2} |Z_{\mu}|^2.
\end{equation}
whereas the field $A_{\mu}$ massless.
This expression now contains explicit formulas for the gauge boson
masses
\begin{eqnarray}
M_{W}^{2} &=& g_{2}^{2} (\frac{v^2}{4}), \\
M_{Z}^{2} &=& g_{0}^{2} (\frac{v^2}{4}) = (g_{1}^2 + g_{2}^2) (\frac{v^2}{4}),
\end{eqnarray}
Hence, we identify the fields $W^{\pm}_{\mu}$ and $Z_{\mu}$ as
 massive phenomenological gauge particles; whereas the field $A_{\mu}$
maintains a local $U(1)$ symmetry and is identified with the photon
field. 
It is straightforward to derive the expression for the electromagnetic 
charge $e=\frac{g_{1}g_{2}}{g_{0}}$ by rewriting the covariant derivative in
terms of the field $A_{\mu}$.  It is also customary to define the 
{\em Weinberg angle,} $\sin{\theta_{W}} \equiv g_{1} / g_{0} $.

\subsection{Fermion Masses and the CKM Matrix}

One may also postulate interactions between the Higgs field and the
fermions of the standard model.   We may write this  Yukawa
interaction  with the fermions: 
\begin{eqnarray} \nonumber
L_{Yukawa}&=&(y_{e}^{nm} \bar{e}^{n}_{R} \Phi^{\dagger}
L^{m}_{L}+(C.C.) )\\ \nonumber
&+&(y_{u}^{nm} \bar{u}^{n}_{R} \Phi^{\dagger} Q^{m}_{L}+(C.C.) ) \\
&+&(y_{d}^{nm} \bar{d}^{n}_{R} \tilde{\Phi}^{\dagger} Q^{m}_{L}+(C.C.) ),
\end{eqnarray}
where $n$ and $m$ run over the fermion families.  
The field $\tilde{\Phi}^{\dagger}$ is related to the complex conjugate
of the Higgs field by
\beq
\tilde{\Phi}= -i \sigma_y \Phi^{*}.
\eeq
This expression is the most general term that we may write.
Note that there is no requirement that the Yukawa matrix $y_{nm}$ be
diagonal or real.  It will, in general, mix electroweak eigenstates
of the fermions. This has the implication that CP symmetry is violated
in a general theory with a Higgs.

First, let's consider the leptons.  Since the neutrinos are
considered to be massless, and have only a
left hand component, we may freely redefine the fields
such that the mass matrix is diagonal. The three diagonal coefficients
of the mass matrices are then interpreted as the masses of the
leptons. The diagonal Yukawa matrix
elements are then the masses of the three leptons with the factor
$v/\sqrt{2}$ scaled out;  that is, $m_l = y_l v/\sqrt{2}$.  Note that
this also implies that the Higgs boson couples to leptons with
strength $y_{l}=\sqrt{2} m_{l}/v$; that is, the coupling is directly
proportional to the lepton mass. This is true of the Higgs coupling to
all particles, and is of great importance in phenomenology, as we shall
see in the next section.

The up and down  mass matrices, due to their additional complexity,
may not  be simultaneously diagonalized.
It is conventional to choose a basis in
which the up type quarks are diagonal, and the down type quarks are
off-diagonal. One then finds that since the $W^{\pm}$ terms mix
quark mass eigenstates, the $W$ Boson violates ``quark family number.''  

Let's consider this situation in a little more detail.  Since in the
Standard Model the
only coupling  which actually mixes isospin-up and
isospin-down fermions is the coupling to  $W^{\pm}$, we find  family mixing
terms in the 
couplings between the fermions and the W boson terms.  That is, the W
couples to ``weak eigenstates'' which differ from mass 
eigenstates by the rotation matrix, $V$:
\begin{equation}
\left( \begin{array}{c} d \\ s \\ b \end{array} \right)_{Weak}
=
\left( \begin{array} {ccc} V_{ud}&V_{us}&V_{ub}\\
			   V_{cd}&V_{cs}&V_{cb}\\
			   V_{td}&V_{ts}&V_{tb}
\end{array} \right)
\left( \begin{array}{c} d \\ s \\ b \end{array} \right)_{Mass}.
\end{equation}
This is the so called Cabbibo-Kobayashi-Maskawa (CKM) matrix \cite{CKM}.
This matrix has the standard parameterization
\begin{equation}
\left( \begin{array} {ccc} c_{12}c_{13}&s_{12}c_{13}&s_{13}e_{13}\\
-s_{12}c_{23}-c_{12}c_{23}s_{13}e_{13}&c_{12}c_{23}-s_{12}s_{23}s_{13}e_{13}
&s_{23}c_{13}\\
s_{12}s_{23}-c_{12}c_{23}c_{13}e_{13}&-c_{12}s_{23}-s_{12}c_{23}s_{13}e_{13}
&c_{23}c_{13}\\ 
\end{array} \right),
\end{equation}
where $c_{ij}=\cos{\left(\theta_{ij}\right)}$,$s_{ij}=\sin{\left(\theta_{ij}\right)}$, and
$e_{ij}=\exp{\left(-i \delta_{ij}\right)}$. 
In the Standard Model, these angles are not constrained by any
symmetry, and must be determined experimentally.  
It may not be possible to make these same arguments in extended
models, and one may 
consider searching for deviations in the CKM matrix as an indication
of physics beyond the SM.  See \cite{timthesis} for an excellent
discussion of these indicators.

If the neutrino sector of the theory contains
a Dirac mass term, we must forego the previous analysis, as one
cannot simultaneously diagonalize the electron and neutrino mass
matrices unless at least one of the two is completely massless.
Recent measurements of neutrino oscillations \cite{nuMass}  do seem to
suggest  that the neutrino
species have a mass difference (and therefore, masses), which of
course must be accounted for
theoretically.   One may postulate the existence of an additional
right-handed component, which will in general have its own coupling
to the Higgs.    
After diagonalizing the charged leptons, the
neutrino components are left with off-diagonal terms. These terms,
while extremely important to astrophysics, do not contribute much to
conventional high energy physics collider experiments, where the
approximation that the neutrino is massless is certainly viable.

\chapter{The Higgs Boson: Phenomenological Overview}
 
The Higgs boson is currently the only
missing component of the Standard Model.  Precise knowledge of Higgs
physics will most likely give the first glimpses of the underlying
structure responsible for electroweak symmetry breaking.  Since the
couplings of the Higgs to the various particles in the Standard Model
are already determined, any deviation from the rates computed using these couplings could be an indication of new physics.

However, in order to examine these rates, we must first find a Higgs
and measure its mass.  The Higgs mass is not fixed by any theoretical
argument in the Standard Model.  In fact, a tree-level analysis
quickly reveals that there are no constraints whatsoever on the Higgs
mass.

At the 1-loop and higher orders, however, constraints do arise on 
the Higgs mass.  A renormalization group analysis reveals that the
Higgs mass cannot be too large or too small relative to the Weak
scale.  The former results in {\it triviality}, while the latter in
{\it vacuum instability}.
The Higgs mass also appears in 1-loop corrections
to $e\bar{e} \to Z \to f \bar{f}$, and so by precisely measuring the
cross section to this process, we may constrain the Higgs mass
indirectly, as was done for the top quark before its discovery in
1995 \cite{top_disc_D0, top_disc_CDF}.


\section{Vacuum Stability and Triviality}

We begin our analysis of Higgs phenomenology by considering some
theoretical issues concerning the mass of the Higgs. As mentioned
before, the classical theory of the Higgs and its couplings to the
other Standard Model particles does not in any way restrict the Higgs
mass.  However,
radiative corrections cause the VEV and the quartic coupling to run,
and can cause problems if the Higgs mass is too far removed from the
other electroweak parameters.  

One-loop corrections to the 
Higgs effective potential lead to vacuum stability constraints.  These
result from the
fact that the asymmetric electroweak-violating vacuum could decay into 
a symmetric one, therefore destroying the universe as we know it.
Typically, these analyses put lower bounds on the Higgs mass which
are far less then the current mass bound from LEP.

Triviality arises from the fact that the quartic coupling $\lambda$
in $\phi^4$ theory exhibits a Landau pole \cite{HHG}, and therefore the theory
becomes strongly coupled at some high energy scale.  To see this, consider 
the expression for $\lambda$ running from some energy scale $\mu_0$ to $\mu$,
%
%
\beq
\lambda(\mu)= \lambda_0 \, \left(1+\frac{3 \lambda_0}{4 \pi} \log \left(
\frac{\m_0}{\mu} \right) \right)^{-1}.
\eeq
Since $\lambda_0$ is related to the Higgs mass through the VEV
(i.e. $m_H^2=\lambda v^2$), the
requirement that the coupling remain finite below some relevant scale
puts an  upper constraint on 
the Higgs mass.  If the theory is to remain perturbative up to the
Planck Scale, then the Higgs mass cannot be more than about 1 TeV.

Since this behavior is essentially nonperturbative,  such
calculations are best done on the lattice. \footnote{Lattice Gauge
Theory is an attempt to formulate quantum field theories numerically
by placing them on a discrete space-time lattice, and evaluating the
path integral numerically.} A recent lattice study
\cite{triviality} has placed an upper bound on the Higgs mass of about 
460  GeV, if no new physics arises before then \footnote{Reference
\cite{HHG} is an excellent introduction to these theoretical issues,
including the theoretical issues surrounding the Higgs phenomenon in
extended models.}.

\section{Electroweak Precision Data and the Rho Parameter}

The masses, decay widths and cross sections for $W$ and $Z$ also
receive corrections from virtual Higgs loops.  The result is that the
$\rho$ 
parameter,
\beq
\rho=\frac{m_W^2}{m_Z^2 \cos^2(\theta_W)},
\eeq 
which is 1 at tree level, picks up the following corrections from
virtual top and Higgs loops:
\beq
\rho=1
+\frac{3\,e^2}{64 \pi^2 \sin^2{\theta_W}}\frac{ m^2_t}{m^2_W}
-\frac{3 g^2}{32 \pi^2}\tan^2(\theta_W)
\log\left(\frac{m_H}{m_W}\right)+ \dots.
\eeq
All of the fermion loop
corrections to $\rho$ vary quadratically with their masses , so it is
easy to see that the top gives the dominant contribution. Therefore, a
precise  knowledge
of $\theta_W$ $m_W$, $m_Z$, and $m_t$ can be used to constrain the Higgs
mass.

The major problem with indirectly measuring the Higgs mass is that the 
corrections to the $\rho$ parameter vary only logarithmically with
$M_H$, rather  than
quadratically, as do the top corrections.  It is hence very difficult 
to have enough statistics and eliminate enough systematic uncertainty
to put a precise constraint on the mass.

As of this writing,  the upper limit from the electroweak precision
data is 224.8  GeV. 
The lower bound on the Higgs mass is most strongly constrained by the
direct search at LEP\cite{ALEPH}.  They find a lower bound of 114.1
GeV at 95\% confidence level. 

\section{Phenomenology at Colliders}

In order to search for the Higgs in a
collider experiment, we must have a precise understanding of
how it is produced, and how it decays. This is the main topic of this
section.

\subsection{Width and Branching ratios}

\begin{figure}
\hspace{.68in}
\scalebox{.6}{\includegraphics[width=6in]{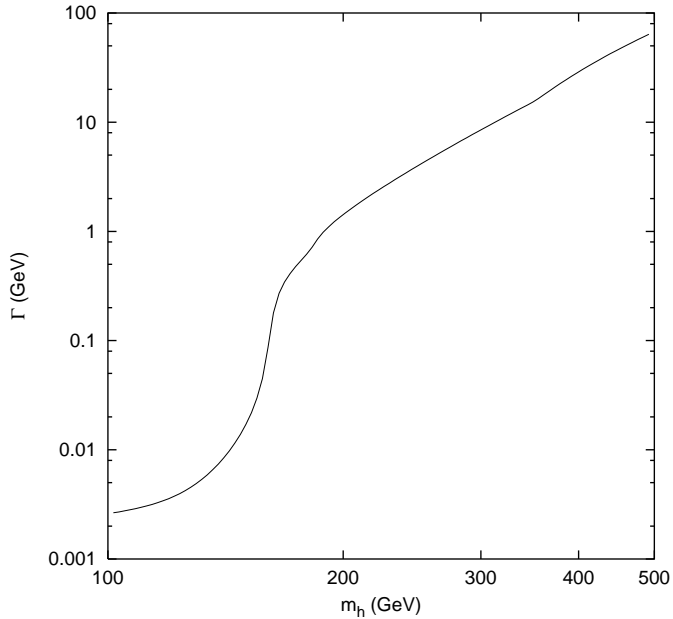}}
\caption{Higgs Width vs Mass}
\label{graph:hwidth}
\end{figure}

\begin{figure}
\hspace{.68in}
\scalebox{1.1}{\input{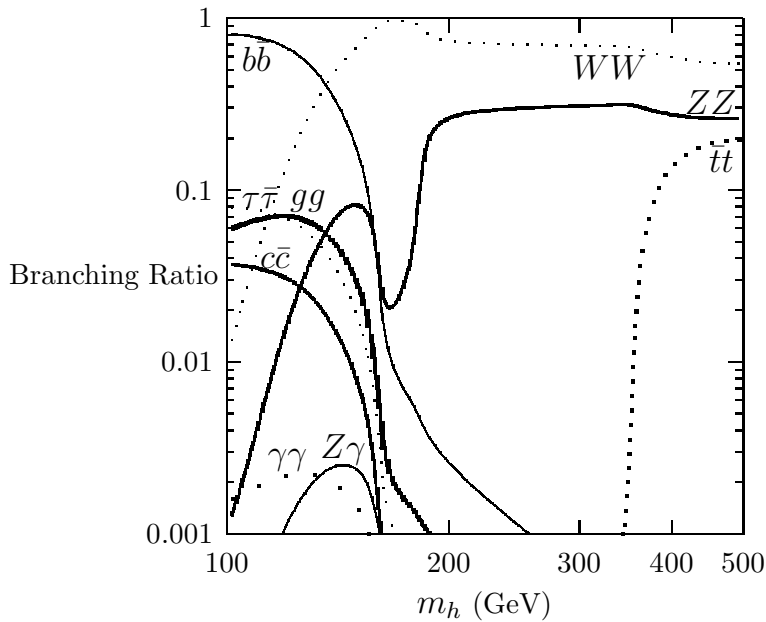}}
\caption{Higgs Branching Ratios}
\label{graph:hbranch}
\end{figure}

A graph of the Higgs width as a function of mass is provided in Figure 
\ref{graph:hwidth} \footnote{This graph and the branching ratio graph
(figure \ref{graph:hbranch}) were generated using the program HDECAY
\cite{hdecay} }.  As can be plainly seen, the width increases
dramatically around 160 GeV.
This dramatic increase in the width
is due to the fact that the Higgs is now massive
enough to decay into weak bosons, with which it has a comparatively
strong coupling. Higgs particles that can decay into weak bosons have
a collider phenomenology dramatically different than those which
cannot decay to weak bosons.  The former are referred to in this work
as {\it heavy Higgs}, and the latter {\it intermediate mass Higgs}.  

Looking at the branching ratios for a  Higgs in the intermediate mass
range (Figure \ref{graph:hbranch})
, we see why its
detection is so problematic. 
As noted in the last section, the Higgs couples to particles with a
coupling strength that is proportional to the particle mass.
Therefore, it prefers to decay to the heaviest particles for which it
is kinematically allowed.  For the intermediate mass Higgs,  it decays
 most of the time  into bottom
quarks, which are swamped by QCD background events. 
If one wishes to look for the Higgs through the $b \bar{b}$
channel, one must be able to efficiently tag bottom quarks, via
vertex detection.  It is not
clear that bottom tagging will have the resolution necessary to see
Higgs events.

About 18\%  of the time,
the Higgs will decay into taus, charms or gluons.  The decay to taus
is also a very promising mode, since taus usually decay
into electroweak particles.  Observations of Higgs decays to charm quarks is
basically out of the question, since the ability of detectors to tag
charm quarks 
is extremely unreliable.  Gluons are out
of the question as well, since they may not be tagged at all.

The mode  $H \to \gamma \gamma$, which arises through loop effects,
is quite rare,  with a branching
ratio  of $10^{-3}$.  This, in particular, is quite unfortunate, since
this mode offers the best chance for detection in a hadron-hadron
collider \cite{Spira, Csaba}.  This
process has been called the ``golden mode'', since the energy
resolution of 
these events in a particle detector is relatively fine and its
background is fairly small.  One can hope to measure the rate to
$\gamma$'s and look for the characteristic enhancement in the
transverse momentum spectrum, provided that one has a  precise
prediction  of the spectrum of  background events.

The situation is  better for a heavy Higgs, which decays
predominately to $W$'s, $Z$'s or top quarks.  While the
production rate for a Higgs in this mass range is just too small to
measure at the Tevatron,  these rates can presumably be measured at
the LHC, provided that the Higgs mass is within this range.

\subsection{Cross Sections}

Overall, the best way to search for a Higgs is with a lepton collider,
such as LEP, since one has a precise knowledge of the initial state
energies of the interacting particles.  At these machines, one begins
to expect Higgs particles 
around 
an energy of $E_{\mathrm{COM}}=M_Z+M_H$, with the Higgs radiating from the
final state $Z$ boson.  The decay into jets or leptons is relatively
clean, allowing the invariant mass of such events to be reconstructed.
If the Higgs mass were within the accessible range of LEP, then it most
surely would have been discovered.  This allows a lower bound to be
placed on the Higgs mass, which is currently  114.1 GeV.  Recently,
a recent analysis by the ALEPH collaboration revealed an excess in  events at LEP
at around 206 GeV \cite{ALEPH}.  These events, while still
preliminary,  are very suggestive of a Higgs with $M_H \approx 115$ GeV.
 
Producing a Higgs boson at a hadron-hadron collider is generally not
difficult.  In fact, even 
at the Tevatron (FNAL), there are most likely dozens of Higgs produced
each year that the laboratory takes data (depending on the Higgs
mass).  The real problem is detection; if 
the mass is less than 135  GeV, the Higgs decays predominately into
$b\bar{b}$ pairs.  Since these particles generally show up in the
detector as jets, this process is swamped by QCD background events.
Therfore, for Higgs bosons below 135 GeV, the most likely channel for
discovery at the Tevatron is the associated production channels
$p\bar{p} \to HZ$ and $p\bar{p} \to HW$. For these channels, tagging
of b-quarks is crucial.  For $m_H>135$ GeV, the $WW^*$ decay mode
(where $W^*$ represents an off-shell $W$ boson) becomes available.  The
actual reach at the Tevatron at Run II depends on the total integrated
luminosity, of course.  A recent study \cite{Carena} found that for 20
$\mathrm{fb}^{-1}$, the Tevatron can exclude a Higgs at 95\% confidence level
up to 180 GeV, and could discover it at the $3\sigma$ level if the mass is
below 125  GeV.

If the Higgs boson is not discovered at the Tevatron, 
then the Large Hadron Collider (LHC) at
CERN, which is due to start runs around 2006, will find the
Higgs particle.  
The main advantage of the LHC is that there will simply be more
events, due to the 
extremely large luminosity of this machine, as well as the enhanced
cross section.  This allows for a much better chance for detection,
especially in the $H \to \gamma \gamma$ mode.  We will therefore focus
our discussion on this device, where the discovery of the Higgs will
most likely happen.

\begin{figure}
\rotatebox{-90}{\scalebox{.6}{\includegraphics[width=6in]{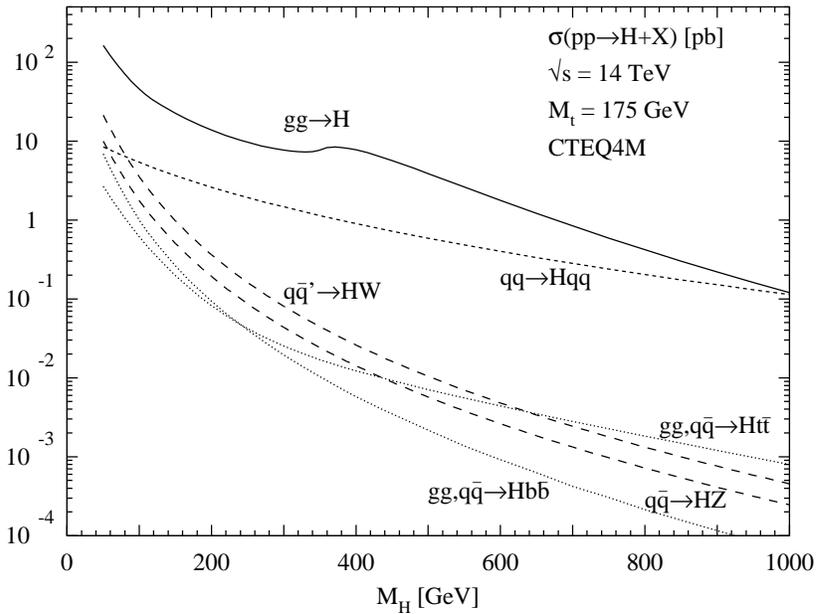}}}
\caption{Higgs production rates at the LHC}
\label{Xsec:all}
\end{figure}

There are a number of  ways that a Higgs boson can be
produced at a hadron collider. One notable mechanism is the production
rate through
the gluon fusion process, $gg \to H $, which dominates the overall
production rate.  Other ways to produce a Higgs generally
fall into the category of radiation from electroweak bosons,
production through electroweak boson fusion, and radiation from
heavy quarks.  The 
expected values of these cross sections at the LHC are given in Figure
\ref{Xsec:all} \footnote{We thank Michael Spira for the generous use of
this plot, as well as the program HDECAY \cite{Spira, hdecay}.}.

Clearly, the process $gg \to H$ dominates the total production rate
at the  LHC.  Therefore, we will concentrate our discussion on this
particular production mechanism.  This mechanism is not only the most
dominant production mechanism, but is the least understood in a
quantitative sense.  In particular, we will show that the QCD corrections
to the overall cross section as well as the differential cross
section are sizeable.


\chapter{Perturbation Theory}

If one wishes to search for a Higgs particle in a high energy
collision, one needs to have an accurate prediction of its
behavior in such an event.   For instance, we would like to know in
what kinematic regions we should should search for decay products.  In
order to do so, we are going to have to calculate observables using
the theoretical apparatus of Quantum Field Theory.

Integrations over momentum in  such theories are plagued with
divergences of several
types.  We will therefore have to come up with a consistent way to
render these integrals finite, so that the divergences  may be dealt
with in a rigorous manner.  

Once the divergences are regularized, we need a systematic
procedure for eliminating them from our predictions.  
Fortunately, in a well defined theory, these divergences will either
cancel amongst  
themselves or we will be able to 
factorize them  into universal parameters of the theory in a process known
as {\it renormalization}.  These procedures, if applied rigorously and
correctly, will allow us to compute precise, divergence-free expressions for
observables in particle accelerators.

\section{Divergences and Regularization Schemes}

\subsection{Divergences}

When one calculates an amplitude in QCD or any other field theory, one
typically gets expressions which are poorly behaved in the large
energy region of phase space.  These are known as {\em ultraviolet}
(UV) divergences, since they correspond to a divergence in the number
 of high energy (frequency) fluctuations.
An example of an integral which contains these types of divergences is
\beq
\lim_{\Lambda \to \infty}
\int_{0}^{\Lambda^2}
\frac{d(p^2)}{p^2-m^2} = 
\lim_{\Lambda \to \infty}
\log{\left(\frac{\Lambda^2-m^2}{-m^2} \right)}.
\label{eqn:diverge}
\eeq
Clearly this expression diverges in the limit that the regularization
parameter $\Lambda$ goes to infinity.

In addition, if the mass of the
particles is small, two additional types of divergences appear, which
are collectively called {\it infrared} (IR) singularities.  
First, there is a proliferation in the number of particles generated at
low energy.  This is known as a 
{\it soft} singularity, called such 
since it causes little recoil when emitted from a more energetic
particle.  In a renormalizable 
quantum theory such as QCD, these infrared singularities will cancel
in the calculation of any properly defined ``infrared safe''
observable quantity.  We will return to the question of what it means
to be an ``infrared safe'' observable later.

The other type of infrared singularity that one may have is called a {\it
collinear} or mass singularity, which occurs, for instance, in QCD.  A
collinear singularity occurs when the momenta of two massless particles
line up. To be more precise, as the invariant mass of the two distinct
particles becomes zero, the resultant jets that are produced when the
particles hadronize become indistinguishable from one another.
Therefore, one does not really know whether a jet of hadrons came from one
particle or two or more that were collinear. We will discuss how to
handle such divergences in the section on perturbative QCD.

\subsection{Historical Regularization Schemes} 

In this subsection, we shall discuss two regularization schemes that were
used historically to quantify the ultraviolet and  infrared
divergences .  We will see
that they both lead to consistency problems.

In the fictitious mass scheme, one introduces a fictitious mass term
into the integration to 
 regularize the soft and collinear divergences.  The
problem with this is that since mass terms violate gauge invariance,
gauge anomalies will in general arise in calculations, and their
removal is quite subtle.  For this reason, this fictitious mass scheme is
rarely used in modern calculations. 

The UV divergences were historically handled
using the {\em Pauli-Villars regularization scheme} (PV).  PV
regularizes 
the UV divergences through the introduction of a massive scalar ghost 
field, which converts the problem terms into logs
of the PV mass term $\Delta_{PV}$.  
The ghost fields must be introduced in a gauge-invariant manner, which
can become quite complicated for all but the simplest theories.



\subsection{Dimensional Regularization}

Almost all current calculations use {\it dimensional regularization} (DR)
to regularize the divergences.  Using dimensional regularization has a
number of advantages. Since no extraneous fields are introduced,
unitarity is manifest, and the gauge invariance of the theory is
preserved.   

Let us explore the basic ideas of dimensional regularization.  In a
divergent field theory, we expect to have terms arise in calculating
observables that are poorly behaved in the large (UV) or small (IR)
energy limit.  Equation \ref{eqn:diverge} contains examples of both of
these types of divergences
First, it  diverges as $\Lambda \to \infty$. This is the
aforementioned {\em
ultraviolet divergence}. Second, the integral also contains an
infrared divergence, as $m \to 0$.

The main idea of dimensional regularization is that divergent
integrals of this type may be rendered finite by analytically
continuing the number of space-time dimensions to $4-2\e$ dimensions.
The $d(p^2)$ terms which come from
the radial piece of the phase space in 4 dimensions, are replaced with
the analogous piece in $4-2\e$ dimensions:
\beq
d^4p \equiv\frac{ p^2}{2} d(p^2) d \Gamma \to d^{4-2\e}p \equiv
(p^2)^{1-\e}\frac{ d(p^2)}{2} d \Gamma^{\e}. 
\eeq
Where $d\Gamma^{\epsilon}$ is the $4-2\epsilon$ dimensional angular measure.
The net result of our efforts is that the divergent $p^2$ integral,
which arises 
from evaluating the phase space in 4 dimensions, picks up an
additional factor $p^{-2\e}$:
\beq
\int_{0}^{\infty}
\frac{d(p^2)}{p^2-m^2}\left( \frac{p^2}{\mu^2} \right)^{-\epsilon},
\label{eqn:DR_Example}
\eeq
which is well defined for $\e>0$. The $\mu$ term, for now,
maintains the dimensionality of the original expression.

We may also regularize the IR
divergence in a similar manner. Eliminating the mass term in equation
(\ref{eqn:DR_Example}), and  noting
that we must carefully separate
the IR divergent and the UV divergent regions of phase space,
we obtain
\beqa \nonumber
\int_{0}^{\infty}
\frac{d(p^2)}{p^2}\left( \frac{p^2}{\Lambda^2} \right)^{-\epsilon}
&\to&
\int_{0}^{m}
\frac{d(p^2)}{p^2}\left( \frac{p^2}{\Lambda^2} \right)^{-\epsilon}
+\int_{m}^{\infty}
\frac{d(p^2)}{p^2}\left( \frac{p^2}{\Lambda^2} \right)^{-\epsilon}\\
&=&\frac{1}{-\e_{IR}}\left( \frac{m^2}{\Lambda^2} \right)^{-\e_{IR}}
-\frac{1}{-\e_{UV}}\left( \frac{m^2}{\Lambda^2} \right)^{-\e_{UV}},
\eeqa
where we have defined $\e_{UV}>0$ and $\e_{IR}<0$ to represent the
ultraviolet and infrared poles, respectively.
In QCD, we also find that poles arise from the angular integration
piece, where the internal partons become collinear.  These poles in
$\e$ also contribute infrared poles to the matrix element. 

In dimensional regularization, then, we find that we have traded a poorly
behaved log for a pole of the form, $-\frac{1}{\epsilon}$.  These poles must
either cancel amongst themselves
(which is the case for the infrared poles), or they must cancel against
renormalization counterterms (as is the case for ultraviolet or
collinear poles), which we will discuss shortly. If everything works
as it should, we should be able  
to analytically continue the complete expression, including
counterterms,  from $\epsilon=\pm \infty 
\to 0$ {\em without hitting any poles}.  That is, all poles of the
form $(-\epsilon - n)^{-1}$, where n is a integer must cancel
\footnote{$n/2$ is the {\em degree of divergence} of the theory, that is,
a quadratically divergent theory will typically have terms like
$(\epsilon - 1)^{-1}$.  All of the interactions of the Standard Model save
that of the Higgs are logarithmically divergent, so all of these
superficially higher order divergences must cancel amongst
themselves.}.  We may therefore set $\epsilon \to 0$ with good
conscience.

Now that we have examined the basic ideas of dimensional
regularization, let us examine the finer details.
Immediately, we see an ambiguity in this scheme.  It is not
immediately obvious how to handle quantities that are only
well-defined  in integer dimensions,  like the number of spins, nor 
is it obvious what to do with antisymmetric tensors like $\gamma^5$ 
(which depends crucially on the integer nature of $N_{dim}$).
The most common prescription is to continue the number of spins of
bosons by setting $N_{spin}=2(1-\epsilon)$, and by demanding that:
\begin{itemize}
\item  $\gamma^5$ {\em anticommutes} with $\gamma^0$, $\gamma^1$,
$\gamma^2$, $\gamma^3$, and 
\item  {\em commutes} with all other $\gamma$'s.
\end{itemize}
This convention was developed by t'Hooft and Veltman
\cite{ThooftVelt}, and is the standard way of handling this
problem.  Gauge invariance is manifest in this scheme, and there are
precisely two chirality states (defined in terms of projection
operators) for the fermions.  Thus, this scheme avoids any anomalous
degrees of freedom which could arise with other ``generalizations'' of 
$\gamma^5$.  By setting the number of gluon helicities equal to
$N_{spin}=2(1-\epsilon)$,  we avoid having to separate out the
$g_{\m\n}$ tensors that arise from the internal spin sums.   This way
of handling the helicities, where the gauge boson spins
are manifestly continued to $2-2\e$ and the number of
fermion helicities is fixed to be 2, is known as {\em Conventional Dimensional
Regularization}.

\section{Renormalization}

Having  regularized the divergent pieces, we now seek to eliminate
them from our predictions in a manner which is consistent with our
theory.  We will take advantage of  the freedom that our theory
provides to absorb 
these divergent terms into the definitions of the wave function normalization,
 the definition of the charge, and the definition of the mass.  That
is, we have the freedom to fix 
our normalizations to whatever we wish.  We may,  for example, 
make the replacements
\beqa \nonumber
\phi &\to& \sqrt{Z_\phi} \phi,\\ 
e &\to& Z_e e,\\
m_\phi &\to& Z_m m_\phi \nonumber.
\eeqa
These reparameterizations will allow us to absorb the
ultraviolet divergences into these definitions.
Provided that we do this consistently, we arrive at a well-defined,
finite prediction for our observable.  The gauge invariance of the
theory will in general impose relations between these normalization
constants, but in general, we still have significant room to absorb
the ultraviolet divergences.

\subsection{Modified Minimal Subtraction}

Even with the gauge restrictions, we find that we have significant
freedom to reparameterize the theory.  We must therefore set down
conventions for the inclusion of
the UV counterterms. These terms,  which arise because of a
redefinition of the normalization constants in the field theory, are
somewhat ambiguous.  This ambiguity arises since renormalization group analysis
only defines how one modifies the charge to $O[\e^{-1}]$.  We
wish to remove the poles in $\e$, but we  also have the freedom
to add  any finite piece to the renormalized charge that we
wish.  Consider
\beq
g^{(0)} \to g^{(1)}=g^{(0)}\left(1-
 \frac{\alpha_s}{4\pi}\beta_0\left(\frac{1}{\epsilon}+O[\e^0]\right)\right),
\eeq
where the coefficient $\beta_0$ is the first {\em renormalization group
coefficient}.  The details of this coefficient depends on the symmetry
of the group, and the number of fermions interacting with the gauge
bosons.   $O[\e^0]$ represents terms that are unspecified by any gauge
or other restrictions.

In modified minimal subtraction ($\overline{\mathrm{MS}}$), we keep
finite terms 
which tend to arise from the finite phase space integration in  the
dimensional regularization procedure outlined above. 
Instead of subtracting just the $\e$ pole, 
we subtract off the piece
\beq
g^{(0)} \to g^{(1)}=g^{(0)}\left(1-\frac{1}{\epsilon} \frac{(4
\pi)^{\epsilon}}{ \Gamma(1-\epsilon)} \frac{\alpha_s}{4\pi}\beta_0\right)
\label{MSbar}
\eeq
in $\overline{\mathrm{MS}}$.
Again, this is a just a convention which is used to remove the extraneous
universal pieces that tend to complicate calculations in dimensional
regularization.  There are a number of other ways of expressing the
$\overline{\mathrm{MS}}$ counterterm, which are technically different
from this one at $O[\e]$.   However, since these $O[\e]$ terms will
eventually be set to zero in a next-to-leading order calculation,  the
same result is obtained.

In principle, the subtraction scheme should not matter
when one is calculating an observable. In practice, however, different
subtraction schemes can cause significant variations in the final
result at a fixed order in perturbation theory.  One must also consistently
use the same subtraction scheme to calculate the parton distribution
functions from the raw deep inelastic scattering (DIS) data \cite{ASTW},
as well as to  calculate the  UV counterterm from the charge renormalization.
If one wishes to do a calculation in a different scheme, one
must either know how to convert the result in that scheme back into
$\overline{\mathrm{MS}}$, or one must be able to show that the scheme
produces the same results at that order of perturbation theory.  Since
it is fairly involved to do either of these things, most calculations
are done using the modified minimal subtraction scheme.

\section{Quantum Chromodynamics}

Quantum Chromodynamics is the theory of the strong interaction at high
energies.  This model introduced dynamics to the quarks of the
Gell-Mann $\mathrm{SU}(3)_{\mathrm{flavor}}$ model, and proposed a carrier of the
strong nuclear force, the gluon.  We shall begin by discussing the
{\it Naive Parton Model}, which explains the behavior of the hadrons
in a qualitative sense, and then move on to {\it perturbative QCD},
which allows one to compute much more precise expressions for collider
observable quantities.

\subsection{The Naive Parton Model}

We shall begin the discussion of Quantum Chromodynamics (QCD) by
introducing the {\em parton model}, due to Bjorken
and Feynman.  The parton model was introduced in an effort to
explain the deep inelastic scattering data, which seemed to indicate that the
proton had a definite substructure.  As originally proposed, the
parton model suggested that hadrons consisted of massless point particles which
could then interact electroweakly with the scattering electrons to
produce {\em jets} (semi-collinear collections of hadrons).  The
model originally had no other interactions built in to explain any of
the strong coupling effects seen.  This model is known as the {\em
naive parton model}, and ignores any perturbative QCD effects.  
Indeed, this model did seem to describe the data in a qualitative fashion, and 
it was later deduced from the DIS data that the components of the
proton that were interacting with the electron were spin 1/2
objects. This is precisely what one would expect it these ``partons''
were to be identified with the quarks of the Gell-Mann
$\mathrm{SU}(3)_{\mathrm{flavor}}$ model.

The naive parton model assumes the high energy limit, in which both
quarks and hadrons are massless.  To calculate a cross section in
DIS, we convolve a partonic cross section with a form factor
$f_{H \to q}(\xi)$.  In other words, the hadronic cross section is
\begin{equation}
d \sigma_{eH \to eX}=\int_0^1 d\xi f_{H \to q}(\xi) 
d\hat{\sigma}_{q}(\xi), 
\label{H}
\end{equation}
where we have introduced $f_{H \to q}(x)$, 
the process-independent {\em parton distribution functions} (PDF).  
These empirically measured functions represent the probability of the
interacting parton having a fraction $\xi$ of the parent hadron's
momentum. In the limit that both particles are massless, the hadron
momentum is given in terms of the proton momentum as
\beq
p_p = \xi p_H.
\eeq

The hadron-hadron cross section is defined in a similar fashion, where
the structure functions are deemed to be {\em universal}, that is,
process independent.  We may then measure the parton distribution
functions through DIS and apply them to any process we wish.  The
hadron-hadron cross section is then given by
\begin{equation}
d \sigma_{HH' \to X}
=
\int_0^1 d\xi_a f_{H \to a}(\xi_a) 
\int_0^1 d\xi_b f_{H \to b}(\xi_b)
d\hat{\sigma}_{ab}(\xi_a,\xi_b).
\label{HH}
\end{equation}
In equation (\ref{H}), $q$ may be a quark or an anti-quark, and in
equation (\ref{HH}), $a$ and $b$ may be any species of parton. These
expressions must be summed over the parton species, $q$, $a$, and $b$,
since
partons may not be directly observed.

\subsection{Perturbative Quantum Chromodynamics: Improved Parton Dynamics}

Ideally, one would like to extend the qualitative successes of the
naive parton model with a dynamic theory.  Such a theory would have the
same qualitative behavior of the naive parton model, but with enough
computational power to make precise analytic predictions of observable
quantities. 

We find that, if we allow the quarks of the Gell-Mann model to
interact dynamically via an unbroken SU(3) Yang-Mills theory, we
succeed in extending the parton model.   This theory contains
generalizations of the hadron structure functions mentioned in the
previous subsection, and allows improvements of the predictions of
that model by the dynamic interaction with the SU(3) gluon field.

This  model based on the SU(3) gauge theory  
is  called {\em Quantum Chromodynamics} (QCD).  QCD is an
(unbroken) SU(3) gauge theory in which the quarks of Gell-Mann's
flavor model interact with spin-1 bosons known as gluons.  There are
eight of these, interacting with the quarks via a three-dimensional
quantum number known as {\em color}. One may show that such a theory
satisfies {\em asymptotic freedom}, in which the coupling becomes very
small at high energy scales. 

This theory is also interesting, since it gives an explanation as to why the
quarks are never seen in nature.  The coupling $\alpha_s$ of the SU(3) gauge
theory actually diverges for a small energy scale, indicating that any
perturbation theory is poorly defined in this limit. In  quantum
mechanics, small energies are equivalent to large distances.  This
implies that any attempt to pull quarks away from one another results
in a steady increase in energy, until the available energy is enough
to create a quark anti-quark pair out of the vacuum.  In order to
examine such behavior, one must be able to calculate exact
expressions, without resorting to perturbation theory.  

If the interaction is over a very small distance (large energy
scale), one may use perturbation theory, since $\alpha_s$ is small
over  these distances. Such an expansion, and the
requisite renormalization of the PDF's and the coupling constant, are
known as {\em perturbative QCD}.  For very ``hard'' scattering
processes, where a significant amount of energy is exchanged among the
hadrons, we may calculate corrections to the naive parton model. This 
consists of calculating a parton-level cross section in standard
perturbation theory, then inserting the results into the convolution
equations above.

QCD, like all gauge theories, is poorly behaved in the ultraviolet
region.  Wave function renormalization removes
the ultraviolet singularities that occur.  In addition, two
other types of singularities  occur, {\em collinear} (or mass)
singularities, and {\em soft} (or infrared) singularities.    The soft
singularities must cancel 
when all of the pieces of the cross section are added together, but
the collinear singularities do
not cancel against any other pieces of the calculation.
However, the collinear poles are universal (process independent), and
may be absorbed into the (universal) structure functions. That is,
\beq
f_{h \to p}(z) \to \int^1_0 dx/x f_{h \to p'}(x/z,\mu_F)  K_{p' \to p}(z),
\label{pdf}
\eeq
where the convolution kernel $K_{p \to p'}(z)$ is
\beq
K_{p' \to p}(z)=\delta(1-z) \delta_{p p'}
	- \left(\frac{\alpha_{s}}{2 \pi \epsilon }  \right)
\lp \frac{4 \pi \mu_R}{\mu_F}\rp^\e
\frac{P_{p \to p'}(z)}{\Gamma(1-\epsilon)}+ {\rm finite.}
\label{pdf2}
\eeq
The convolution kernel may be thought of as being analogous to the
normalization constants $Z$ in the ultraviolet renormalization procedure.
The {\em splitting kernels} $P(z)$ in equation (\ref{pdf2}) are
\beqa
 P_{q \to q }(z)&=&\frac{4}{3}\left( \frac{1+z^2}{(1-z)_{+}} \right) + 
2 \delta(1-z)\\ 
 P_{q \to g }(z)&=&\frac{4}{3}\left( \frac{1+(1-z)^2}{z} \right) \\
 P_{g \to g }(z)&=&6\left( \frac{1-z}{z}+ \frac{z}{(1-z)_{+}}+ 
                           z(1-z) \right)
		+\left( \frac{11}{2}-\frac{n_f}{3} \right) \delta(1-z) \\
P_{g \to q}(z)&=&\frac{1}{2}(z^2+(1-z)^2).
\eeqa
In the preceding relations, we have introduced the ``+'' prescription,
which is
defined through the relation
\beq
\int^{1}_{0} dz \frac{1}{(1-z)_{+}}g(z)=\int^{1}_{0} dz
\frac{g(z)-g(1)}{(1-z)} 
\eeq
for any arbitrary function $g(z)$.

In equation (\ref{pdf2}) we have introduced another
dimensional byproduct of the 
renormalization process, the factorization scale $\mu_F$.  We have also
made explicit the separate scales introduced in ultraviolet and infrared
regularization processes; the ultraviolet renormalization scale will
be referred to as
$\mu_R$.  In practice, one generally takes these scales to be the
same, though there is no theoretical argument that they {\it should}
be taken to be the same.  When we have set these scales equal to each
other, we will refer to the resulting parameter as $\mu$, to
emphasize it's generic nature.

The physical interpretation of the factorization scale is somewhat
different than that of the renormalization scale.  Indeed, the factorization
scale represents the cutoff scale for so-called ``hard processes.''
In other words, it quantifies exactly how much of the soft QCD
radiation is to be included in the partonic matrix element, and how
much is to be absorbed into the parton distributions.  The
renormalization scale represents the cutoff scale at both the large
energy end, where we expect new physics to come into play, and the low
end, where we absorb the remaining IR singularity into the quantum
mechanical excitations. 

Now, in order to cancel the soft poles that occur in the matrix
element, we must carefully define our observables.  In general, an
observable $O$ may be a function of the number of particles, as well
as the kinematics of the system.  In order for an observable to be
{\em infrared safe}, we guarantee that (for $N$ and $N+1$ parton
systems):
\beq
O_{N+1} (E_{1} \to 0) =O_{N}.
\eeq
That is, $O(N+1)$ is infrared safe if, as the energy of one of the partons
vanishes, $O(N+1)$ becomes identical to the analogous function with one
less parton, $O(N)$.  Similarly, 
an observable is {\em collinear safe} if for two particles with 
momenta $p_{1}$, $p_{2}$, we have that
\beq
O_{N+1} (S_{12} \to 0) =O_{N},
\eeq 
where $(S_{12}=(p_{1}+p_{2})^2)$.
If $O$ satisfies these two conditions, then the cross-section is
guaranteed to be finite to NLO in perturbative QCD.

This process may be generalized. We may calculate cross
sections and other observables to any 
order in perturbation theory that we wish.  These calculations tend to
be quite complicated, as the additional complexity of the SU(3)
gauge group tends to generate large numbers of contributing graphs.
This is especially true in processes that involve large numbers of
gluons, since gluons can couple not only to fermions, but to other
gluons as well.

\section{Helicity Amplitudes and Color Ordering}

In recent years, many techniques have been developed to simplify the
computation of multi-parton amplitudes in QCD \cite{MaPa, BeDiKoSi}.
By using techniques such as helicity
amplitudes and gauge-invariant color orderings, one may simplify a
calculation greatly.  This technique  significantly reduces
the number of diagrams that one has to evaluate in computing, for
instance,  gluon amplitudes. 

In order to avoid calculating things redundantly, one may make use of
some of the symmetry properties of QCD.  In the approximation that all of
the partons are massless, the spinor states of the partons become true
helicity eigenstates.  Since some of these states are related to other
states by  
charge conjugation, Bose symmetry, or parity, one avoids calculating
redundant helicity states.  Also, each non-cyclic color projection is gauge
invariant from other non-cyclic projections, so we may reduce the
number of terms by calculating a single cyclically related
color projection.

Another simplification lies in the fact that these projections of
color and helicity are gauge invariant, so that one may choose a
different gauge to evaluate the partonic matrix element for each
different helicity state.  It is generally convenient to express
polarization vectors in terms of massless Weyl spinors
\cite{Kleiss:helicity,Xu:helicity}.  The
polarization vectors for a gluon of momentum $p$ may  be written
\beq
\e_\m (p, q)=\mp \frac{\left<\pm p | \gamma^\m | q  \pm \right>}
{\sqrt{2}\left<\mp q |  p  \pm \right>},
\eeq
where
\beq
\left|  p  \pm \right> = \frac{1\pm\gamma^5}{2}u(p)
\eeq
is a massless spinor, and $q$ is some arbitrary massless reference
momentum such that $q \cdot p \neq 0$.
Since a change in $q$ is equivalent to adding a piece proportional to
$p$, we have basically absolute freedom to choose $q$.  The terms
proportional to $p$ will then cancel in gauge
invariant amplitudes via the Ward identities.  

Using these expressions, the number of
terms one needs to calculate a QCD observable is reduced considerably.
However, ambiguities
arise at NLO, where we would like to analytically continue the number
of dimensions.  The number of spins states in a conventional
$\overline{\mathrm{ MS}}$ calculation is generally taken to be $2(1-\e)$
which is quite clearly an abstraction.  If we consider actual helicity
states, the number of external particles {\em must be two exactly}.
Internal (loop) particles may have either, but are generally taken to
have $2(1-\e)$ helicity states.  Therefore, when calculating
expressions using helicity states, one
must know how to alter the resultant expressions in order to obtain the
conventional $\overline{\mathrm{ MS}}$ expression.  We will encounter
this issue later, during our calculation of the Higgs differential
cross section at NLO in QCD.

\chapter{The Higgs Boson Cross Section and Transverse Momentum
Spectrum: Leading Order Calculation}  
One of the most important quantities that will be measured  at the LHC
is the 
production rate of the Higgs \cite{Spira, Kunszt:Review,
Dawson:Review}. Therefore, a precise theoretical
understanding of the rate of Higgs production is critical to any
attempts to search for the particle. Since the cross section  is
dominated by the gluon-fusion process \cite{Spira}, this process must
be calculated as precisely as 
possible, in order to precisely determine how many of
these particles will be produced in a year of running.

Aside from having a good idea of the overall production rate of the
Higgs, the transverse momentum spectrum and the rapidity spectrum of the Higgs boson is
of great interest to the LHC project. Since the detectors themselves
must ideally be designed around the kinematic regions where one expects to see
the Higgs, having a reliable prediction for these differential cross
sections is of particular importance.

Unfortunately, QCD observables are notoriously
unreliable at leading order, and these must generally be calculated to
 NLO (at least)
in order to have a reliable prediction for their behavior.  
Moreover, leading order QCD calculations tend to exhibit a strong dependence on
the factorization/renormalization scale, $\mu$, which reflects this
theoretical uncertainty.

The focus of this work is to calculate a reliable numerical estimate
of the 
transverse momentum/rapidity spectrum.  In order to improve the
existing calculation, we will add to it radiative corrections which will
hopefully have less scale dependence, and therefore less theoretical
uncertainty.  Since the procedure for calculating this differential
cross section  at NLO
includes calculating the LO spectrum, it is instructive to begin with
the much simpler LO calculation.  We will then show in the next chapter how to
modify this expression in 
order to obtain the differential cross section at NLO.

We  will assume in this calculation that the dominant
contributions to the transverse momentum spectrum come from the
aforementioned gluon-fusion process.  We will calculate the LO production rate
both for a finite top quark mass, and in the limit that the top mass
goes to infinity (the {\it large $m_t$ limit}) .  We will then argue
that in the Higgs mass range under consideration, the finite mass
calculation and the infinite-mass
approximation give virtually the same result, whereas the QCD
corrections give very large contributions to both the transverse
momentum spectrum as well as the overall production rate.
\section{The Gluon Fusion Cross Section}
\begin{figure}
\begin{picture}(32000,10000)
\drawline\gluon[\E\REG](12900,9000)[6]
\drawline\fermion[\SE\REG](\gluonbackx,\gluonbacky)[6000]
\drawline\fermion[\SW\REG](\fermionbackx,\fermionbacky)[6000]
\global\gaplength=300
\global\seglength=900
\drawline\Scalar[\E\REG](\fermionfrontx,\fermionfronty)[6]
\drawline\fermion[\N\REG](\fermionbackx,\fermionbacky)[8550]
\drawline\gluon[\W\REG](\fermionfrontx,\fermionfronty)[6]
\put(12000,8000){$g$}
\put(12000,1000){$g$}
\put(30500,4500){$H$}
\put(18000,4500){$f$}
%
%
\end{picture}
\caption{Feynman Diagram for Higgs Production through Gluon Fusion}
\label{XsecGraph}
\end{figure}
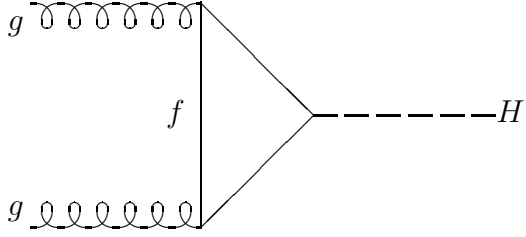

The gluon fusion process proceeds through a virtual top quark loop as
represented by the Feynman diagram in figure \ref{XsecGraph}.  As this
process is well known in the literature \cite{ Wilczek},  we
will just
present the result
\beq
\hat{\sigma} = \frac{ \alpha_s^2}{256 \pi v^2} |A|^2 \d
\lp1-\frac{m_H^2}{s} \rp .
\label{eqn:xsecfin}
\eeq
The form factor $A$ is
\beq
A(\tau) = \tau \left[ 1+\lp 1-\tau \rp f(\tau) \right]
\eeq
where
\beqa
f(\tau)& \hspace{-1.5cm} = \mathrm{arcsin^2} \lp \sqrt{\frac{1}{\tau}}
\rp&\ \ \mathrm{for}  \ \tau < 1, \\
f(\tau)&  = -\frac{1}{4} \lp \log \lp \frac{1+\sqrt{1-\tau}}{1-\sqrt{1-\tau}
}
\rp -i\pi \rp
&\ \ \mathrm{for} \ \tau >1,
\eeqa
and $\tau=4 m_{\mathrm{top}}^2 / m_H^2$.  
In the limit of a large quark mass, this form factor converges to a
constant
\beq
A(\tau \to \infty) = -\frac{2}{3},
\eeq
which results in the cross section in the limit of a large quark mass;
\beq
\hat{\sigma}(\tau \to \infty) 
= \frac{ \alpha_s^2}{576 \pi v^2}  \d \lp1-\frac{m_H^2}{s} \rp .
\eeq

\begin{figure}
\hspace{1cm}
\scalebox{.8}{\includegraphics[width=14cm, height=14cm]{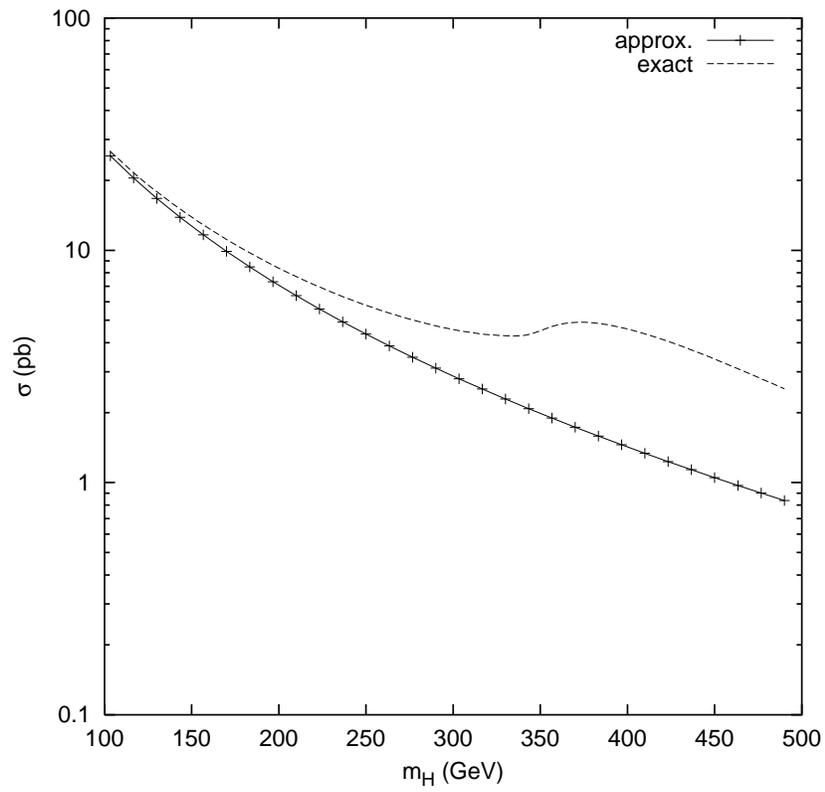}}
\caption{Comparison of the infinite quark mass approximation with the
$m_\mathrm{top} =$174 GeV calculation.}
\label{graph:xsec}
\end{figure}
A graph of the exact hadronic cross section is given in Figure \ref{graph:xsec} for a
top mass of 174 GeV.  In addition, 
the corresponding infinite mass limit is plotted on the same graph as
a reference.  
Clearly, both
graphs are qualitatively the  same for an intermediate mass Higgs 
($m_H \approx $ 100-200 GeV), which is the region that we will be
considering. If we assume that the LHC will run at a luminosity of
approximately  
$\approx 300\ \mathrm{fb^{-1}\ yr^{-1}}$ \cite{RPP}, then  we may
compute the expected 
number of Higgs produced in one year of continuous running by 
multiplying this figure by the cross section. 
Looking at the data on this graph, we see that the LHC
will produce approximately 
$0.4-4 \times 10^6$ Higgs bosons in a year of running, depending on the Higgs 
mass.  We can conclude that  Higgs production will be quite prolific at this
machine. 

The close correspondence of the finite calculation with the
infinite mass limit calculation seems to suggest that in the 
infinite quark mass limit,
the dynamics of the loop freezes out, allowing us to replace the loop
with an effective coupling. As
long as the relevant energy scales in the problem do not become too
large, we can be confident that the results of this calculation give a
fairly good approximation to the true cross section.

Even in the region where the energy scale is approximately the top quark
mass, the infinite mass approximation is quite good.
The $O[m_{\mathrm{top}}^{-2}]$ give additive corrections \cite{Sally} which  behave as
\beq
\sigma(m_{\mathrm{top}}) \approx \sigma(m_{\mathrm{top}}=\infty)\left(1+\frac{7}{15}\left (\frac{m_H}{2
m_{\mathrm{top}}}\right )^2\right)
\eeq
so that for a Higgs mass with a mass as large as 200 GeV, the exact
leading-order expression differs  
from the large $m_{\mathrm{top}}$ approximation by $\approx$ 15\%.
Since every experimental indicator tells us that the Higgs is very likely
below this limit, we can with good faith apply this effective theory
to the inclusive cross section.

We may also note that the cross section calculated in the large
$m_{\mathrm{top}}$ limit  {\it underestimates} the Higgs mass over the entire
range of plotted masses. As we argued earlier, it is highly improbable
that the Higgs mass is outside of this range.  We therefore expect
that the infinite mass effective theory will be quite good over the
entire range of applicable masses.
Moreover, recent work \cite{Spira:Xsec, Graudenz:Xsec} seems to
suggest that the 
 ratio of the next-to-leading order calculation to the leading order
calculation  is more 
or less independent of the top quark mass. Therefore, one may 
compute the next-to-leading order corrections using the infinite top mass
limit, divide that value by the leading order, infinite mass limit
result, and apply the resulting enhancement to the leading order
expression with confidence.
\section{Low Energy Effective Lagrangian}
Since the large $m_{\mathrm{top}}$ approximation is quite valid in
our leading order calculation of the cross section, a   
systematic way to calculate the various Higgs observables using this
approximation is desirable.  We may therefore propose a low energy effective
theory to do gluon fusion calculations \footnote{Aside from the
original leading order calculation \cite{LO:xsec}, The radiative
corrections using this infinite $m_\mathrm{top}$ limit have also been
computed \cite{Sally, Djouadi:Xsec}. 
Moreover, progress has recently been made on the NNLO calculation
\cite{Harlander:NNLO, Harlander:NNLO2}.}.

This idea of freezing the top quark inside the loop, and replacing it
with an effective coupling leads to the following effective Lagrangian,
\beq
\mathcal{L}=-\frac{1}{4}\left(\frac{\alpha_s}{3 \pi v} \right) \lp
1+\Delta \rp H G^{\m \n} G_{\m \n}, 
\eeq
where the top-loop dynamics have been replaced by an effective
coupling 
\footnote{The
effective coupling $\alpha_s/3 \pi v$ receives a finite
renormalization at each order in 
perturbation theory.  We represent this with a constant $\Delta$,
which is zero at leading order. there is a contribution from $\Delta$
at NLO which one must take care to include.},
$\frac{\alpha_s}{3 \pi v^2}$.  In order to be 
valid, the relevant energy scales of the process must be less than the 
top mass, and even outside that neighborhood this approximation
generally 
underestimates the cross section, so that one errs on the
conservative side.  We may derive the following
Feynman rules from this effective low energy Lagrangian:
\\
The Higgs-gluon-gluon coupling is
\beq
H_{ab}^{\mu \nu}(p_1, p_2)= -i A \delta_{ab} \left(p_1 p_2\,g^{\mu \nu}-p_1^\nu p_2^\mu \right),
\eeq
\\
the Higgs-three gluon coupling is
\beq
H_{abc}^{\mu \nu \sigma}(p_1, p_2, p_3)
= g A f_{abc}
\left((p_1- p_2)^{\sigma}\,g^{\mu \nu}
+(p_2- p_3)^{\mu}\,g^{\sigma \nu}
+(p_3- p_1)^{\nu}\,g^{\mu \sigma}\right),
\eeq
\\
and the Higgs-four gluon coupling is
\beqa
H_{abcd}^{\mu \nu \sigma \rho}(p_1, p_2, p_3, p_4)
= i g^2 A [& f_{abe}f_{cde}& 
	\left( g^{\mu \rho}g^{\nu \sigma}-g^{\mu \sigma}g^{\nu \rho}\right)  
\nonumber\\ & f_{ace}f_{bde}& 
	\left( g^{\mu \nu}g^{\rho \sigma}-g^{\mu \sigma}g^{\nu \rho} \right)
\nonumber\\  & f_{ade}f_{bce}& 
	\left( g^{\mu \nu}g^{\rho \sigma}-g^{\mu \rho}g^{\nu \sigma} \right)
       ],
\eeqa
where  $A=\lp \frac{\alpha_s}{3 \pi v}\rp \lp 1+\Delta \rp$ is the
effective coupling and $f_{abc}$ are the  $SU(3)$ group coefficients
in the adjoint representation.
These are represented graphically as: \\
\hspace{2in}
\begin{picture}(10000,18000)
\drawline\gluon[\S\REG](5500,16000)[6]
\global\gaplength=300
\global\seglength=875
\drawline\Scalar[\W\REG](\gluonbackx,\gluonbacky)[5]
\drawline\gluon[\S\REG](\gluonbackx,\gluonbacky)[6]
\drawline\gluon[\E\REG](\gluonfrontx,\gluonfronty)[6]
\put(8700,5800){$H^{\mu \nu \sigma}_{abc}$}
\put(6100,15500){$a,\mu$}
\put(10000,10100){$b,\nu$}
\put(6100,3000){$c,\sigma$}
\end{picture}
\hspace{1.5cm}
\begin{picture}(10000,18000)
\drawline\gluon[\E\REG](0,9000)[6]
\global\gaplength=300
\global\seglength=900
\drawline\Scalar[\N\REG](\gluonbackx,\gluonbacky)[6]
\drawline\gluon[\E\REG](\gluonbackx,\gluonbacky)[6]
\put(5200,5800){$H^{\mu \nu}_{ab}$}
\put(200,9500){$a,\mu$}
\put(11000,9500){$b,\nu$}
\end{picture}
\hspace{1.5cm}
\begin{picture}(10000,18000)
\drawline\gluon[\SE\REG](0,16000)[5]
\drawline\gluon[\SW\REG](\pbackx,\pbacky)[5]
\drawline\gluon[\NE\REG](\pfrontx,\pfronty)[5]
\drawline\gluon[\SE\REG](\pfrontx,\pfronty)[5]
\global\gaplength=300
\global\seglength=900
\drawline\Scalar[\E\REG](\pfrontx,\pfronty)[5]
\put(4400,5000){$H^{\mu \nu \sigma \rho}_{abcd}$}
\put(1200,4800){$a, \mu$}
\put(1200,15600){$b, \nu$}
\put(8400,15600){$c, \sigma$}
\put(8400,4800){$d, \sigma$}
\end{picture}. \\
In the preceding expressions and diagrams, the gluon $a$ has color
$a$, momentum $p_1$, and helicity index $\mu$, with analogous
relations for the remaining particles. Note that these rules become
those for QCD if one simply takes the 
limit $p_H \rightarrow 0$ and replaces the coupling $A$ with 1.
\section{The Leading Order $p_\perp$ spectrum in the large $m_{\mathrm top}$ limit}
By far, the most significant contribution to Higgs production is through
the gluonic coupling via a virtual top quark loop.   Not only is the
rate enhanced by the large top Yukawa coupling,  it is also enhanced
considerably by the large number of gluons (in comparison to quarks)
inside the proton.  Therefore, in computing the transverse momentum
spectrum, we shall restrict our calculation to
radiative corrections of the $gg \to H$ cross section. In particular,
those that produce a nontrivial transverse momentum will be
considered.   When we refer
to the {\em Leading Order} or {\em Tree Level } contribution, we mean
the lowest order in perturbation theory which produces this
nontrivial contribution.

At a proton-proton collider such as the LHC, by far the most dominant
contribution is the process $gg \to Hg$.  This is simply because the
gluon distributions are much larger than the quark distributions
at these energies.  The quark anti-quark mechanism is also small,
since the antiquark distributions are also very small.  

We would like to use our effective theory to calculate this process,
but  we have provided no arguments that the effective theory
will be any good whatsoever in this markedly different calculation.
Therefore, we will compare it to the finite top mass calculation done
in reference \cite{ElHi}.  Indeed, we will present the argument that
the effective theory does quite nicely as long as the mass scales in
the problem (in this case, $m_H$ and $p_\perp$) are fairly low in
comparison to the top quark mass.  
\subsection{Kinematics and Notation}
We begin by defining our kinematics, and establishing our notation for
the calculation.  As is the convention of parton model, we relate the
hadron momenta, $P_a$, $P_b$ to the parton momenta $p_b$, $p_b$ via
momentum fractions, $\xi_a, \xi_b$; 
\beq
p_a= \xi_a\hspace{.1in} P_a,\hspace{1in}  p_b= \xi_b\hspace{.1in} P_b.
\eeq
We will sometimes use the convenient shorthand $a=p_a$, $b=p_b$ for
the parton level momenta. It is important to note that the  final
state parton momentum will be called $Q$ in 
this work.  This differs from many similar calculations in which the
momentum of the 
final-state massive particle is referred to as $Q$. We therefore have
the following momentum conservation
relation;
\beq
p_a+p_b=p_H+Q.
\eeq
We shall define the parton Mandelstam invariants in the usual fashion:
\beqa
&s=(p_a+p_b)^2,&\nonumber \\
 t=(p_a-p_H)^2,&& u=(p_b-p_H)^2. 
\eeqa
If we define the hadronic Mandelstam variables analogously
\beq
S=(P_a+P_b)^2,\hspace{.5 in} T=(P_a-p_H)^2,\hspace{.5 in} U=(P_b-p_H)^2, 
\eeq
these are related to the partonic variables via
\beqa
& s=\xi_a \xi_b S,&\nonumber  \\
t=m_H^2(1-\xi_a)+\xi_b T ,&& u=m_H^2(1-\xi_a)+\xi_b U .
\eeqa

If we choose to evaluate the Higgs momentum in the hadron-hadron center of momentum
(COM)
frame, as is usually the convention, we may re-express the hadronic variables
as;
\beqa
&S= E_{COM}^2,& \nonumber \\ T=m_H^2-m_\perp E_{COM} \exp{(-y_H)}
,&& U=m_H^2-m_\perp E_{COM} \exp{(y_H)}. \nonumber \\
\eeqa
Here, the rapidity of the Higgs in the COM frame is denoted $y_H$, and
$m_\perp$ denotes the transverse mass ($=\sqrt{m_H^2+p_\perp^2}$).
Therefore the partonic momenta can be expressed in terms of the
traditional external variables and the momentum fractions;
\beqa
&s=\xi_a \xi_b E_{COM}^2,&\nonumber \\ \nonumber
 t=m_H^2-\xi_a m_\perp E_{COM} \exp{(-y_H)},&&
 u=m_H^2-\xi_b m_\perp E_{COM} \exp{(y_H)}. \\
\eeqa
\subsection{Analytic Procedure}
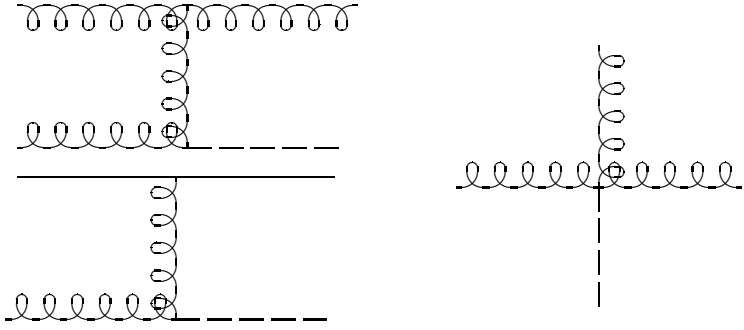
\begin{figure}
\begin{picture}(25000,12000)
\drawline\gluon[\E\REG](5000,11500)[6]
\drawline\gluon[\E\REG](\gluonbackx,\gluonbacky)[6]
\drawline\gluon[\S\REG](\gluonfrontx,\gluonfronty)[5]
\global\gaplength=300
\global\seglength=900
\drawline\Scalar[\E\REG](\gluonbackx,\gluonbacky)[5]
\drawline\gluon[\W\REG](\gluonbackx,\gluonbacky)[6]
%
%
\drawline\gluon[\S\FLIPPED](27000,10000)[5]
\drawline\gluon[\W\REG](\gluonbackx,\gluonbacky)[5]
\drawline\gluon[\E\FLIPPED](\gluonfrontx,\gluonfronty)[5]
\global\gaplength=300
\global\seglength=900
\drawline\Scalar[\S\REG](\gluonfrontx,\gluonfronty)[4]
%
%
\drawline\fermion[\E\REG](5000,5000)[6000]
\drawline\fermion[\E\REG](\fermionbackx,\fermionbacky)[6000]
\drawline\gluon[\S\REG](\fermionfrontx,\fermionfronty)[5]
\global\gaplength=300
\global\seglength=900
\drawline\Scalar[\E\REG](\gluonbackx,\gluonbacky)[5]
\drawline\gluon[\W\REG](\gluonbackx,\gluonbacky)[6]
%
%
\end{picture}
\caption{Feynman diagrams for Higgs production in the large $m_\perp$
limit.}
\label{LO_Graphs}
\end{figure}
The Higgs $p_\perp$ spectrum has three contributing partonic processes at
leading order,  $gg \to Hg$,   $q\bar{q} \to Hg$, and $gq \to
Hq$. These may be derived from the graphs in Figure \ref{LO_Graphs}
using the Feynman rules outlined above.  We find the following 
 expressions for the (helicity-averaged) matrix elements
in $d=4-2\e$ dimensions
\beqa
|\overline{M}_{gg \to gH}|^2=
\frac{24}{256(1-\e)^2}
\frac{4\alpha_s^3}{9\pi v^2}
\left\{(1-\e) \frac{s^4+t^4+u^4+m_H^8}{s t u}-4\e\, m_H^2 \right\},
\eeqa
\beq
|\overline{M}_{q\bar{q} \to gH}|^2=\frac{16}{9(36)}\frac{\alpha_s^3}{\pi\,v^2}
 \left\{\frac{t^2+u^2}{s}(1-\e)-2\e\,\frac{t u}{s}\right\},
\eeq
and
\beq
|\overline{M}_{gq \to qH}|^2= \frac{16}{9(96)(1-\e)}\frac{\alpha_s^3}{\pi\,v^2}
 \left\{\frac{s^2+u^2}{-t}(1-\e)-2\e\,\frac{s u}{-t}\right\}.
\eeq
Because of CP invariance, the 
process $gq\to qH$ is identical to the expression for
$g\bar{q}\to \bar{q}H$, and the matrix element for $q\bar{q}\to
gH$ may be derived from the latter expression using crossing
symmetry.  The
appropriate spin and color averaging  factors must, of course, be
substituted.  These have been included in the preceding expression. 


The general expression for a 2 to 2  partonic differential cross
section is given by 
\beq
\frac{d\hat{\sigma}}{d p_\perp^2 \, d y}=\frac{\d \lp{\tiny{Q^2}}\rp}{16 \pi \, s}|\overline{M}|^2.
\label{DCS_LO}
\eeq
The hadronic cross section is related to the partonic cross section
via the double convolution, as outlined in Chapter 3;
\beq
d\sigma_{A \, B} = \sum_{i} \sum_{j} \int_0^{1} d\xi_a f_{A \to
i}(\xi_a, \mu_F)
 \int_0^{1} d\xi_b f_{B \to j}(\xi_b, \mu_F)
d\hat{\sigma}_{ij}(\xi_a, \xi_b).
\eeq
Using the expression for cross section found in the
appendix D and the expression for the partonic cross section,
Eq. (\ref{DCS_LO}), we may derive the following expression for the
differential cross section: 
\beq
 \frac{d\sigma_{pp \to HX}}{d p_{\perp}^2\, d y_H}=
 \sum_{i} \sum_{j} \int_{0}^{1} d\xi_a f_{A \to i}(\xi_a, \mu_F)
 \int_{0}^{1} d\xi_b f_{B \to j}(\xi_b, \mu_F)
\,\delta(Q^2)\,\frac{1}{16 \pi s}|\overline{M_{ij}}|^2 .
\eeq
Noting that the double convolution of the delta function may be
written (also Appendix D)
\beq
\int_0^{1} d\xi_a  \int_0^{1} d\xi_b \delta(Q^2) \to
\int_{x_+}^{1} \frac{d\xi_a}{\xi_a S +T-M_H^2}+ 
\int_{x_-}^{1} \frac{d\xi_b}{\xi_b S +U-M_H^2},
\eeq
where 
\beq
x_{\pm}=\frac{m_\perp+p_\perp}{\sqrt{S}}\,\exp{(\pm y_H)},
\eeq
we arrive at our final expression for the differential cross section,
\beqa \nonumber
 \frac{d\sigma_{pp \to HX}}{d p_{\perp}^2\, d y_H}&=&
 \sum_{i} \sum_{j}
\left(\int_{x_+}^{1} \frac{d\xi_a}{\xi_a S +T-M_H^2}+ 
\int_{x_-}^{1} \frac{d\xi_b}{\xi_b S +U-M_H^2}\right)
\\&& \hspace{1 in}\times f_{A \to i}(\xi_a, \mu_F) f_{B \to j}(\xi_b, \mu_F)
 \frac{1}{16 \pi s}|\overline{M_{ij}}|^2.
\label{LOXS}
\eeqa
Note that there is only one degree of freedom in the integration, as
the momentum fractions are fixed by the on-shell condition in each term.
\subsection{Numerical Procedure and Results}
The convolution (\ref{LOXS}) must be evaluated numerically since the
parton distribution functions are empirically determined.  We
shall use the CTEQ5L parton distributions \cite{CTEQ}, and we shall
choose our scale to be the transverse energy,
$m_\perp$.  The number of (light) quarks $n_f$ is 5 throughout
the calculation. 

The finite-mass calculation proceeds identically to the
one outlined above.  One
simply replaces the effective theory expression for the matrix element
with the exact result \cite{ElHi}. Because the mass of the top quark
is so large, we expect the large $m_{\mathrm{top}}$ limit to be quite
good out to very large transverse momentum.  

\begin{figure}
\includegraphics[width=10cm,angle=-90]{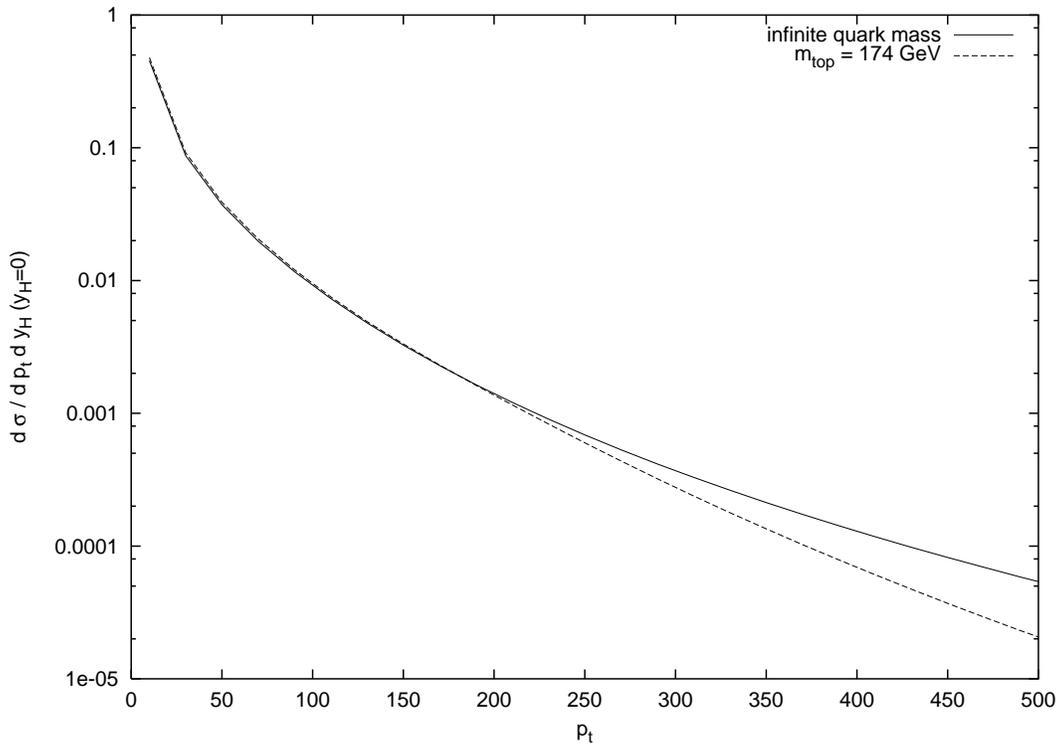}
\caption{Comparison of the exact leading-order $p_\perp$ spectrum
with the effective theory calculation.}
\label{finiteMtPt}
\end{figure}

Figure \ref{finiteMtPt} displays a graph of the transverse momentum
spectrum at zero rapidity for  a Higgs of
mass $m_h=115$ GeV calculated using the full
$m_{\mathrm{top}}$ dependence (dotted line). Superimposed over this
graph is the identical calculation performed with the simplification
that the fermion mass is taken to be infinite.  It is plain to see that
the agreement is quite good out to the very large $p_\perp $ region,
which is roughly about 300 GeV.
There is really no reason to suspect that the behavior of any of the
higher-order corrections will behave any differently than this.

\begin{figure}
\includegraphics[width=10 cm,angle=-90]{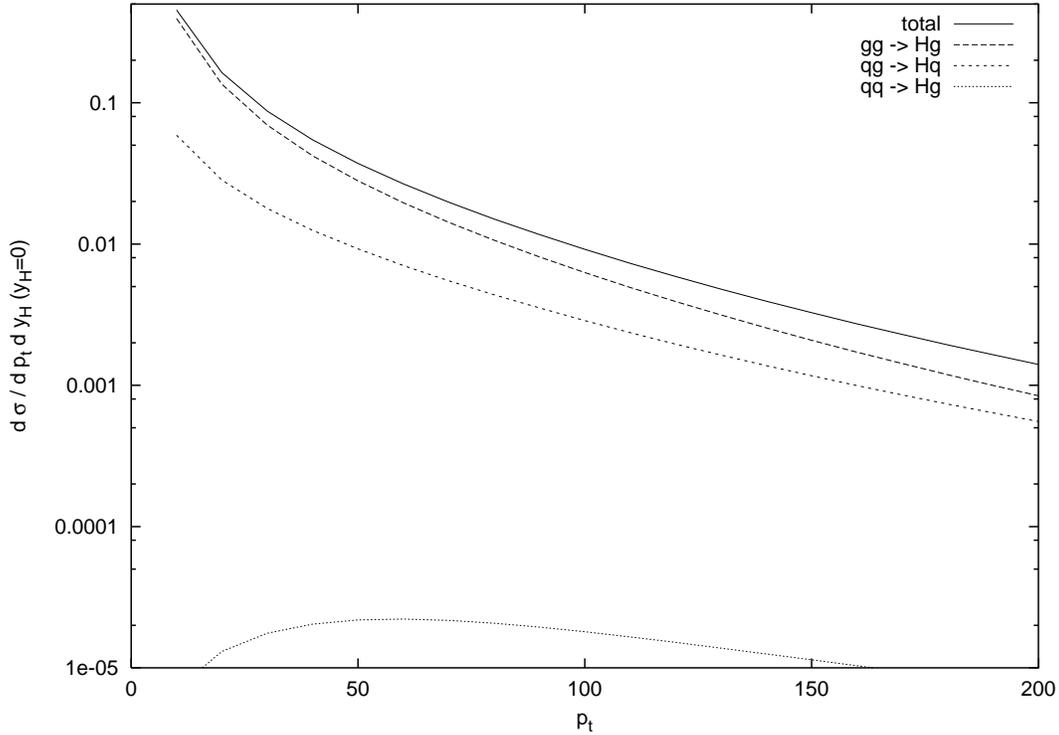}
\caption{The Higgs $p_\perp$ spectrum at leading order}
\label{LO_Full}
\end{figure}

It is advantageous to break the infinite-mass limit calculation up
into the various 
contributions from its partonic constituents, so that we may see which
partonic graphs contribute the most to the cross section.
A plot of the various contributions to the  $p_\perp$ spectrum at
leading order  for a Higgs with a mass of 115 GeV is given in Figure
\ref{LO_Full}. 
All of the terms  exhibit the expected divergent behavior as
$p_\perp$ goes to zero, with the exception of the s-channel piece,
$q\bar{q} \to Hg$.

As can be plainly seen in the graph, the dominant contribution to the
total cross section (solid black line) comes
from the gluon initiated process, $gg \to Hg$ (long dashes).
This contribution is roughly  an order of magnitude larger than the
next largest 
contribution, $qg \to Hg$ (dashed line), in the small transverse
momentum region.  Even in the large $p_\perp$ region, the gluon
initiated process is twice as large as the gluon-quark initiated one. 
 The process $q\bar{q} \to
Hg$ is completely negligible compared to these other two processes. 

To summarize, the gluon fusion process dominates at the LHC, and
working with an effective field theory in which the heavy top loop is
replaced with an effective vertex is a quite good approximation.  Even
for the inclusive cross section, the NLO pieces are found to produce
enhancements which are much larger than the difference between the
finite $ m_\mathrm{top} $  calculation and this infinite limit, at least for
Higgs masses in the expected mass range, 100-200 GeV.  Since the
differential cross sections at leading-order tend to be even more
unreliable than the  leading-order cross sections, it is
imperative that we include some of these next-to-leading order
terms in order to improve our estimate for the differential cross
section. 


\chapter{The Higgs Boson Transverse Momentum Spectrum: Gluonic
Radiative Corrections}
In the previous chapter, we found that the dominant production
mechanism for producing a Higgs boson at the LHC was that of gluon
fusion, and that this process is well represented by an effective
theory in which the massive quark in the loop is replaced by an
effective coupling between the Higgs boson and two or more gluons.

Moreover, the dominant partonic process is the gluon initiated $gg \to
Hg$ process, which is responsible for about 90\% of the total cross
section.  This is primarily due to the large number of gluons in the
protons compared to the quarks.

As we have mentioned previously, leading-order calculations in QCD 
 are at best an order of magnitude approximation to
the actual process.  This is reflected in a large scale dependence in
the leading order cross section.  If one wishes to have a more
precise prediction for the cross section, one needs to carry out the
calculation of the 
differential cross section  to next-to-leading order in
perturbative QCD. It is hoped that the resulting calculation will have
less scale dependence, and therefore, less theoretical uncertainty.

There are several different methods for doing this.  One notable
method is using a Monte-Carlo event generator to do the entire
integration \cite{CaSe}.  Since the matrix element contains soft and
collinear divergences, one must take care to extract the divergent
pieces and integrate them in $4-2 \epsilon$ dimensions in order to
cancel the divergences in the virtual pieces.  A calculation of this
sort has been done \cite{evl}.

Nonetheless, it is advantageous to do the entire calculation analytically
for several reasons.  First, one would like to calculate the process
using conventional $\overline{\mathrm {MS}}$, and show that it agrees
with the Monte Carlo calculation.  Second, one needs an analytic
formula in order to expand the matrix element around small transverse
momentum.  This is necessary to calculate the $B^{(2)}$ coefficient in
Collins-Soper-Sterman resummation scheme, and compare it to the 
result calculated in a different manner in reference \cite{B2}.  To
have a complete next 
to leading order calculation, good for all transverse momentum,  the
intermediate  $p_\perp$ calculation
included herein should be combined with a resummed curve good in the
small $p_\perp$ limit \cite{Kauffman:resum1, Kauffman:resum2,
Yuan:resum, Hinchliffe:resum}.

In this chapter, we will be computing the dominant piece of the NLO
corrections, the partonic process represented by gluonic radiation from
 the leading order process, $gg \to Hg$.  This, we will argue,
dominates the NLO cross section in the intermediate $p_\perp$ region
with which we are concerned.

To do this analytically, we will use the procedure outlined
in Chapter 3.  Since, in these corrections,  we will be required to
renormalize 
ultraviolet divergences and factorize collinear divergences, 
we will have to carefully compute the matrix elements in
$d=4-2\e$ dimensions.
The parton distributions that we will be using are calculated
from deep inelastic scattering data using the $\overline{\mathrm{MS}}$
factorization/renormalization scheme, so we must use the same scheme
in order to 
obtain a meaningful result.
Once we have a finite expression, the convolutions with the structure
functions may be evaluated numerically to obtain the cross section as
a function of the various kinematic parameters.
\section{Notational Conventions}
The notations that we will be using are in complete analogy to those of
the previous section.
The 4-momenta of the hadrons shall be taken to be $P_a$ and $P_b$, and
the 4-momenta of the  
partons shall be denoted $p_a,\ p_b\ (= \xi_{a,b}\hspace{.1in} P_{a,b} )$ .
The 
Higgs boson momentum shall be denoted $H$ and the mass $m_H$,
the transverse momentum $p_\perp$, and the rapidity $y_H$.  Therefore, the
hadronic ($S,\ T,\ U$) and partonic ($s,\ t,\ u$) Mandelstam
variables have precisely the same form as the  previous chapter.

It is also convenient to define crossed momenta,  $p_{1,2}
=-p_{a,b}$, and to label the final state partons $p_3, p_4$.  This
allows us to express the invariants of the $2 \to 3$ process in terms
of generalized Mandelstam variables, 
\beq
S_{ij}=(p_i+p_j)^2,\hspace{.5in} S_{ijk}=(p_i+p_j+p_k)^2.\hspace{.5in}
\eeq
The invariant mass $Q^2$ of the final state gluons and the Higgs
invariants $t=(a-H)^2$, $u=(b-H)^2$ are related to these via
\beq
Q^2 = S_{34}=(p_3+p_4)^2, \hspace{.5in}
t= S_{234}=(p_2+p_3+p_4)^2,
\eeq
and
\beq
 u= S_{134}=(p_1+p_3+p_4)^2.
\eeq
One may then show that the Higgs mass $M$ is related to the remaining
variables by 
\beq
M^2=s+t+u-Q^2,
\eeq
and that the transverse momentum $p_\perp$ of the Higgs may be expressed as
\beqa
p_\perp^2&=&\frac{(t-Q^2)(u-Q^2)}{s} -Q^2,
\\&=& \frac{t u -Q^2 M^2}{s}.
\eeqa

We will find it convenient to define the parameters $Q_\perp$, $\n$
as follows:
\beqa \nonumber
Q_\perp&=&\sqrt{Q^2+p_\perp^2}, \\
\n&=&Q^2/Q_\perp^2.
\eeqa
The physical interpretation of $Q_\perp$ is that it is the transverse
mass of the gluon pair.
Finally, we will also use $S_t$ and $S_u$, which are defined as
\beqa
S_u&=&S_{13}+S_{14}=u-Q^2, \\
S_t&=&S_{23}+S_{24}=t-Q^2,
\eeqa
and we will occasionally use the parameter $S_H$,
\beq
S_H=S_{123}+S_{124}=M^2+Q^2-s.
\eeq
\section{Cross Section at NLO}
To next-to-leading order, the two-to-two ($2 \to 2$) differential cross
section is: 
\beq
d\sigma_{NLO}=d\sigma^{(0)}_{2 \to 2}+d\sigma^{(1)}_{2 \to 2}
+ d\sigma^{(0)}_{2\to 3}.
\eeq
Here, $d\sigma^{(0)}_{2 \to 2}$ is the tree-level contribution,
$d\sigma^{(1)}_{2\to 2}$ is the expression for the 1-loop virtual
corrections, and
$d\sigma^{(0)}_{2 \to 3}$ is radiative corrections to the leading
order  process.  If we include a collinear counterterm 
$d\sigma^{AP}_{2 \to 2}$ and a UV counterterm $d\sigma^{UV}_{2 \to
2}$, then the cross section may be written as
\beqa
\frac{d\sigma}{dp_\perp^2 \, dy_H}& \nonumber
=&\int \int d\xi_a f(\xi_a, \mu_F) d\xi_b f(\xi_a, \mu_F) s \times \\&&
\left\{ 
\delta(Q^2) \left(\frac{d\sigma_{2 \to 2}^{(0)}}{dt} +
\frac{d\sigma_{2 \to 2}^{(1)}}{dt } 
+ \frac{d\sigma_{ 2 \to 2}^{UV}}{dt } \right)
+\frac{d\sigma_{2 \to 3}^{(0)}}{dt \, du} + 
\frac{d\sigma_{2 \to 3}^{AP}}{dt \, du} 
\right\}.
\eeqa

Since these are to be computed using the conventional dimensional
regularization,  we
must be careful to include all $\epsilon$ dependence in the latter two
terms.  Note that the expression for the $2\to 2$ differential cross
section  becomes
\beq
\frac{d \sigma_{2 \to 2}}{dt} = \left( \frac{4 \pi \mu^2}{p_\perp^2} \right)^{\e}
\frac{|\overline{M}|^2}{16 \pi s^2 \,\Gamma(1-\e)}.
\eeq 
The expressions from the other two $2 \to 2$ terms (the virtual
corrections and UV counterterm) are identical to this one with the
exception that one must substitute  the requisite matrix elements. 

The contribution coming from the radiative corrections (those
involving 5 particles) may be written
\beqa
\frac{d\sigma^{(0)}_{2 \to 3}}{dt \, du}
&=&\left( \frac{4 \pi \mu^2}{p_\perp^2} \right)^{\e}
\left( \frac{4 \pi \mu^2}{Q^2} \right)^{\e}
\frac{1}{\Gamma(1-2\e)}
\frac{1}{(2\pi) (4\pi)^2}  \nonumber \\ & \times &
2 \int_{0}^{\pi} d\theta (\sin{(\theta)})^{1-2\e}
\int_{0}^{\pi} d\phi (\sin{(\phi)})^{-2\e}\frac{ |\overline{M}|^2}{16
\pi s^2}.
\eeqa 
The contribution from the Altarelli-Parisi term is
\beqa
 \frac{d\sigma^{AP}_{2 \to 3}}{dt \, du}&=&
 \left( \frac{4 \pi \mu}{p_\perp^2} \right)^{\e}
\frac{\alpha_s}{2 \, \pi} \frac{(4 \pi)^\e}{\e \, \Gamma(1-\e)^2}
\frac{1}{16 \pi s^2 }\times 
\nonumber \\&&
\int^{1}_{0} \frac{dz_a}{z_a} P(z_a) \delta( z_a (s+t-M^2)+u)
|\overline{M(\xi_a z_a,\xi_b)}|^2 +(t \leftrightarrow u).
\eeqa
In the preceding expressions, we must note that we have suppressed the
sum over parton species for clarity. 

The angles $\theta$, $\phi$ in the expression for the $2 \to 3$ cross
section will be the polar angles of the final state partons in
the $Q^2$ rest frame.  The $z$ axis in this frame will be
defined by either initial state parton or the Higgs boson, whichever
is convenient.   See Appendix B for a further discussion on the
explicit evaluation of these angular integrals.  Once these
integrations are calculated, we need only concern ourselves with the
end result.
\section{Angular Integration of Real Emission Piece}
To calculate the real contribution, we must integrate the 
two angular degrees of freedom occuring in the matrix element.
 While there
are a  large number of terms to integrate, the
calculation lends itself to an algorithm very nicely.   The basic
procedure for the angular integration is the following:
\begin{enumerate}
\item Partial fraction the amplitudes using the relationship amongst
the various Mandelstam invariants.  This will reduce all of the
integrals to the form $(\mathrm{const}) \times S_{ijk}^a S_{mn}^b$ or
$(\mathrm{const} ) \times S_{ij}^a S_{mn}^b$, for which standard formulae
exist.  Here, const is some collection of kinematic invariants that does not
depend on the angular variables $\theta$ and  $\ \phi$.
\item Extract the terms proportional to $1/(1+\e)$ and verify
that they cancel amongst themselves.  This verifies that there are no
quadratically divergent pieces in the amplitude (which is true in
general for QCD).
\item Extract the universal $-1/\e$ piece and verify that it
cancels algebraically against the collinear counterterm.
\item Regularize the soft divergences ($Q^2 \to 0$), and extract the
universal soft pole. This will cancel against the soft pole in the
virtual piece.
\end{enumerate}
Once this is done, the real matrix element is integrable over the
convolution space outlined in Appendix D. The final integral must, of
course, be  done numerically.

As discussed in chapter 3, the helicity amplitudes obtained in the
literature are computed using gluons with two polarizations.  To use
$\mathrm{\overline{MS}}$, we must have expressions for the matrix
elements with  $2-2\e$ gluon polarizations. Therefore, in
order to use $\mathrm{\overline{MS}}$, we will
need to  calculate the
$O[\e]$ corrections to these expressions due to higher dimensional
gluon spin effects.  This calculation is much
simpler than the original matrix element calculation, since we need
only to calculate the matrix element in the collinear limit.  We will
explain this shortly.
\subsection{$gg \to Hgg$}
We now restrict ourselves to the process {$gg \to Hgg$}.
To begin, we first break the cross section into two sections, a piece
which represents the sum over the $(++++)$ and $(--++)$ helicity
configurations, and a piece which represents a sum over the
odd-helicity components, $(-+++)$.  Let
\beq
\frac{d\hat{\sigma}^{(0)}_{gg \to Hgg}}{dt du} =
\left(\frac{d\hat{\sigma}_{gg \to Hgg}}{dt du}\right)_1+
\left(\frac{d\hat{\sigma}_{gg \to Hgg}}{dt du}\right)_2,
\label{eqn:break}
\eeq
with
\beqa
\left(\frac{d\hat{\sigma}_{gg \to Hgg}}{dt du}\right)_1 &=&
\Gamma(1-\epsilon)
\left( \frac{p_\perp^2}{4\pi \, \mu^2}\right)^{-\e} \nonumber
\int \frac{d\Phi}{16 \pi^2 s^2} \times \\ && \nonumber
 2 \left(|M(1^+2^+3^+4^+)|^2+|M(1^-2^+3^+4^-)|^2+ \right. \\&&  \left.
|M(1^+2^-3^+4^-)|^2+|M(1^+2^+3^-4^-)|^2
\right),
\eeqa
and
\beqa 
\left(\frac{d\hat{\sigma}_{gg \to Hgg}}{dt du}\right)_2 & =&
\Gamma(1-\epsilon)
\left( \frac{p_\perp^2}{4\pi \, \mu^2}\right)^{-\e} \nonumber
\int \frac{d\Phi}{16 \pi^2 s^2} \times \\ &&
2 \left(|M(1^-2^+3^+4^+)|^2 +{\mathrm{3\ cyclic\ spin\ perms}}\right).
\eeqa
In these expressions, the  remaining parts of the $2 \to 3$ phase
space $d\Phi$ may be
succinctly written
\beq
d\Phi =  \frac{1}{8 \pi^2} 
\frac{\Gamma \lp 1-\e \rp}{\Gamma \lp 1-2\e \rp} \lp \frac{4 \pi \mu^2}{Q^2} \rp^\e 
 d\Omega^{(\epsilon)},
\eeq
where $d\Omega^{(\e)}$ is the $4-2\epsilon$ angular integration measure,
\beq
 d\Omega^{(\epsilon)} = \frac{1}{2 \pi} 
 \int_{0}^{\pi} d\theta (\sin{(\theta)})^{1-2\e}
\int_{0}^{\pi} d\phi (\sin{(\phi)})^{-2\e},
\eeq
which is defined in Appendix B.

First, we define
\beq
 \nonumber \hspace{-.5in}
\sigma_\e = 
\frac{\a_s^2}{256 \pi \, s^2}
\frac{N_c^2}{N_c^2-1}
\left( \frac{ \a_s}{3  \p \, v_e^2 } \right)^2
 \left( \frac{4 \p \m}{Q^2} \right)^{\e} 
 \left( \frac{4 \p \m}{p_\perp^2} \right)^{\e} 
\frac{1}{\Gamma(1-2\e)}.
\eeq
Then, the angular integral of the first contribution in
(\ref{eqn:break}) yields 
\beqa
\lp \frac{d\hat{\sigma}_{gg \to Hgg}}{dt du} \rp_1 &=& \nonumber
4 \sigma_\e \times
\left\{ -\left( \frac{1}{\e}\right) \frac{M^8+\s^4+\su^4+\st^4+
Q^8}{{\hspace{-.3mm} \s^2}\, Q^2 \, p_\perp^2}\times
\right. \\&& \nonumber\left.
 \left[
(1+\n)(1-\n)^{-\e}
+ (1-\n)\n^{-\e} (1+\frac{\e^2 \pi^2}{6}) 
\right] + \right. \\&& \nonumber 
\frac{\su^4+\st^4}{{\hspace{-.3mm} \s^2}\, Q^2 \, p_\perp^2} \left(-\frac{67 \e}{18 }
+\frac{1-\n}{6}(40\n^3-42\n^2+30\n-11)
\right.  \\  && \hspace{.7in} \left. \nonumber 
-\frac{(\n^4 + (1-\n)^4-1)}{\e}
\right)
\\ &&
+ \left.  4 \left( -\frac{2\n}{1-\n} +\ln(1-\n ) 
\right)+\frac{\e}{3 \n}+ \frac{17}{3}
\right\}. 
\eeqa
Noteworthy are the terms that behave like
$\frac{\ln(1-\n)}{1-\n}$ and $\frac{1}{p_\perp^2}$, which diverge as
$p_\perp \to 0$, as well as 
the universal $\frac{1}{\e}$ pole.  

For the $(-,+,+,+)$ terms, we write the results as a sum of the
universal divergent pieces and an analytic piece which is finite as
$Q$ and $p_\perp$ go to zero.  We find from the symmetries of the
matrix element that there are in actuality only 4 independent terms,
and that the sum over the odd-helicity components may written
\beqa
\sum_{\mathrm{colors,}(-,+,+,+)} |M|^2 &\rightarrow& \nonumber
4 \left\{ \left( m(1-,2+,3+,4+) + \frac{1}{2} m(1-,4+,2+,3+) 
+(u \leftrightarrow t) \right) 
\right. \\&& \left. +2 \hspace{.1in} m(3-,4+,1+,2+)+m(3-,2+,4+,1+) 
\right\}.
\eeqa
Using this relationship, we may write
\beqa
\left(\frac{d\hat{\sigma}_{gg \to Hgg}}{dt du}\right)_2 &=& 2\, \sigma_\e
\left\{ 
\left[
2\, \Omega(1,2,3,4)+2\, \Omega(3,4,1,2)+\Omega(1,4,2,3)+\Omega(3,2,4,1)
\right] \nonumber  \right. \\ && \left. \hspace{1in}+(u \leftrightarrow t)
\right\},
\eeqa
where the $\Omega$'s are defined as the angular integrations of the
helicity subamplitudes, e.g.
\beq
\Omega(1,2,3,4) = \int \frac{d\Omega}{2\pi} |m(1^+,2^-,3^-,4^-)|^2,
\eeq
with similar relationships for the other subamplitudes.

The angular integration over the (1-, 2+, 3+ 4+) helicity
configuration may be written as a sum over two terms, a term that is
finite as $Q$, $\e \to 0$ which we shall call $\Omega_{0}(1,2,3,4)$,
and a term which contains the pole structure, which we shall call 
$\Omega_{div.}(1,2,3,4)$;  i.e.,
\beq
\Omega(1,2,3,4)=\Omega_{0}(1,2,3,4)
	 + \Omega_{\mathrm div.}(1,2,3,4).
\eeq
The second term in this equation is
\beqa \nonumber
\Omega_{\mathrm div.}(1,2,3,4)&=& 
-\frac{1}{\e} \frac{1}{Q^2 s^2 p_{\perp}^2}
 \left[ \left( \frac{M^2 Q^2}{u}\right)^4+ \left( \frac{s
p_\perp^2}{u} \right)^4  +t^4 -2
\frac{M^4 t^3 s Q^2}{u (M^2-u)(M^2-t)}\right]  
\\ &&
-\frac{1}{\e} \frac{2 M^4 t^3}{s \hspace{.05in} Q_\perp^2 u (
M^2-t)(M^2-u)} (1-\nu)^{-1-\e} ,
\eeqa
and the term $\Omega_{0}(1,2,3,4)$ may be found in Appendix C.
The matrix element with the same color ordering, but with the
final-state gluons having the opposing helicities, integrates to
\beqa \nonumber
\Omega_{\mathrm div.}(3,4,1,2)&=&  \left\{ 
-\frac{1}{\e} \frac{1}{Q^2 s^2 p_\perp^2} \left[ \left(\frac{s
u}{u-Q^2} \right)^4 + \left(\frac{M^2 (Q^2-t)}{ t} \right)^4 + 
\left( \frac{s Q^2 p_\perp^2}{(t-Q^2) t} \right)^4 \right. \right. \\
\nonumber 
&&\left. \left. -\frac{2 Q^2 s M^4
(M^2-u)^3}{(M^2-t) t u} \right] 
 - \frac{1}{\e}\frac{2 M^4(M^2-u)^3}{(M^2-t) s t u
Q_\perp^2}(1-\n)^{-1-\e}  \right. \\ && \left. +
\left(-\frac{11}{6}-\frac{67 \e}{18}
\right)\frac{M^8+s^4}{Q^2 s^2 Q_\perp^2}-\frac{\e}{3}\frac{M^4}{Q^2
Q_\perp^2} \right\}.
\eeqa
The term with the complicated pole structure (1324) yields
\beqa \nonumber
\Omega_{\mathrm div.}(1,3,2,4)&=& \left\{
-\frac{2}{\e} \frac{1}{Q^2 s^2 p_{\perp}^2}
 \left[ \left( \frac{M^2 Q^2}{u}\right)^4+ \left( \frac{s
p_\perp^2}{u} \right)^4  +t^4(1-2 \frac{s p_\perp^2}{t u})
\right. \right.\\ && \left. \left.
 -2 \frac{M^4 t^3 s Q^2}{u (M^2-u)(M^2-t)}  \right]  
-\frac{2}{\e} \frac{2(1-\nu)^{-1-\e} M^4 t^3}{s \hspace{.05in} Q_\perp^2 u (
M^2-t)(M^2-u)}  \right. \nonumber \\ && -\frac{2}{\e}
\left. \frac{2 t^3}{u s Q_\perp^2} \n^{-1-\e} {\tiny {(1+ \e^2 { \pi^2 \over 6})}}
\right\}
\eeqa
for the term where the initial state particles have the opposing
helicities, and
\beqa \nonumber
\Omega_{\mathrm div.}(3,2,4,1)&=& \left\{ 
-\frac{1}{\e} \frac{1}{Q^2 s^2 p_\perp^2} \left[ \left(\frac{s
u}{u-Q^2} \right)^4 + \left(\frac{M^2 (Q^2-t)}{ t} \right)^4 + 
\left( \frac{s Q^2 p_\perp^2}{(t-Q^2) t} \right)^4 \right. \right. \\
\nonumber 
&&\left. \left. -\frac{2 Q^2 s M^4
(M^2-u)^3}{(M^2-t) t u} - \frac{s p_\perp^2}{tu}(M^8+Q^8+s^4)\right]
\right. \\ && \left.
 - \frac{1}{\e}\frac{2 M^4(M^2-u)^3}{(M^2-t) s t u
Q_\perp^2}(1-\n)^{-1-\e} -\frac{1}{\e} 
\frac{M^8+Q^8+s^4}{s Q_\perp^2 t u}\n^{-1-\e} (1+ \e^2 { \pi^2 \over 6})
\nonumber  
\right.  \\&& \left. +(u \leftrightarrow t)   \right\}  
\eeqa
for the terms that have the final-state gluons with opposing
helicities.  Again,  the expressions for the $\Omega_{0}$'s may be
found in Appendix C. 

In using the helicity amplitudes to calculate the amplitude squared,
we have neglected terms of $O[\e]$ which appear naturally from the sum
over spins.  If these $O[\e]$ spin  pieces are proportional to a collinear
singularity, they will contribute to the finite pieces left
over from the collinear cancellation, and hence, the cross section.
Therefore, it is necessary to calculate these terms in order to
complete the next-to-leading order calculation.

While it may seem formidable to calculate these terms, one realizes
that one only needs the terms proportional to  collinear poles
in the matrix 
element, since the remaining terms vanish when the limit $\epsilon \to
0$ is taken.
The correction term may be calculated in this manner.  For the (1234)
color ordering, we find the correction term to be
\beq
|\overline{M_{cor.}}|^2=
 - 2 \e \lp A \rp^2 g^2 \frac{ N^2}{(N^2-1)}
  \lp \frac{ S_{14} S_{23}-S_{13} S_{24} }{S_{34}} \rp^2 
	\lp \lp\frac{M^2}{S_{134} S_{234} }\rp^2 + \frac{1}{S_{12}^2} \rp.
\eeq
In this expression, everything save the 
$  \lp  S_{14} S_{23}-S_{13} S_{24} \rp^2 / S_{34}^2 $
is evaluated in the limit  
$S_{34} \to$~$0$.  The remaining divergent pieces (as $S_{23}\to0$ and
$S_{14}\to 0$) may be obtained from this one by cyclically permuting 
the momenta in the above expression for the amplitude, and taking the
subsequent limit.  Upon integrating this term, we find the following
correction to the (1234) piece,
\beq
\sum_{poles}\int \lp \frac{d\Omega}{2 \pi} \rp \frac{1}{2} 
|\overline{M_{cor.}}|^2_{1234} =
\lp - 2 \e \lp A \rp^2 g^2 \frac{ N^2}{(N^2-1)} \rp 
\lp T_1+ T_2 +T_3 \rp,
\eeq
where
\beq
T_1=
\frac{1}{3} 
\frac{p_\perp^4}{s Q^2}
\lp 1+\frac{M^4}{p_\perp^4} \rp,
\eeq
\beq
T_2=\frac{-2}{\e} \lp \frac{Q^2}{p_\perp^2} \rp 
\lp 1+\frac{M^4 p_\perp^4}{u^4} \rp,
\eeq
and
\beq
T_3 = T_2 (u \leftrightarrow t).
\eeq
The integral over the (1342) ordering yields precisely the same
result.  The (1423) piece simply gives
\beq
\sum_{poles}\int \lp \frac{d\Omega}{2 \pi} \rp \frac{1}{2} 
|\overline{M_{cor.}}|^2_{1423} =
\lp - 2 \e \lp A \rp^2 g^2 \frac{ N^2}{(N^2-1)} \rp 
\lp 2 T_2 +2 T_3 \rp.
\eeq
Note that, when added to the previous expression, all of the
extraneous pole behavior of the form $\e/Q^2$  is eliminated.
\section{Extraction of Soft Divergences and Collinear
Renormalization}
The expression for the real piece contains collinear singularities,
which must be absorbed into the PDF's. The expression for the
splitting function is
\beqa \nonumber
s \frac{d \sigma}{dt}_{\mathrm AP}=
\frac{1}{ 16 \pi s} \int^{1}_{0}\frac{ dz}{z}\, \frac{1}{\e}
\frac{\a_s \,  P_{b' \to b}(z) }{2 \pi \, \Gamma[1-2\e]}
 \, |{\large M^{(0)}_{a, b}}|^2(\xi_a, z \xi_{b'})\,
(4 \pi)^{-\e} \, \d (z (t-Q^2)-t)\\+
 \frac{1}{ 16 \pi s} \int^{1}_{0}\frac{ dz}{z}\, \frac{1}{\e}
\frac{\a_s \,  P_{a' \to a}(z) }{2 \pi \, \Gamma[1-2\e]}
 \, |{\large M^{(0)}_{a, b}}|^2(z \xi_{a'}, \xi_b)\,
(4 \pi)^{-\e} \, \d (z (u-Q^2)-u).
\eeqa
Using the $\d$ function to evaluate the $z$ integrals gives the
following:
\beqa \nonumber
s \frac{d \sigma}{dt}_{\mathrm AP}&=&
 \,
\frac{(4 \pi)^{-\e}}{ 16 \pi  s} \, \frac{1}{\e}
\frac{\a_s \,  P_{b' \to b}(z_t) }{2 \pi \, \Gamma[1-2\e]}
 \,\frac{ |{\large M^{(0)}_{a, b}}|^2(\xi_a, z_t \xi_{b'})}{-t}
 \\&+& \,
 \frac{(4 \pi)^{-\e}}{ 16 \pi s} \, \frac{1}{\e}
\frac{\a_s \,  P_{a' \to a}(z_u) }{2 \pi \, \Gamma[1-2\e]}
 \, \frac{|{\large M^{(0)}_{a, b}}|^2(\xi_u z_{a'}, \xi_b)}{-u},
\label{eqn:APpiece}
\eeqa
where the parton momentum fractions are defined via the delta
functions to be
\beq
z_{t,u}=\frac{t,u}{t,u-Q^2}.
\eeq


For the process in question, the matrix elements above become
\beq
|{\large M^{(0)}_{gg \to gH}}|^2(\xi_a, z_t \xi_{b'})
=
\frac{g^2 A^2 N_c (N_c^2-1)}{4(N_c^2-1)^2}
\frac{M^8+(z_t s)^4+t^4+\lp \frac{ z_t s p_\perp^2 }{t} \rp ^4} {z_t^2 s^2 p_\perp^2},
\eeq
while the splitting kernel $P_{g\to g}(z_t)$ is
\beq
P_{g\to g}(z_t) = \frac{\lp 1+(1-z_t)^4 +z_t^4 \rp }{z_t \lp 1-z_t \rp_+}
+ \beta_0 \d(1-z_t).
\eeq
The second term in (\ref{eqn:APpiece}) is related to this one by $u
\leftrightarrow t$.

The entire pole behavior of the matrix element may be expressed as
\beqa \nonumber
s \frac{d\sigma}{dt \, du}
&=& \sigma_\e \left\{
	-\frac{1}{N_c\e} \frac{P^{\e}_{g\to g}(z_t)}{-t} 
\frac{M^8+(z_t s)^4+t^4+\lp  z_t s p_\perp^2 / t  \rp ^4} {z_t^2 s^2 p_\perp^2}
\right.  \\&& \left. \nonumber 
-\frac{1}{\e} \lp \n^{-\e} \lp 1+ \frac{\pi^2 \e^2}{6} \rp -1\rp
\frac{M^8+(z_t s)^4+t^4+\lp z_t s p_\perp^2 / t  \rp ^4} {z_t^2 s^2
p_\perp^2 Q^2}
\right.  \\&& \left.
-\frac{1}{2}\lp\frac{11}{6} +\frac{67 \e}{18} \rp 
\frac{M^8+(z_t s)^4+t^4+\lp z_t s p_\perp^2 / t  \rp ^4} {z_t^2 s^2
p_\perp^2 Q^2} +(t \to u) 
\right\} \nonumber
\\&& + {\mathrm{Reg.}},
\label{eqn:RealPole}
\eeqa
where ``Reg.''  represents all of the remaining nonsingular (as $Q^2 \to 0$) 
pieces of the
cross section.  The functions $P^{\e}_{g\to g}$ are defined to be
\beq
P^{\e}_{g\to g}(z) = N_c \frac{(1+z^4+(1-z)^4)}{z(1-z)}.
\eeq
This term, when multiplied by the $Q^{-2\e}$ implicit in $\sigma_\e$,
contains both soft and collinear poles.  We must take care to extract
the poles from this term, and cancel them accordingly. In addition,
the term in the second line is also poorly behaved as $Q^2 \to 0$,
although it is free of any collinear divergences.

%
In order to extract the infrared pole from the expression
(\ref{eqn:RealPole}), so that we may cancel the remaining (collinear) pole
against the Altarelli-Parisi expression (\ref{eqn:APpiece}), we make use
of the identity
\beqa \nonumber
\left( \frac{1}{Q^2} \right)^{1+n\e}=-\frac{1}{ n \e}(-\t)^{-n\e} \d(Q^2)
&+&  \left( \frac{z_t}{-t} \right)^{1+n \e} \left\{ \left(
\frac{1}{1-z_t} \right)_{+}-n \e
\left(\frac{\ln(1-z_t)}{1-z_t} \right)_{+} \right\}.
\eeqa

%
%
This allows us to write the pole term as
\beqa
s \frac{d\sigma}{dt \, du} \nonumber
&=& \sigma_\e \left\{
\lp \frac{\pi^2}{12} +\frac{1}{\e^2} \left[ 1+\frac{1}{2} \lp\frac{p_\perp^2}{-t}  \rp^{\e} \right] +\frac{\beta_0}{\e N_c} + \frac{1}{2\e}\frac{11}{6} \rp 
\times
\right.  \\&& \left. \nonumber 
\hspace{.75in}
 \lp\frac{ -t}{\mu^2}\rp^{-\e} \d(Q^2) 
 \frac{M^8+s^4+t^4+  u ^4} {s^2 p_\perp^2} 
\right.  \\&& \left. \nonumber 
	-\frac{1}{N_c\e} \frac{P_{g\to g}(z_t)}{-t} \frac{M^8+(z_t s)^4+t^4+\lp  z_t s p_\perp^2 / t  \rp ^4} {z_t^2 s^2 p_\perp^2}
\right.  \\&& \left. \nonumber 
+\frac{z_t}{-t} \left[ \lp\frac{\log(1-z_t)}{1-z_t}\rp_+ - \log\lp -\frac{z_t \mu^2}{ t} \rp \lp\frac{1}{1-z_t} \rp_+  \right] 
 \times
\right.  \\&& \left. \nonumber 
\hspace{.75in}	\frac{1}{N_c\e} \frac{P^{(0)}_{g\to g}(z_t)}{-t} \frac{M^8+(z_t s)^4+t^4+\lp  z_t s p_\perp^2 / t  \rp ^4} {z_t^3 s^2 p_\perp^2}
\right.  \\&& \left. \nonumber 
+\frac{z_t}{-t} \left[ \lp\frac{\log(1-z_t)}{1-z_t}\rp_+ - \log\lp -\frac{z_t Q_\perp^4}{\mu^2 t} \rp \lp\frac{1}{1-z_t} \rp_+  \right] \times
\right.  \\&& \left. \nonumber 
\hspace{.75in} \frac{M^8+(z_t s)^4+t^4+\lp z_t s p_\perp^2 / t  \rp ^4} {z_t^2 s^2
p_\perp^2 }
\right.  \\&& \left.
-\frac{1}{2}\frac{11}{6}\lp\frac{1}{1-z_t} \rp_+ 
\frac{M^8+(z_t s)^4+t^4+\lp z_t s p_\perp^2 / t  \rp ^4} {z_t s^2
p_\perp^2 (-t)} +(t \to u) 
\right\} \nonumber
\\&& + {\mathrm{Reg.}},
\eeqa
where $P_{g\to g}(z)$ is the gluon splitting kernel, and
\beq
P^{(0)}_{g\to g}(z) = N_c\frac{ 1+z^4+(1-z)^4 }{z}.
\eeq
The second line now
cancels identically against the gluonic contribution to the
Altarelli-Parisi counterterm; equation (\ref{eqn:APpiece}). The
remainder is finite with the exception of the infrared pole
proportional to the delta function. It should be noted that while this
expression is now well behaved for any invariant mass $Q^2$, it is poorly
behaved as $p_\perp \to 0$.
\section{Virtual Contribution}
The second term that contributes to the NLO cross section is the
one-loop virtual correction \cite{CarlVirt}.  This expression contains both
ultraviolet and soft divergences.  The ultraviolet divergence cancels
against the charge renormalization counterterm, whereas the soft
divergence cancels against the soft divergence in the real
contribution.  

The complete 1-loop correction, including the ultraviolet
renormalization piece, may be written as follows:
\beqa \nonumber
s \frac{d \sigma}{dt du}& = & 4\sigma_\e \lp 1+\frac{\pi^2}{3} \rp U_\e
 \frac{M^8+s^4+t^4+u^4}{s t u} + \\&&
\frac{4}{3}\lp 1-\frac{n_f}{N_c} \rp \sigma_\e \frac{s^2 p_\perp^2 M^2+M^4 s (M^2-s+p_\perp^2)}{s^2 p_\perp^2}.
\eeqa
In this expression, $U_\e$ is the universal singular piece, which has
been renormalized as described in Chapter 3 and
Appendix B. Explicitly, it is
\beqa \nonumber
U_\e &=& -\frac{1}{\e^2} \left[ \lp \frac{\mu^2}{-t} \rp^{\e} +\lp
\frac{\mu^2}{-u} \rp^\e +\lp \frac{\mu^2}{s} \rp^\e \right]
+\frac{\pi^2}{6} 
+\frac{11}{N_c}+ \frac{3}{\e} \frac {\beta_0}{N_c}\\&& \nonumber
+\log\lp\frac{m_H^2}{s} \rp^2
+\log\lp\frac{m_H^2}{m_H^2-t} \rp^2
+\log\lp\frac{m_H^2}{m_H^2-u} \rp^2
\\&& \nonumber
-\log\lp\frac{s}{m_H^2}\rp\, \log\lp\frac{-t}{m_H^2}\rp
-\log\lp\frac{s}{m_H^2}\rp\, \log\lp\frac{-u}{m_H^2}\rp
-\log\lp\frac{-u}{m_H^2}\rp\, \log\lp\frac{-t}{m_H^2}\rp \\&&
+2  \mathrm{Li}_{2} \lp 1-\frac{m_H^2}{s} \rp
+2  \mathrm{Li}_{2} \lp \frac{m_H^2}{m_H^2-t} \rp 
+2  \mathrm{Li}_{2} \lp \frac{m_H^2}{m_H^2-u} \rp.
\label{eqn:Uep}
\eeqa
In this expression, we have rewritten the dilogarithm
functions $\mathrm{Li}_{2}$ using several transformation formulae.  The above expression
is completely real over the physical range of the parameters.
\section{Cancellation of Soft Singularities and\\ Renormalization of
Ultraviolet Singularities}
When the contribution from the virtual matrix element (\ref{eqn:Uep}) is
combined with 
the term proportional to $\delta(Q^2)$ from  (\ref{eqn:RealPole}),
we find that all of the singular terms cancel, up to terms of order
$n_f$, which we are neglecting.  Therefore the finite remainder of the
cancellation of the soft singularities $U_{\mathrm{reg}}$ is
\beqa
\nonumber
U_{\mathrm{reg}} &=& 
+\frac{\pi^2}{3}+ \frac{67}{18}+\frac{11}{N_c}
        + \frac{1}{2}\log\lp\frac{u}{t}\rp^2
        +\frac{11}{12}\log\lp \frac{\mu^4 }{s p_\perp^2} \rp
\\&& \nonumber
+\log\lp\frac{m_H^2}{s} \rp^2
+\log\lp\frac{m_H^2}{m_H^2-t} \rp^2
+\log\lp\frac{m_H^2}{m_H^2-u} \rp^2
\\&& \nonumber
-\log\lp\frac{s}{m_H^2}\rp\, \log\lp\frac{-t}{m_H^2}\rp
-\log\lp\frac{s}{m_H^2}\rp\, \log\lp\frac{-u}{m_H^2}\rp
-\log\lp\frac{-u}{m_H^2}\rp\, \log\lp\frac{-t}{m_H^2}\rp \\&&
+2  \mathrm{Li}_{2} \lp 1-\frac{m_H^2}{s} \rp
+2  \mathrm{Li}_{2} \lp \frac{m_H^2}{m_H^2-t} \rp 
+2  \mathrm{Li}_{2} \lp \frac{m_H^2}{m_H^2-u} \rp \nonumber
\\&& + O[ n_f ],
\label{eqn:Ureg}
\eeqa
where we have set  $n_f$ equal to zero.
To complete the cancellation of the soft poles, the soft singularities
resulting from 
the process $gg \to Hq\bar{q}$ must be also
included. This piece  contributes
the remaining $n_f/6$ piece required to cancel the (complete) $\beta_0$ pole.

For the present calculation, we will assume that these
fermionic pieces do cancel, and that the finite remainder
contributes very little to the overall cross section.  With this
simplification, we take $\beta_0 \to N_c 11/6$, and the pole cancellation is
complete.  While this will probably give us a very good approximation
to the cross section, the gluon initiated process $gg \to H q \bar{q}$
is probably of the same order of magnitude as the pure gluonic
process \cite{KaDeRi}, and must eventually be included in a complete
analysis.
\section{Numerical Procedure and Results}
In order to generate a transverse-momentum spectrum that may be
observed in nature, we must numerically evaluate the convolutions of
the preceding quantities with the parton distribution functions.  We
choose to use the CTEQ5 structure functions \cite{CTEQ}.  The
convolutions of the matrix elements with these structure functions
must be evaluated numerically.  We use a simple Gaussian integration
routine, as found in Numerical Recipes for C \cite{NumRes}.

With this code, we may generate the differential cross section for any
transverse momentum or rapidity (assuming of course that these
are kinematically allowed). The process
is to be plotted as observed at the LHC, a proton-proton collider
running at a COM 
energy of $\sqrt{S}=14$ TeV. Unless otherwise stated, we shall use a
Higgs mass of 115 GeV in our evaluations. 




The leading order cross section, when shown, was computed using the
 CTEQ5L distributions, together with
the leading order definition of $\alpha_s$.  
The next-to-leading order cross section will always be computed using
the NLO definitions of the CTEQ PDF's, namely the CTEQ5M1 distributions.
In particular, we have
taken care to define the K factor as it is traditionally defined, as
the ratio of the NLO cross section (with NLO PDF's and $\alpha_s$) to
the LO cross section (with LO PDF's and $\alpha_s$).  Also, all
contributions involving quarks have been ignored in this calculation
\footnote{For the sake of consistency, we use only the gluon-initiated graphs
in the LO piece.}.  We do however, keep $n_f=5$ in the parton
distributions and the strong coupling $\a_s$.

We shall begin with a discussion of the transverse momentum spectrum,
focusing on how it differs from the leading order behavior.  Then, the
scale dependence is addressed.  We close by looking at the dependence
on rapidity and the behavior of the cross section at small transverse
momentum.
\subsection{$p_\perp$ Spectrum and the $K$ Factor}
%

\begin{figure}
\begin{picture}(2000,20000)
\put(0,16000){$\mathtt{\frac{d\sigma}{d p_\perp\, d y_H},}$}
\put(900,14000){{$\mathtt{\frac{pb}{GeV}}$}}
\end{picture}
\scalebox{.8}{\includegraphics{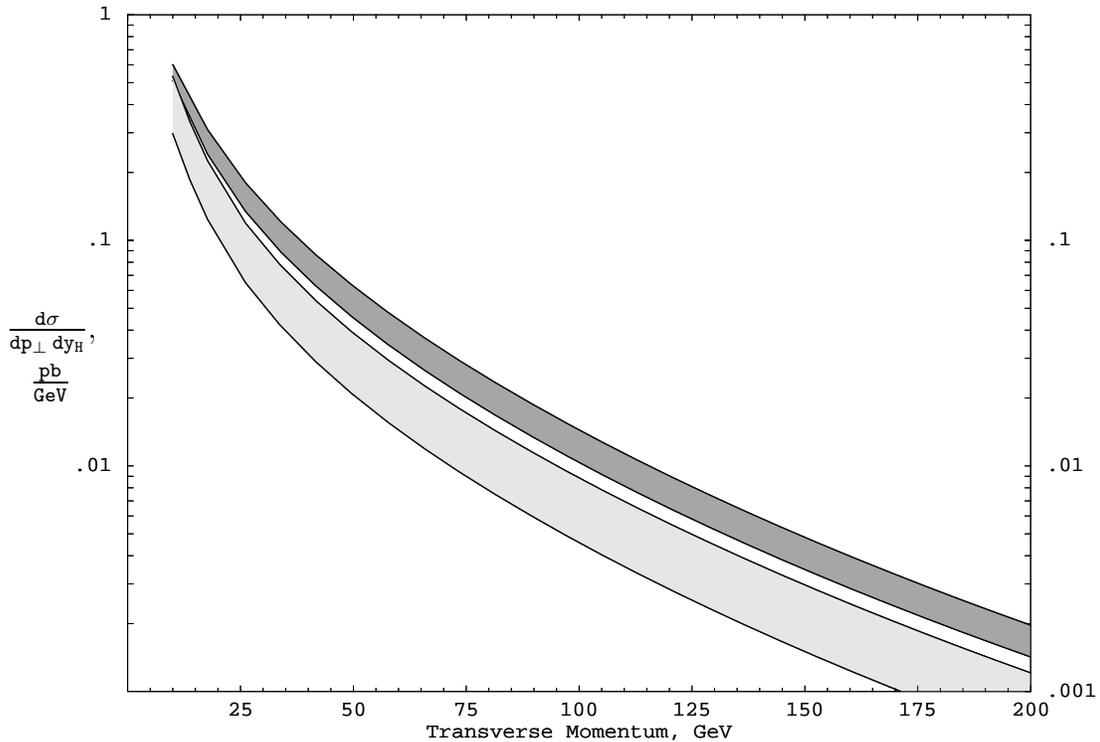}}
\caption{Variation of the $p_\perp$ spectrum with scale}
\label{graph:ptscale}
\end{figure}

Theoretically, the dramatic scale dependence of the cross section at
LO should be reduced significantly at NLO.  Figure \ref{graph:ptscale}
shows the cross section at NLO superimposed over the cross section at
leading order over a wide range of rapidity.  We have chosen to plot
bands in this case in which the scale has been varied from $.5 m_\perp
\to 2 m_\perp$.   The
leading order graph (light grey) clearly varies more over the chosen
scale range  than the next-to-leading
order graph.  Again, this improvement is exactly what one expects for
an NLO calculation, as the scale dependence represents the theoretical
uncertainty in the calculation.  As we improve our calculation
order-by-order in QCD, our dependence on this artificial parameter should
decrease.

\begin{figure}
\scalebox{.9}{\includegraphics[angle=-90,width=15cm]{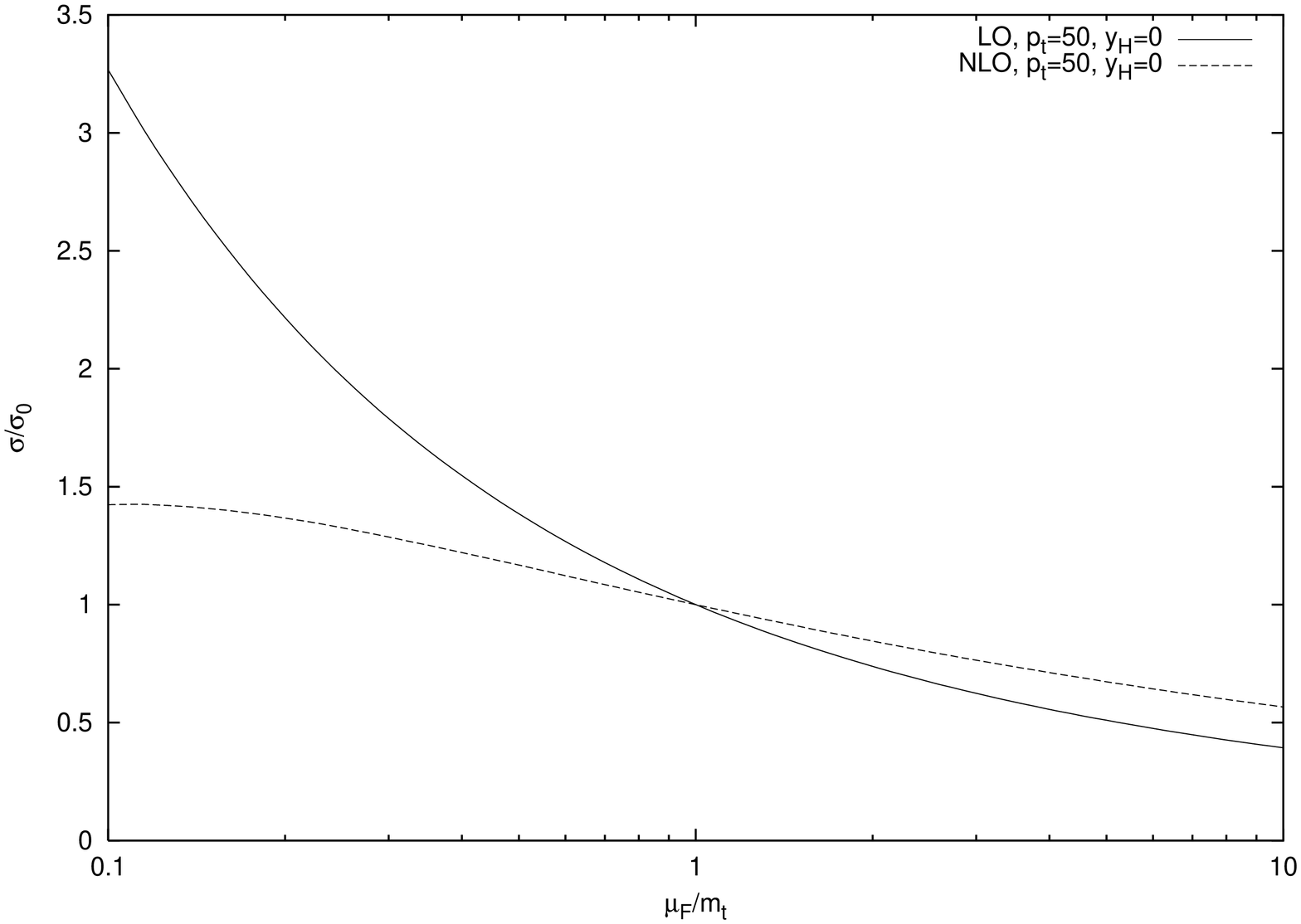}}
\caption{Scale dependence at $p_\perp =50 \mathrm{GeV}$}
\label{graph:scale}
\end{figure}

The graph of Figure \ref{graph:scale} displays the dependence of the
next-to-leading-order cross section on scale (dashed line)
compared with  the scale dependence of the leading-order
calculation (solid line).  We display the
ratio of the cross section at a scale $\mu=\lambda m_{\perp}$,
 normalized to the  cross section for $\lambda = 1$. The quantity
plotted, then, is basically the fractional uncertainty in each
distribution which is purely due to theory.

As can be plainly seen in the graph, 
the leading order cross section  increases monotonically as $\lambda
\to 0$, whereas the NLO cross section appears to be approaching a
peak around $d\sigma/d\sigma_0=1.4$.  Eventually, the NLO graph will
turn over and begin to drop to zero as the scale approaches
$\Lambda_{QCD}$.  However, it is very clear the NLO calculation
represents a dramatic improvement in the theoretical uncertainty due
to scale dependence.  For instance, for a typical range
$.5<\lambda<2,0$, the leading order cross section varies from 75\% to
140\% of the $\lambda=1$ value, whereas the NLO cross section only
varies from 85\% to 115\%.

\begin{figure}
\scalebox{.9}{\includegraphics[angle=-90,width=15cm]{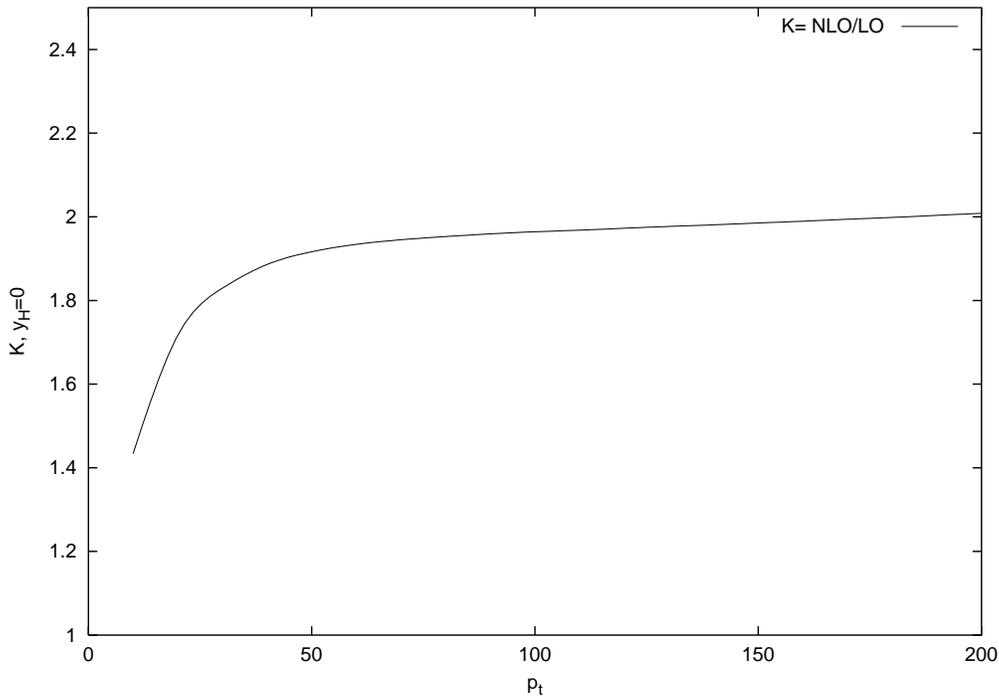}}
\caption{K factor at the LHC ($\sqrt{S}=14$  $TeV$)}
\label{graph:Kfactor}
\end{figure}

In looking at Figure \ref{graph:ptscale}, it is clear that the
enhancement over the leading order approximation is very significant.
The graph of 
the NLO cross section divided by the LO cross section (called the {\em
K Factor}) is, perhaps more interesting, since it entails explicitly
the magnitude of these corrections relative to the LO approximation.
This is displayed in Figure \ref{graph:Kfactor}.  


The $K$ Factor as a function of $p_\perp$ 
is fairly flat down to  about 25 GeV, where it begins a dramatic
decline.  This is to be expected, since for small $p_\perp$, the NLO
$p_\perp$ spectrum
behaves roughly like $\log^3{(p_\perp/m_H)}/p_\perp^2$, which diverges
to negative infinity, while the leading order expression goes to positive
infinity for small $p_\perp$.

We should note that our expression disagrees substantially with the
previous Monte Carlo calculation \cite{evl}.  Whereas their K factor
is about 
1.5, ours is roughly 1.9.  The most likely reason for this discrepancy
is that we have not included the quark graphs in our calculation, whereas
the Monte Carlo calculation has these pieces included.
It should be noted that our K factor is computed at $y_H=0$, whereas
the Monte Carlo calculation  integrates over $y_H$.  As we shall soon
see, the K factor appears to exhibit a minimum at $y_H=0$, and
therefore the integral over rapidity is unlikely to be the source of
the disagreement.  Were we to
compute the integrated K factor, it would most likely disagree by an
even wider margin. However, before claiming a discrepancy
with the calculation in reference
\cite{evl}, we must include the quark pieces in our calculation, or
at least establish that their contribution is non-negative.
\subsection{Rapidity Spectrum}
\begin{figure}
\scalebox{.9}{\includegraphics[angle = -90, width = 15cm]{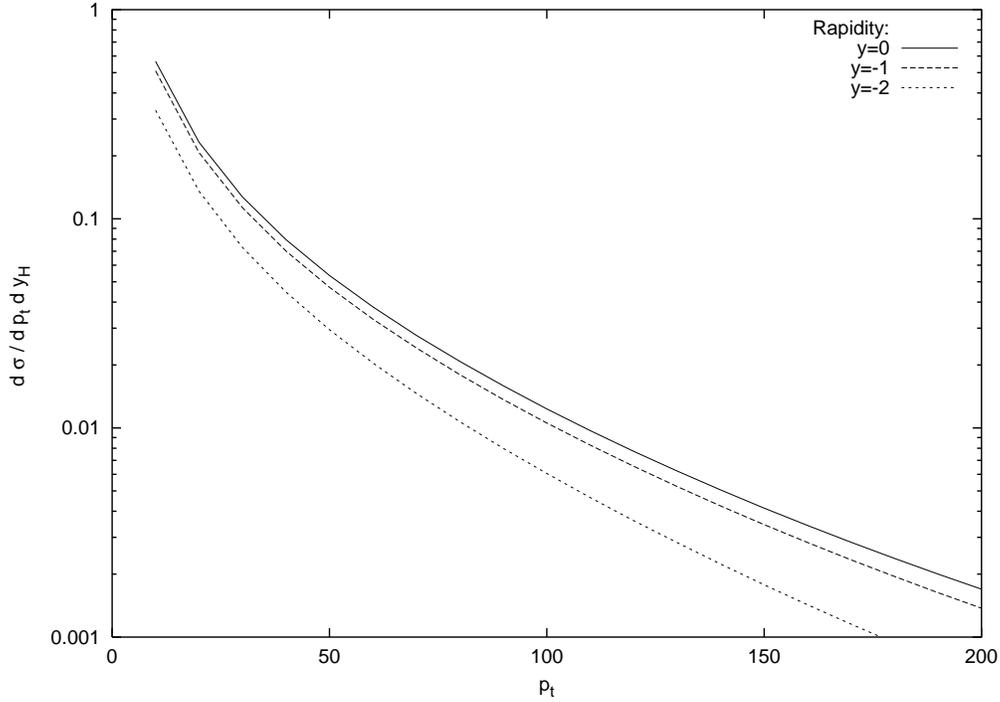}}
\caption{NLO Higgs $p_\perp$ spectrum at the LHC ($\sqrt{S}=14$  TeV)
for various values of the Higgs rapidity, $y_H$}
\label{graph:ptplot}
\end{figure}

Kinematically, it is simply less probable to produce a massive
particle  at large rapidity unless the transverse momentum is small.
Therefore, the cross section should fall off rapidly for large
rapidity.  
Figure \ref{graph:ptplot} is a graph of the
complete $p_\perp$ spectrum at next-to-leading order for various
rapidities, $y_H = 0,1,2$, which are the solid, dashed, and dotted
lines respectively.  The graphs indeed decrease rapidly as the
absolute value of the rapidity is increased.

\begin{figure}
\scalebox{.9}{\includegraphics[angle=-90,width=15cm]{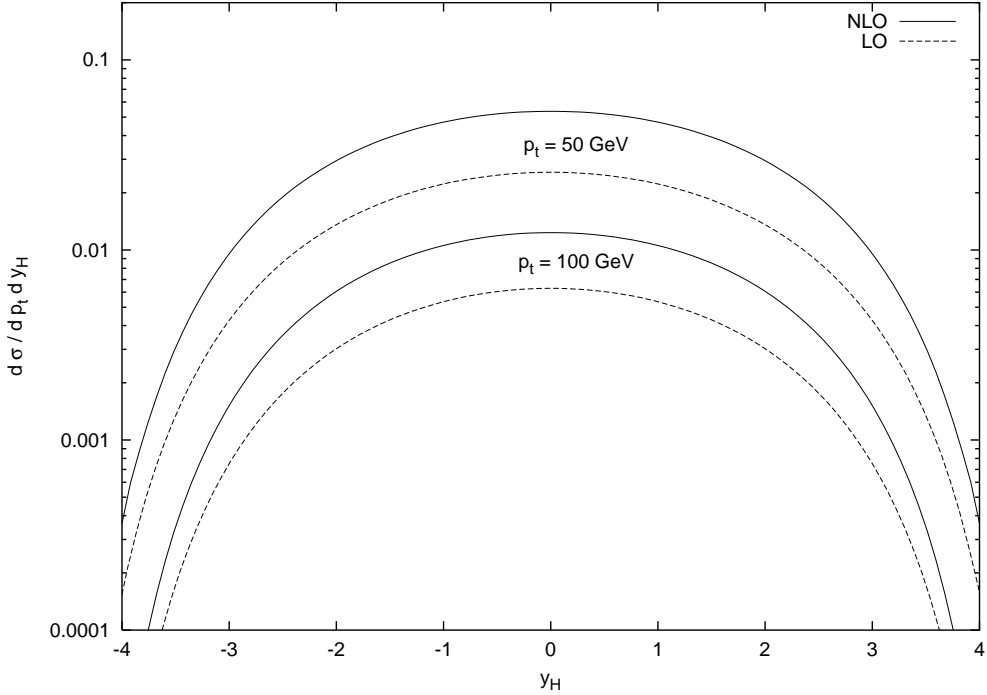}}
\caption{Rapidity Spectrum at the LHC ($\sqrt{S}=14$  TeV).  The
dotted lines are the leading order calc, and the solid lines are the
NLO calc.}
\label{graph:rap}
\end{figure}

Figure \ref{graph:rap} is a plot of the
complete next-to-leading order rapidity spectrum at the LHC for
$p_\perp$ values of 50 GeV and 100 GeV.   
This  graph displays the classic peak at $y_H=0$, and rapidly
drops to zero as the edges of accessible phase space is approached.
That is, the maximum rapidity of a Higgs Boson is governed by
\beq
(p_\perp+m_\perp)\exp{|y_H|}< \sqrt{S}
\eeq
beyond which it is kinematically forbidden to produce a Higgs.

\begin{figure}
\scalebox{.9}{\includegraphics[angle=-90,width=15cm]{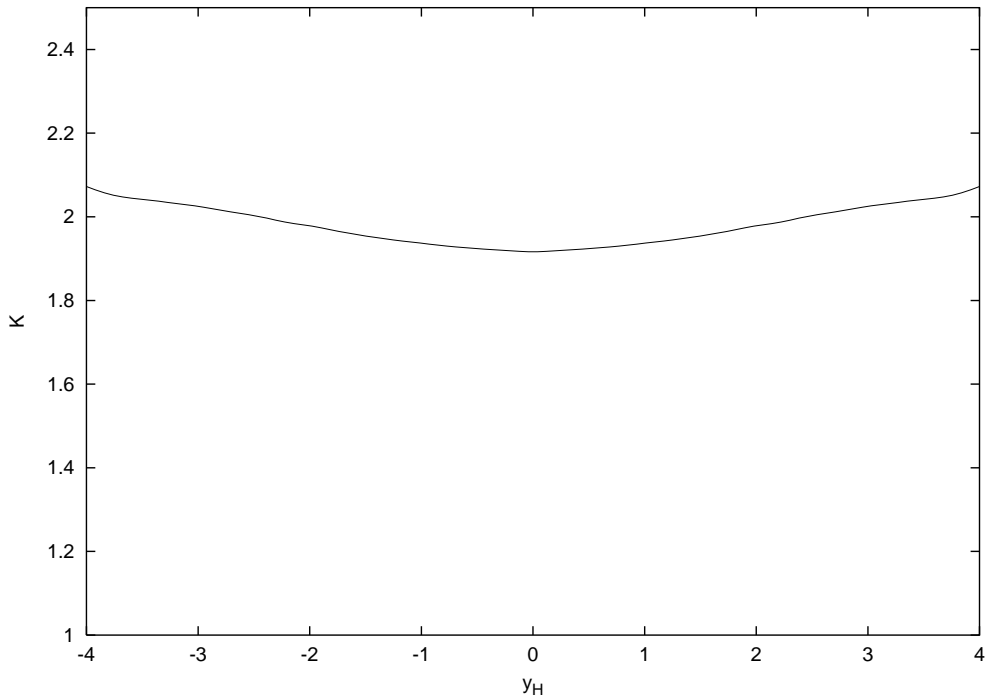}}
\caption{Rapidity dependence of the K factor ($\sqrt{S}=14$  $TeV$)}
\label{graph:rapK}
\end{figure}

The rapidity dependence of the K factor is plotted in Figure
\ref{graph:rapK}.  The first thing to note is that the graph is very
flat as a function of rapidity, indicating that the phase
space limitations more or less govern the shape of this graph.  Also,
note the minimum that we  mentioned earlier at
$y_H=0$, indicating that an integral over the rapidity will further
enhance our expression for the K factor.  Were we to integrate over
 $y_H$, this would most likely increase our K factor slightly.  Again,
this is interesting due to our disagreement with \cite{evl}.

%
%
\subsection{Small $p_\perp$ Behavior}
The transverse momentum spectrum calculation at any fixed order in
perturbation theory is poorly behaved for very small
$p_\perp$, as it is dominated by logarithms of the transverse momentum
divided by the Higgs mass in this limit.  At each order in
perturbation theory, the coefficient of the leading divergence changes sign,
therefore rendering the predictive power of traditional perturbation
theory useless in this limit.

For our calculation, we expect to see exactly this.  The cross section
should display a peak in the small $p_\perp$ region, after which the
cross section starts to diverge to negative infinity.  We expect that
our calculation will be a fairly  good approximation beyond this
peak. 

\begin{figure}
\scalebox{.9}{\includegraphics[angle=-90,width=15cm]{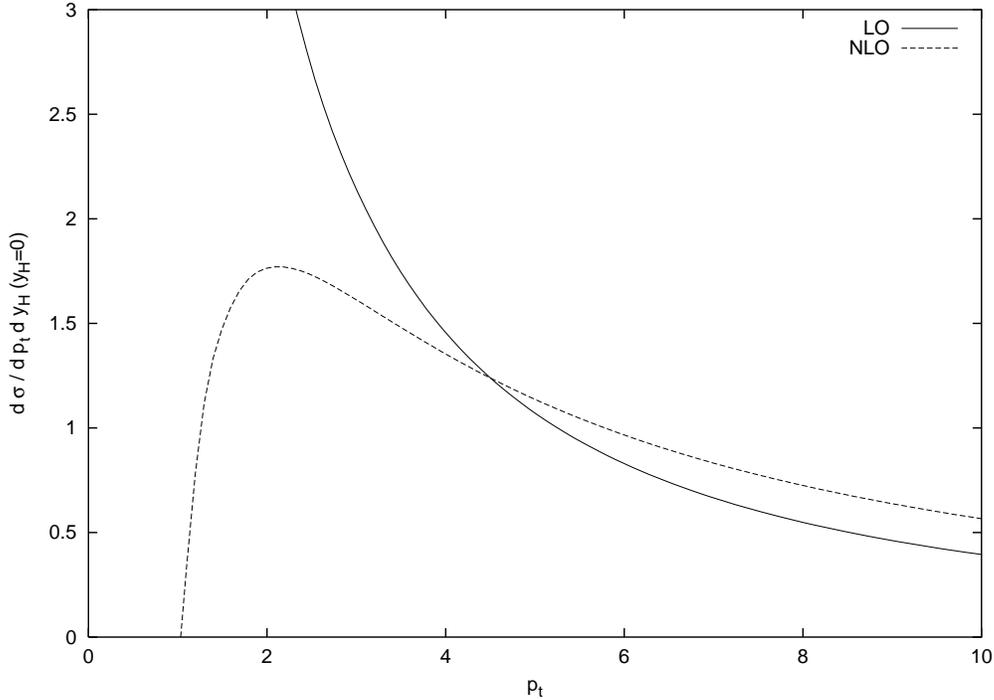}}
\caption{Small $p_\perp$ dependence of the cross section, $y_H=0$.}
\label{graph:smallPt}
\end{figure}

Our plot, Figure \ref{graph:smallPt}, of the small $p_\perp$ region
displays just this behavior (dotted line).  Again we have superimposed
the leading order calculation (solid line) as a reference.  Our
calculation displays a prominant peak around 2.1 GeV for our standard
choice of the Higgs mass, 115 GeV.



If one wishes to have a good prediction of the cross section in the
small $p_\perp$ 
region, one may resum the large logarithmic behavior via the
techniques developed by Collins, and Soper
\cite{Collins},  and Sterman \cite{George}. 
 The first order of business for a calculation of this
nature would be to expand our analytic result, and extract the $B^{(2)}$
resummation coefficient.  This could then be used to verify the result
of reference \cite{B2}.  In order to have a complete small transverse
momentum calculation, one needs to analytically calculate the
contributions from the process $gg \to q\bar{q} H$ in the small
$p_\perp$ limit.

This work is a major part of a larger calculation.
Whereas we have the
dominant part of the cross section, the gluonic piece,  completed, we
need the quark contributions to have a complete and reliable
prediction.  One also wishes to have a reliable calculation at
small transverse momentum, which involves calculating the resummation
coefficient mentioned above.  Including the quarks is  work in
progress, as is the 
computation of the resummation coefficient from the asymptotic
expansion of our analytic results.



\chapter{Conclusions}

In this work we have seen how the Standard Model of particle physics,
while close to being complete from a phenomenological standpoint,
still lacks a crucial piece, the Higgs Boson.  The advent of
the LHC, which in effect will be a ``Higgs factory'', will allow us to
study this component of the SM in considerable detail.  If we are to
understand the mechanism behind electroweak symmetry breaking, we must
understand the behavior of a pure Standard Model Higgs as much as possible.

One of the most important properties to understand is the transverse
momentum spectrum of the Higgs at the LHC.  Primarily, one needs to
know what to expect in a detector as far as the decay product energies,
among other things.
Since higher order corrections to $p_\perp$  
spectra from QCD radiation tend to be quite significant, one may not
consider the LO process to be reliable.  

As we have seen, the results herein have verified that perturbative
corrections to the Higgs boson transverse momentum spectrum are very
significant.  The terms that we have calculated enhance the cross
section by a factor of approximately 2 in the large transverse
momentum region.  In addition, we find that the theoretical
uncertainty due to varying the renormalization scale is far less than
the leading order result, indicating, that our calculation is more
reliable. This result also agrees with the expected asymptotic
behavior in the small $p_\perp$ region, and agrees qualitatively with the
Monte Carlo calculation by DeFlorian et.al. \cite{evl}. In order to
precisely compare their the results with this calculation, the NLO quark
terms, which have been neglected, must be included.  

In the next ten to twenty years, much will change as to our
understanding of the fundamental building blocks of nature.  In
particular, many clues to the fundamental phenomenon behind
electroweak symmetry breaking lie in wait to be uncovered.  The
discovery of the phenomenon behind the electroweak symmetry breaking
mechanism  will probably be the first significant
increase in our understanding of particle physics since the
development of the Standard Model in the 1970's.  Discovering the
Higgs, and understanding its properties, is the first crucial step.

\appendix
\chapter{Matrix Elements}
In this appendix, we include the expressions for the (unintegrated)
matrix elements for the $2 \to 2$ and $2 \to 3$ gluonic parton
processes, computed in the large quark mass limit. Moreover, we
include the expressions for the $2 \to 2$ process $H \to gq\overline{q}$, so
that the complete leading-order transverse momentum spectrum may be
computed in the large $m_{\mathrm{top}}$ limit.  
$H \to ggg$ and $H \to gq\overline{q}$ amplitudes have been known in
the literature  for a while
\cite{ElHi, Sally}, as have the expressions entailing the process
$H \to gggg$ \cite{DaKa}.  Moreover, the 
amplitudes for the virtual corrections to the $2 \to 2$ partonic
processes $H \to ggg$ and $H \to gq\overline{q}$ may be found in
reference \cite{CarlVirt}.
For simplicity, we give the matrix elements for all particles, except
the Higgs, in the final state.   The matrix elements relevant to our
calculation can be obtained from these by crossing symmetry.

Since we will be using color-ordered helicity amplitudes in our
calculation we will begin by reviewing the standard notation and
identities associated with the decomposition of QCD amplitudes into
color-ordered, gauge-invariant subamplitudes.  We will then present the
matrix elements relevant to this calculation in their entirety.  

\section{Helicity Amplitudes}

We begin by presenting the standard notation used in calculations of
amplitudes, and list a series of identities useful in simplifying
these same amplitudes \footnote{Those who are new to helicity spinor
techniques will find references \cite{MaPa}
useful.}.    We introduce the following
notation for (massless) Dirac spinors,
\beq
u_{\pm}(p)=|p \pm \ra.
\eeq
In addition, we shall use the following notation for spinor products,
\beqa
\bar{u}_-(p) u_+(q) &=& \la p-  | q+  \ra = \la p q \ra, \\
\bar{u}_+(p) u_-(q) &=& \la p+  | q-  \ra =  [ p q ].
\eeqa
The momenta $p,q$ are assumed to be massless, and therefore these may
also be identified with relativistically invariant chirality states.

The polarization vectors of the gluons are parameterized in
a dual basis \cite{Kleiss:helicity, Xu:helicity}
\beqa
\e(p, k)^{\mu}_\pm =\frac{1}{\sqrt{2}} \frac{\la k\pm | \gamma^{\mu} |
	p\pm  \ra} {\la p\mp  | k \pm \ra},
\eeqa
where $p$ is the momentum of the gluon, and $k$ is some arbitrary
massless reference vector.

Products of these spinors are found to satisfy the following
identities (for massless $p, q$):
\beqa
\la pq \ra [qp]=2 p \cdot q \\
\la pq \ra^*={\mathrm{sgn}}({p \cdot q})[q p] \label{eqn:signpq}\\
\frac{\left(1 \pm \gamma_{5}\right)}{2}\slash{p} =| p \pm \ra \la p
\pm |.  
\eeqa
From this last relation, we see that we may derive the following
identity,
\beq
\la +p_1 | -p_2 \ra \la- p_2 |+ p_3 \ra \la +p_3| -p_4 \ra \dots
\la -p_n | +p_1 \ra = \frac{1}{2}
\mathrm{Tr} \left\{ (1+\gamma_5)\slash{p_1} \slash {p_2} \slash{p_3}
\slash{p_4} \dots \slash{p_n}\right\}.
\eeq
This last relation is particularly useful in computing absolute
squares of matrix elements, and converting them into traces over
products of gamma matrices.
\section{Real Contributions}

In order to calculate the $p_\perp$ spectrum to next-to-leading order,
we need to write down all the separate partonic contributions to this
process.  They can be broken up into three separate terms;  the
{\em leading order} or tree-level process which consists of Higgs to
3-partons , the {\em virtual} or 1-loop correction to this
process, and the 4-parton process, which we refer to as the 
{\em real corrections}.  The (unintegrated) expressions for these have
been computed in works outlined below.

One may compute the {\it spin- and color-averaged} matrix element
squared by simply
summing over colors, adding up the $2^N$ spin terms, then dividing by the 
number of spins and colors of the particles in the initial state.
That is, if $N_{a,b}$ are the number of spins and
$C_{a,b}$ are the number of colors of
the initial-state  partons $a$ and $b$, then
\begin{equation}
|\overline{M(a,b;1,2,\dots,n)}|^2=\frac{1}{C_a C_b N_a N_b}\,\sum \,
|M(a^{\lambda_a},b^{\lambda_b};1^{\lambda_1}, 2^{\lambda_2}, \dots,
n^{\lambda_n})|^2,
\end{equation}
where the $\lambda_i$ represent the helicity of particle $i$, and the
sum is implicitly  over the spins and colors, unless noted otherwise.
Generally, the amplitudes for the process involving initial state
partons may be computed using crossing symmetry, and being careful to
note that certain terms can pick up a sign via equation
(\ref{eqn:signpq}).

\subsection{$H \to ggg$}
We shall begin by writing down our expressions for the Higgs
production  matrix  elements.  The matrix element with the full
$\epsilon$-dependence was first computed by S. Dawson \cite{Sally}.  It is
fairly straightforward to compute from the conventional Feynman rules
(together with the non-conventional ones from our low energy Higgs
model).

There are 2 independent subprocesses that contribute to the leading-order
expression.  The $H \to ggg$ piece is given by 
\begin{eqnarray}
M(1^+,2^+,3^+) &=& { g A f_{abc} \mh^4
 \o \sqrt{2} \la 12 \ra \la 23 \ra \la 31 \ra }, \\
M(1^-,2^+,3^+) &=& { g A f_{abc} [23]^3
 \o \sqrt{2} [12] [13]  }.
\label{eq:hgggamp}
\end{eqnarray}
The remaining contributions may be obtained through parity and charge
conjugation operations.
If we sum this over spins and colors, we find
\begin{equation}
\sum |M(H\to ggg)|^2 = { g^2 A^2 N_c(N_c^2-1) \o S_{12} S_{13} S_{23} }
\l( S_{12}^4 + S_{13}^4 + S_{23}^4 + \mh^8 \r).
\label{eq:hgggsum}
\end{equation}

The remaining  subprocess that we will consider for the leading-order
transverse momentum spectrum is $H \to q\bar{q}g$.  The
matrix element for this subprocess is given by
\beq
M(1^+,2^-,3^+) = -{ig T^a_{ij} A \o \sqrt{2} } {[13]^2 \o [12]}.
\label{eq:hqqgamp}
\eeq
This may be also be easily summed over spins and colors
\beq
\sum |M(H\to gq\bar{q})|^2 = {g^2 A^2 \left( N_{C}^2-1\right) \o 4}
								{u^2 +t^2 \o s}.
\eeq
It should be noted that each of these amplitudes were computed using the
conventional normalization for the SU(3) generators, 
${\mathrm Tr}[T_a T_b]={1 \o 2} \d_{ab}$.

\subsection{$H \to gggg$}

In order to facilitate the calculation, it is beneficial to break up
the subamplitude into color-ordered {\it helicity subamplitudes}
\beq
M(1^{\lambda_1},2^{\lambda_2},3^{\lambda_2},4^{\lambda_4})
=\,2A\, g^2 \,\sum_{\mathrm{colors}}
\,{\mathrm Tr}[ T^a T^b T^c T^d] \, m(1^{\lambda_1},2^{\lambda_2},3^{\lambda_3},4^{\lambda_4}).
\eeq
These amplitudes are cyclically symmetric in the colors, they are
invariant under reversal of the color indices, and they satisfy the
dual Ward Identities \cite{MaPa}.  
There are three independent helicity subamplitudes for this process:
\beqa
&&m(1^+,2^+,3^+,4^+)={\mh^4\over \la 1 2\ra \la 2 3\ra
   \la 3 4\ra \la 4 1\ra},
\label{eq:pppp}\\ 
&&m(1^-,2^+,3^+,4^+) = -
{\la 1{-}|\slash \ph |3{-} \ra^2 [24]^2 \over S_{124} S_{12} S_{14}}     
-{\la 1{-}|\slash \ph | 4{-} \ra^2 [2 3]^2 \over S_{123} S_{12} S_{23}} \no\\
&&\qq
-{\la 1{-}|\slash \ph | 2{-} \ra^2 [3 4]^2 \over S_{134} S_{14} S_{34}} 
 +{[2 4] \over 
[ 1 2 ] \la 2 3\ra \la 3 4 \ra   [ 4 1]} 
\biggl\{ S_{23} {\la 1{-} |\slash \ph | 2{-} \ra \over \la 4 1\ra } \no \\
&&\qq\qq\qq\qq\qq\qq\qq +       S_{34} 
      {\la 1{-} |\slash \ph | 4{-} \ra \over \la 1 2\ra }
-[2 4 ] S_{234}\biggr\} ,
\label{eq:mppp}\\ 
\eeqa
and
\beq
m(1^-,2^-,3^+,4^+)=-{\la 1 2\ra^4\over \la 12 \ra 
\la 23 \ra \la 34\ra \la 41\ra}
 -{[34]^4 \over [1 2] [23] [34] [41]} .
\label{eq:mmpp}
\eeq
The remaining terms can be related to these by using the charge
conjugation and time reversal transformations, and the dual Ward identities.

\section{Virtual}

The expression for the virtual $H \to ggg$ amplitude was calculated in 
reference \cite{CarlVirt}.  We define the amplitudes in precisely the
same way that we defined the leading-order pieces, in
terms of helicity amplitudes.  The resulting expressions are then
related to the leading-order amplitudes via a coefficient.  These are
outlined in the following subsections.  To construct the contribution
to the amplitude squared, use the formula
\beq
|M^{(0)}+M^{(1)}|^2= 
\sum_{\mathrm{spins}}\, |M^{(0)}|^2 +
2\,\sum_{\mathrm{spins}} \, {\mathrm Re}[M^{(0)}\, (M^{(1)})^{*}]
+\dots
\eeq
In this expression, $M^{(0)}$ and  $M^{(1)}$ represent generic LO and
1-loop amplitudes.  For clarity, the spin dependence has been
suppressed in this equation.

\subsection{$H \to ggg$}

The expressions for the virtual $H \to ggg$ subamplitudes are
\begin{eqnarray}
M^{(1)}(1^+,2^+,3^+)&=&M^{(0)}(1^+,2^+,3^+)\,{\alpha_{s}\over4\pi}
\,r_{\Gamma}\,\biggl({4\pi\mu^2\over-M_H^2}\biggr)^\epsilon
\,\Biggl[N_c U\nonumber\\
&&\qquad\qquad\qquad
+\,{1\over3}(N_c-n_f)\,{S_{31}S_{23}+S_{31}S_{12}+S_{12}S_{23}
\over M_H^4}\Biggr]\nonumber\\
M^{(1)}(1^-,2^+,3^+)&=&M^{(0)}(1^-,2^+,3^+)\,{\alpha_{s}\over4\pi}
\,r_{\Gamma}\,\biggl({4\pi\mu^2\over-M_H^2}\biggr)^\epsilon
\,\Biggl[N_c U\label{loop}\\ 
&&\qquad\qquad\qquad+\,{1\over3}(N_c-n_f)\,{S_{31}S_{12}
\over S_{23}^2}\Biggr]\ ,\nonumber
\end{eqnarray}
where the prefactor is
\begin{equation}
r_{\Gamma}\ =\ {\Gamma(1+\epsilon)\Gamma^{2}(1-\epsilon)
\over\Gamma(1-2\epsilon)}\ =\ 
\frac{\Gamma(1-\e)}{\Gamma(1-2\e)} \lp 1+\e^2 \frac{\pi^2}{3} \rp ,
\label{rgamma}
\end{equation}
and
\begin{eqnarray}
U&=&
{1\over\epsilon^2}\Biggl[
-\biggl({-M_H^2\over-S_{12}}\biggr)^\epsilon
-\biggl({-M_H^2\over-S_{23}}\biggr)^\epsilon
-\biggl({-M_H^2\over-S_{31}}\biggr)^\epsilon
\Biggl]\,+\,{\pi^2\over2}\nonumber\\ 
&&-\,\ln\biggl( {-S_{12}\over-M_{H}^{2}}\biggr)
\ln\biggl({-S_{23}\over-M_{H}^{2}}\biggr)
\,-\,\ln\biggl({-S_{12}\over-M_{H}^{2}}\biggr)
\ln\biggl( {-S_{31}\over-M_{H}^{2}}\biggr)
\, -\,\ln\biggl( {-S_{23}\over-M_{H}^{2}}\biggr)
\ln\biggl( {-S_{31}\over-M_{H}^{2}}\biggr)
  \nonumber\\
&&-\,2\,{\rm Li}_2\biggl(1-{S_{12}\over M_H^2}\biggr)
 \,-\,2\,{\rm Li}_2\biggl(1-{S_{23}\over M_H^2}\biggr)
 \, -\,2\,{\rm Li}_2\biggl(1-{S_{31}\over M_H^2}\biggr)\ .
 \label{Universal}
\end{eqnarray}
It should be noted that all of the singular behavior (including as $
p_{\perp} \to 0$) is contained within terms proportional to $U$.

Since we are working with an effective theory, we obtain a finite
renormalization at each order of
perturbation theory,
\begin{equation}
M^{(1)}\ \rightarrow\ M^{(1)}+(\Delta+3\delta_g)\,M^{(0)}\ ,
\end{equation}
where $\Delta$ is the finite renormalization of the effective $Hgg$ operator, at this order in perturbation theory,
and $\delta_g$ is the gauge-coupling counterterm.  
Using the $\overline{\rm MS}$ subtraction scheme, the counterterm is
\begin{equation}
\delta_g\ =\ -\,{1\over\epsilon}\,{\alpha_s\over4\pi}\,\Gamma(1+\epsilon)
\,(4\pi)^\epsilon\,\biggl[{11N_c\over6}-{n_f\over3}\biggr]\ ,
\end{equation}
and the finite renormalization coefficient is
\beq
\Delta = 11 \lp \frac{\alpha_s}{2 \, \pi}  \rp.
\eeq

\chapter{Formulas for Angular Integration}
In this appendix, we give detailed formulas for the angular
integration of all the terms appearing in the partial-fractioned
matrix element.  There are two different cases:
\begin{itemize}
\item {when the two invariants involve massless particles, and}
\item {when one of the invariants contains a mass term.}
\end{itemize}
Momentum conservation may be used to reduce all of the matrix elements
into sums of terms of this form, multiplied by a coefficient which has no
angular dependence.

\section{General Expression for terms of the form $\frac{1}{S_{13}^m
S_{23}^n}$}

We begin by evaluating integrals involving two massless invariants.
There are two forms that occur, and these are related by a 
$\nu \to (1-\nu)$ symmetry:
\beqa
\Omega^{(m,n)}(\nu) = S_{u}^{m}  S_{t}^{n}
\int d\Omega^{(\e)} \frac{1}{S_{13}^m S_{24}^n},\\
\Omega^{(m,n)}(1-\nu) = S_{u}^{m}  S_{t}^{n}
\int d\Omega^{(\e)} \frac{1}{S_{13}^m S_{23}^n}.
\label{eqn:nomass}
\eeqa
We recall the explicit defintion of $\nu$, 
\begin{eqnarray} \nonumber
\nu&=&\frac{1}{2}(1-\cos(\theta_{0})) \\
   &=&\frac{Q^2}{{Q_\perp}^2}.
\end{eqnarray}
Those that involve the other combinations of the invariants are
related to these via the freedom to relabel the final state parton
indices, that is exchange $3 \leftrightarrow 4$.

First, we evaluate the phase space in the $Q^2$ rest frame.  
Then without losing any generality, we perform this integration assuming
that the invariant $S_{13}$ is independent of the azimuthal angle $\phi$
The following  definitions are useful: 
\begin{eqnarray}
\omega_{13}&=&S_{13}/(S_{13}+S_{14}), \\
	   &=&S_{13}/S_u, \\
\omega_{14}&=&S_{14}/(S_{13}+S_{14}), \\
	   &=&S_{14}/S_u, \\
\omega_{23}&=&S_{23}/(S_{23}+S_{24}), \\
 	   &=&S_{23}/S_t, \\
\omega_{24}&=&S_{24}/(S_{23}+S_{24}), \\
	   &=&S_{24}/S_t.
\end{eqnarray}
Taking the gluon $a$ to lie along the z axis, we find
\begin{equation} \nonumber
\omega_{13} = \frac{1}{2}(1-\cos(\theta_{13})), 
\end{equation}
and
\begin{equation}
\omega_{23} = \frac{1}{2}(1-\cos(\theta_0) \cos(\theta)
	-\sin(\theta_0) \sin(\theta) \cos(\phi) ).
\end{equation}
The integrals (\ref{eqn:nomass}) become
\begin{eqnarray}
\Omega^{(m,n)}(\nu)&=&\int d\Omega^{(-\epsilon)}
	 {\omega_{13}}^{-m}  {\omega_{24}}^{-n} \\
\Omega^{(m,n)}(1-\nu)&=&\int d\Omega^{(-\epsilon)}
	 {\omega_{13}}^{-m}  {\omega_{23}}^{-n}.
\end{eqnarray}
These may be evaluated in general in terms of hypergeometric functions 
\cite{vanNeerven:angular}.  The explicit expression (for arbitrary
$\epsilon$) is
\begin{equation}
\Omega^{(m,n)}(\nu)=\frac{\Gamma^2 (1-\e)}{\Gamma(1-2\e)} 
\frac{\Gamma(1-n-\e)\Gamma(1-m-\e)}{\Gamma(2-n-m-2\e)} 
	F_{1,2}(m,n,1-\epsilon,1-\nu).
\end{equation}
One may easily prove that the relations for $3 \leftrightarrow 4$
yield precisely the same result.
\section{Expressions for terms of the form $\frac{1}{S_{13}^n
S_{123}^m}$} 
For these expressions, a considerable simplification is attained if
one rotates the preceding frame
so that the Higgs momentum lies in the direction of the z axis.  In
the $Q^2$ rest frame, this is equivalent to a frame where ${\bf a}+{\bf
b}=\hat{z}\,|{\bf a}+{\bf b}|$.
The rotation angle
is then given by
\begin{equation}
\cos(\chi)= \frac{- B}{\sqrt{B^2+C^2}},
\end{equation}
and therefore the invariant $S_{123(4)}$ is given by
\begin{equation}
S_{123(4)}=A \mp \sqrt{B^2+C^2}\cos{\theta_H},
\end{equation}
where we have defined
\begin{eqnarray}
A&=&\frac{1}{2}(2 s + S_u+S_t), \\
B&=&\frac{1}{2}(S_u+S_t \cos{\theta_0}), \\
C&=&\frac{1}{2}(S_u+S_t \cos{\theta_0}).
\end{eqnarray}
The kinematic invariants $S_{13}$, etc. take on the requisite form
\beqa
S_{13}=\frac{S_{u}}{2} \left( 1-\cos{\chi} \cos{\theta_H}-
	\sin{\chi}\sin{\theta_H} \cos{\phi} \right), \\
S_{23}=\frac{S_{t}}{2} \left( 1-\cos{\chi} \cos{\theta_H}-
	\sin{\chi}\sin{\theta_H} \cos{\phi} \right). \\
\eeqa
where the other terms are related by $B \to -B$.

 We therefore seek to evaluate expressions of the form
\begin{equation}
\Omega_{H}^{n,m}=\int \, d\Omega S_{13}^{-n} S_{123}^{-m}.
\end{equation}
All of the other angular form factors may be obtained from these by either
exchanging $B \leftrightarrow -B$ or $u \leftrightarrow t$. Their
explicit expressions are (to $O [ \e^0 ]$)
\beqa
\Omega_{H}^{1,1}&=&\frac{1}{A+B' \cos{\chi}} \Omega_{\e}^{0,1}(\nu)
	+\left( \frac{1}{2 \, a} \right) \frac{1}{A+B' \cos{\chi}}
         \log{\frac{(A+B' \cos{\chi})^2}{A^2-B'^2}} \\
\Omega_{H}^{2,1}&=&\frac{1}{(A+B' \cos{\chi})^2} \Omega_{\e}^{0,1}(\nu)
+ \left( \frac{1}{2 \, a} \right) \frac{1}{(A+B' \cos{\chi})^2} \times
	\\ && \left[ 
	\log{\frac{(A+B'\cos{\chi})^2}{A^2-B'^2}}+\frac{2 B'^2 +2 A B'
\cos{\chi}}{A^2-B'^2} 
	\right] \nonumber \\ 
\Omega_{H}^{1,2}&=& \frac{1}{A+B' \cos{\chi}} \left[
\Omega_{\e}^{0,2}(\nu) +\left( \frac{A B' \cos{\chi}+B'^2}{(A+B'
\cos{\chi})^2} \right) \frac{\Omega_{\e}^{0,1}(\nu)}{a} \right] \\
&&+ \frac{1}{2 a^2} \left[ \frac{A
B'\cos{\chi}+B'^2}{(A+B'\cos{\chi})^3} \log{\frac{(A+B'
\cos{\chi})^2}{A^2-B'^2}}-\frac{2 (B' \sin{\chi})^2}{(A+B'\cos{\chi})^3}  \right] \\
\Omega_{H}^{2,2}&=& \frac{\Omega_{\e}^{0,2}(\nu)}{(A+B' \cos{\chi})^2}
 +\left( \frac{3(B' \cos{\chi})^2+2
B'\cos{\chi}(A+B'\cos{\chi})}{(A+B'\cos{\chi})^4} \right) \times \nonumber \\&&
\left( \frac{\Omega_{\e}^{0,1}(\nu)}{2  a}
+\frac{\log{\frac{(A+B' \cos{\chi})^2}{A^2-B'^2}}}{a^2}
\right)
-\frac{4 (B'\sin{\chi})^2}{a^2 (A+B'\cos{\chi})^4}
\nonumber \\&&
+\frac{B'^2}{a^2(A^2-B'^2)(A+B'\cos{\chi})^2}
 \\
\Omega_{H}^{1,0}&=&  \frac{1}{2 B'} \log{\frac{A+B'}{A-B'}}\\
\Omega_{H}^{2,0}&=& \frac{1}{A^2-B'^2}\\
\Omega_{H}^{1,-1}&=& \frac{a}{B'} \left[ (B'+A
\cos{\chi})\Omega^{1,0}_{H}- \cos{\chi} \right]  \\
\Omega_{H}^{2,-1}&=&  \frac{a}{B'} \left[ (B'+A
\cos{\chi})\Omega^{2,0}_{H}- \cos{\chi} \Omega^{1,0}_{H} \right]  \\
\Omega_{H}^{1,-2}&=&  \frac{a}{B'} \left[ (B'+A
\cos{\chi})\Omega^{1,-1}_{H}- \cos{\chi} \Omega^{0,-1}_{H} \right]  \\
&& + \frac{(a \sin{\chi})^2}{2 B'^2} \left[  (B'^2-A^2)
\Omega^{1,0}_{H} + A   \right]\\ 
\Omega_{H}^{2,-2}&=&  \frac{a}{B'} \left[ (B'+A
\cos{\chi})\Omega^{2,-1}_{H}- \cos{\chi} \Omega^{1,-1}_{H} \right]  \\
&& + \frac{(a \sin{\chi})^2}{2 B'^2} \left[  (B'^2-A^2)
\Omega^{2,0}_{H} -1+2 A  \Omega^{1,0}_{H} \right]\\ 
\Omega_{H}^{1,-3}&=&\frac{a}{B'} \left[ (B'+A
\cos{\chi})\Omega^{1,-2}_{H}- \cos{\chi} \Omega^{0,-2}_{H} \right]
 \\ &&+ \frac{3}{2} \frac{(a \sin{\chi})^2}{B'^2} \left[ (B'^2-A^2)
\Omega^{1,-1}_{H} + A \Omega^{0,-1}_{H} +B' (\frac{a}{3})\cos{\chi}  \right]
\\
\Omega_{H}^{2,-3}&=&\frac{a}{B'} \left[ (B'+A
\cos{\chi})\Omega^{2,-2}_{H}- \cos{\chi} \Omega^{1,-2}_{H} \right]
 \\ &&+ \frac{3}{2} \frac{(a \sin{\chi})^2}{B'^2} \left[ (B'^2-A^2)
\Omega^{2,-1}_{H} + A \Omega^{1,-1}_{H} +B'
(\frac{a}{3})\cos{\chi}\Omega^{1,0}_{H}   \right],
\eeqa
where $B'=\mp \sqrt{B^2+C^2}$.  

\chapter{Results of Angular Integration}
\section{Expression for Regular Pieces of $H \to gggg$ Matrix element}

In this section, we give the detailed results for the non-singular
parts of complicated $(+---)$ helicity configurations.  Recall that
the sum over spins of the $H \to gggg$ amplitude may be written
{\small
\beqa
|\overline{M}|^2 & = & (4\pi \, \a_s)^2 \,
\lp \frac{\a_s}{3 \pi v} \rp^2
\frac{N^2\,(N^2-1)}{16\,(N^2-1)^2}\,
\sum_{\mathrm hel.} 
\left\{ 
|m(1^{\lambda_1}, 2^{\lambda_2},3^{\lambda_3},4^{\lambda_4})|^2+
\right. \nonumber \\&& \left. \hspace{1in}
|m(1^{\lambda_1}, 3^{\lambda_2},4^{\lambda_3},2^{\lambda_4})|^2+
|m(1^{\lambda_1}, 4^{\lambda_2},2^{\lambda_3},3^{\lambda_4})|^2
\right\}.
\eeqa
}
We write the expression as
\beq
|\overline{M}|^2  =  (4\pi \, \a_s)^2 \,
\lp \frac{\a_s}{3 \pi v} \rp^2
\frac{N^2\,(N^2-1)}{8\,(N^2-1)^2}
\left\{ m_{++++}+m_{++--}+m_{+---} \right\},
\eeq
where
\beq
m_{++++}=|m(1^{+}, 2^{+},3^{+},4^{+})|^2+|m(1^{+}, 3^{+},4^{+},2^{+})|^2+|m(1^{+}, 4^{+},2^{+},3^{+})|^2,
\eeq
where
\beqa
m_{++--}&\hspace{-.4cm}=
&\hspace{-.4cm}
|m(1^{+}, 2^{+},3^{-},4^{-})|^2+|m(1^{+},3^{+},4^{-},2^{-})|^2+|m(1^{+}, 4^{+},2^{-},3^{-})|^2+\nonumber\\
&&\hspace{-.4cm}
|m(1^{+}, 2^{-},3^{+},4^{-})|^2+|m(1^{+},3^{-},4^{+},2^{-})|^2+|m(1^{+}, 4^{-},2^{+},3^{-})|^2+\nonumber\\
&&\hspace{-.4cm}
|m(1^{+}, 2^{-},3^{-},4^{+})|^2+|m(1^{+},3^{-},4^{-},2^{+})|^2+|m(1^{+}, 4^{-},2^{-},3^{+})|^2.
\eeqa
The angular integrals of the preceding expressions are presented in their
entirety in chapter 4.  The remaining pieces, which we shall refer to as
the {\em odd helicity pieces}, are
\beqa
m_{+++-}&\hspace{-.4cm}=
&\hspace{-.4cm}
|m(1^{+}, 2^{+},3^{+},4^{-})|^2+|m(1^{+},3^{+},4^{+},2^{-})|^2+|m(1^{+}, 4^{+},2^{+},3^{-})|^2+\nonumber\\
&&\hspace{-.4cm}
|m(1^{+}, 2^{+},3^{-},4^{+})|^2+|m(1^{+},3^{+},4^{-},2^{+})|^2+|m(1^{+}, 4^{+},2^{-},3^{+})|^2+\nonumber\\
&&\hspace{-.4cm}
|m(1^{+}, 2^{-},3^{+},4^{+})|^2+|m(1^{+},3^{-},4^{+},2^{+})|^2+|m(1^{+}, 4^{-},2^{+},3^{+})|^2+\nonumber\\
&&\hspace{-.4cm}
|m(1^{+}, 2^{-},3^{-},4^{-})|^2+|m(1^{+},3^{-},4^{-},2^{-})|^2+|m(1^{+}, 4^{-},2^{-},3^{-})|^2.
\eeqa
We define the normalized angular integrals of these expressions by
\beq
\Omega_{+++-}=2 \int\, d\Omega^{(\e)} \, m_{+++-}.
\eeq
$d\Omega^{\e}$ is the angular measure in $4-2\e$ dimensions from appendix B,
\beq
\int\,d\Omega^{(\e)}=
\frac{1}{2\pi}
\int_{0}^{\pi} \, d\theta \,(\sin(\theta))^{(1-2\e)} 
\int^{\pi}_{0} d\phi \,(\sin(\phi))^{(-2\e)} .
\eeq

Using the various symmetries, as well as the fact that we may freely exchange
the partons 3 and 4 (since this does not affect the outcome of the
angular integration), we may rewrite this as
\beq
\Omega_{+++-}=
2\, \Omega(1,2,3,4)+2 \, \Omega(3,4,1,2)+\Omega(1,3,2,4)+\Omega(3,2,4,1) 
+(t \leftrightarrow u),
\eeq
where
\beq
\Omega(i,j,k,l) = \int \, d\Omega^{\e} \ ,| 
m(i^-,j^+,k^+,l^+)|^2.
\eeq

Finally, we break these into a fairly simple singular term
$\Omega_{\epsilon}(i,j,k,l)$
(given in chapter 4), and a nonsingular piece $\Omega_{0}(i,j,k,l)$, which
contains no $\epsilon$ or $Q^2$ singularities.  We have chosen to
include a number of (universal) nonsingular factors in the expressions
for $\Omega_{\epsilon}(i,j,k,l)$.

\subsection{$m(1^-, 2^+,3^+,4^+)$}

First, let us define 
\beq
\eta = \sqrt{1- \frac{4\, s\, M^2}{S_H^2}}.
\eeq
Then we may write the integral over the angular phase space as
\beqa
\Omega_{0}(1,2,3,4)&=&F^{(0)}(s,t,u)+F^{(1)}(s,t,u),
\eeqa
where
{\small
\beqa
F^{(0)}(s,t,u) &=& \nonumber
 - \frac{8\,M^4\,Q^2\,{{p_\perp}}^2}{u^4} - \frac{8\,M^4\,{S_u}}{s\,u^2}
 +   \frac{4\,M^2\,{S_H}}{s\,u}  + \frac{M^2\,\left( -4\,M^2\,s + {{S_H}}^2 \right) }{s\,u^2} +
 \\&& \nonumber
  \frac{\log (\frac{{\left( Q^2\,s + \left(Q^2 -M^2  \right) \,{S_u} \right) }^2}{M^2\,s\,{{S_u}}^2})\,
     \left( {M}^8 + {\left( s + {S_u} \right) }^4 \right) }{s\,u\,\left( s + {S_u} \right) 
 \,{S_H}} +\\&& \nonumber
  \frac{t^3\,\log (\frac{{\left( Q^2\,s + \left(Q^2 -M^2  \right) \,{S_t} \right) }^2}{M^2\,s\,{{S_t}}^2})\,
}{s\, \left( s + {S_t} \right) 
\,{S_H}} 
+ 
\frac{t^3\,\log (\frac{s\,t^2}{M^2\,{{S_t}}^2})\,
     \left( {Q}^2\,u +s\,{S_H} + u\,{S_H} \right) }{Q^2\,s\,u\,
     \left( s + {S_u} \right) \,{S_H}}
 + \\&& \nonumber
  \log (\frac{s\,u^2}{M^2\,{{S_u}}^2})\,\left( \frac{{M}^8\,Q^8 + {t}^4\,u^4 + 
        {s}^4\,{{p_\perp}}^8}{{Q}^2\,s^2\,u^4\,{{p_\perp}}^2} - 
     \frac{2\,M^4\,t^3}{s\,u\,{{p_\perp}}^2\,\left( s + {S_u} \right) \,\left( s + {S_t} \right) } 
\right. \\&&  \left. \hspace{3.5cm}- \nonumber
     \frac{{M}^8 + {\left( s + {S_u} \right) }^4}{s\,u\,\left( s + {S_u} \right) \,{S_H}} - 
     \frac{{t}^3}{s\,\left( s + {S_t} \right) \,{S_H}}
\right. \\&&  \left. \hspace{4.5cm}
 - 
     \frac{{t}^3\,\left( Q^2\,u + s\,{S_H} + u\,{S_H} \right) }
      {{Q}^2\,s\,u\,\left( s + {S_u} \right) \,{S_H}} 
\right),
\eeqa
}
and
{\small
\beqa
F^{(1)}(s,t,u) &=&\frac{M^2\,\left( 7\,s + M^2\,\left( -1 + \frac{2\,s}{S_H} \right)  \right) }{s^2}
+\nonumber
\frac{\log (\frac{1 + { {\eta }}}{1 - { {\eta }}})\,{{C}_{\eta }}}{{ {\eta }}}
+\frac{2\,M^4\,\Gamma _{(-2,2)}(u,t)}{s^2}
\\&&\nonumber
-\frac{2\,M^4\,\left( s + {S_u} \right) \,\left(\Gamma_{(-1,1)}(u,t) - 
      \Gamma_{(-1,1)}(t,u) \right) }{s^2\,S_H}\\&&
+\frac{2\,M^4\,\left( \Gamma _{(-2,1)}(u,t) + \Gamma_{(-2,1)}(t,u) \right) }{s^2\,{S_H}}+
\frac{6\,M^4\,\Gamma_{(-1,2)}(u,t)}{s}.
\eeqa
}
The angular form factors in $F^{(1)}(s,t,u)$ are
{\small
\beqa
\Gamma_{(-2,2)}(u,t)&=&
  \frac{\log (\frac{1 + { {\eta }}}{1 - { {\eta }}})}{2\,S_H\, { {\eta }}}\times \left(
\frac{4\, s\,\left(S_u^2 + s\, p_t^2 - s \, Q^2\right)\,\left( 2\,M^2\,Q^2 - u\,\left( t + u \right)  \right)}{\left({S_H}^2-4\, s \,M^2\right)^2}
\right.
\nonumber\\&&
\left.
\hspace{+.4in}+\frac{4\, S_H\,s^2\, Q^2\, p_t^2}{\left({S_H}^2-4\, s \,M^2 \right)^2}
\right)+ \frac{\left(S_u^2 + s\, p_t^2 - s \, Q^2 \right)^2-4\,s^2\, Q^2\, p_t^2}{\left({S_H}^2-4\, s \,M^2\right)^2}
+\nonumber \\&&
\frac{ s\,\left( 2\,M^2\,Q^2 - u\,\left( t + u \right)  \right)^2}{\left({S_H}^2-4\, s \,M^2\right)^2 {{M}^{2}}}
\eeqa
\beqa
\hspace{-.4in}\Gamma_{(-1,2)}(u,t)&=&
\frac{\left(  2\,M^2\,Q^2 - u\,\left( t + u \right)  \right)}{\left({S_H}^2-4\, s \,M^2\right){ {M}^{2}}}+
\left(\frac{S_u^2 + s\, p_t^2 - s \, Q^2}{{S_H}^2-4\, s \,M^2}\right)
  \frac{\log (\frac{1 + { {\eta }}}{1 - { {\eta }}})}{S_H\, { {\eta }}}
\eeqa
\beqa
\Gamma_{(-2,1)}(u,t)&=&
  \frac{\log (\frac{1 + { {\eta }}}{1 - { {\eta }}})}{S_H\, { {\eta }}} \times \left(
\frac{s^2\, \left(2\, M^2\, Q^2- u\, (t+u) \right)^2 }{\left({S_H}^2-4\, s \,M^2\right)^2}
-
\frac{2\,s^3\,p_t^2\,Q^2\,M^2}{\left(-4\,M^2\,s + {{S_H}}^2 \right)^2}
 \right) \nonumber\\&& \hspace{-.25in}
+\frac{S_u^2\,S_H-2\,s\,u \, S_u}{\left(-4\,M^2\,s + {{S_H}}^2 \right)}
+S_H\,\frac{2 \, s^2 \, Q^2 \, p_t^2 -(S_u\, S_H-2\, s\, u)^2}{2\, \left({S_H}^2-4\, s \,M^2\right)^2},
\eeqa
}
and
{\small
\beqa
\Gamma_{(-1,1)}(u,t)=
\frac{s\,\left( 2\,M^2\,Q^2 - u\,\left( t + u \right)  \right) \,\log (\frac{1 + { {\eta }}}{1 - { {\eta }}})}
   {S_H\left( -4\,M^2\,s + {{S_H}}^2 \right)\, { {\eta }} } + \frac{-2\,s\,u + {S_u}\,{S_H}}{-4\,M^2\,s + {{S_H}}^2}
.
\eeqa
}
Finally, the coefficient $C_\eta$ is in this case
{\small
\beqa
C_\eta&=&\frac{-2\,t^3\,{\left( t + u \right) }^2 }{Q^2\,s\,u\,
    {{S_H}}^2}-
\frac{2}{{Q^2\,s\,u\,{{S_H}}^2}}
\,\left[ - 2\,M^2\,t^3\,\left( 2\,Q^2 + t + u \right) \right. \nonumber \\&&\left. + 
      M^6\,\left( -4\,Q^4 + 6\,Q^2\,u \right)  + 
      2\,M^4\,Q^2\,\left( 4\,t^2 - 2\,t\,u - u^2 + Q^2\,\left( t + u \right)  \right)  \right].
\eeqa
}
Note the existence of a ``naive'' $Q^2$ pole $\Omega_{0}(1,2,3,4)$.
It is superficial, since the limit as  $Q^2 \to  0$ yields a finite
result.

\subsection{$m(3^-,4^+, 1^+, 2^+)$}

Again, we write the integral over the angular phase space as a sum of logs,
\beqa
\Omega_{0}(3,4,1,2)&=&C_0 \nonumber
+C_1 \log{\frac{s t^2}{M^2 S_t^2}}
+C_\eta \frac{\log[\frac{1+\eta}{1-\eta}]}{ {\eta}}
\\&&+ 
\frac{\left( M^2-Q^2+s \right)^3}{s (M^2-t) u} \,
 \log{\frac{{\left( Q^2\,s + \left( -M^2 + Q^2 \right) \,{S_u}
\right) }^2}{M^2\,s\,{{S_u}}^2}}.
\label{omega:3412}
\eeqa
The coefficients $C_i$ in this expression are
\beqa
C_0&=& \nonumber
\frac{63\,Q^2 + 170\,s}{3\,s} + \frac{8\,Q^4\,s^2}{t^4} + \frac{16\,Q^4\,s}{t^3} + 
  \frac{27\,Q^4 - 12\,Q^2\,s + 13\,s^2}{3\,t^2} + \\ && \nonumber
  \frac{3\,Q^6 - 18\,Q^4\,s + 52\,s^3}{3\,Q^2\,s\,t} + \frac{s^2}{3\,u^2} + 
  \frac{4\,s^2}{3\,Q^2\,u} - \frac{11\,
     \left( Q^6 - 3\,Q^4\,s + 3\,Q^2\,s^2 - s^3 \right) }{3\,Q^2\,t\,u} + \\ && \nonumber
  \frac{11\,Q^4\,s^2}{3\,{{S_t}}^4} - \frac{38\,Q^2\,s^2}{3\,{{S_t}}^3} - 
  \frac{2\,Q^2\,s^3}{3\,u\,{{S_t}}^3} + \frac{-9\,Q^2\,s + 74\,s^2}{3\,{{S_t}}^2} + 
  \frac{-9\,Q^2\,s^2 + 5\,s^3}{3\,u\,{{S_t}}^2} + \\ && \nonumber
  \frac{22\,Q^6 + 66\,Q^4\,s + 129\,Q^2\,s^2 - 52\,s^3}{3\,Q^2\,s\,{S_t}} + 
  \frac{-18\,Q^4\,s + 27\,Q^2\,s^2 - 11\,s^3}{3\,Q^2\,u\,{S_t}} + \\ && \nonumber
  \frac{23\,{S_t}}{s} + \frac{2\,s\,{S_t}}{3\,u^2} + \frac{2\,s\,{S_t}}{Q^2\,u} + 
  \frac{{{S_t}}^2}{3\,u^2} + \frac{2\,{{S_t}}^2}{3\,Q^2\,u} + 
  \frac{11\,Q^4\,s^2}{3\,{{S_u}}^4} + \frac{34\,Q^2\,s^2}{3\,{{S_u}}^3} - 
  \frac{2\,Q^2\,s^3}{3\,t\,{{S_u}}^3} +\\ && \nonumber
 \frac{2\,Q^2\,s\,{S_t}}{{{S_u}}^3} + 
  \frac{s\,\left( 3\,Q^2 + 26\,s \right) }{3\,{{S_u}}^2} + 
  \frac{3\,Q^2\,s^2 - 7\,s^3}{3\,t\,{{S_u}}^2} + \frac{3\,s\,{S_t}}{{{S_u}}^2} + 
  \frac{{{S_t}}^2}{2\,{{S_u}}^2} +\\ && \nonumber
  \frac{22\,Q^6 + 66\,Q^4\,s + 93\,Q^2\,s^2 - 4\,s^3}{3\,Q^2\,s\,{S_u}} + 
  \frac{-6\,Q^4\,s + 15\,Q^2\,s^2 - 11\,s^3}{3\,Q^2\,t\,{S_u}} + \\ && \nonumber
  \frac{11\,\left( Q^8 + 4\,Q^6\,s + 6\,Q^4\,s^2 + 4\,Q^2\,s^3 + 2\,s^4 \right) }
   {6\,Q^2\,s\,{S_t}\,{S_u}} + \frac{\left( 11\,Q^4 + 26\,Q^2\,s - 2\,s^2 \right) \,
     {S_t}}{Q^2\,s\,{S_u}} +\\ && \nonumber\frac{2\,\left( 11\,Q^2 - s \right) \,{{S_t}}^2}
   {3\,Q^2\,s\,{S_u}} + \frac{25\,{S_u}}{s} + \frac{24\,Q^4\,s\,{S_u}}{t^4} - 
  \frac{10\,Q^2\,s\,{S_u}}{{{S_t}}^3} + \frac{31\,s\,{S_u}}{{{S_t}}^2} + 
  \frac{24\,Q^4\,{{S_u}}^2}{t^4} +\\ && \nonumber\frac{17\,{{S_u}}^2}{2\,{{S_t}}^2} + 
  \frac{8\,Q^4\,{{S_u}}^3}{s\,t^4} - \frac{8\,Q^2\,{{S_u}}^3}{s\,t^3} + 
  \frac{{{S_u}}^3}{s\,t^2} + \frac{8\,\left( 4\,Q^4\,{S_u} - Q^2\,s\,{S_u} \right) }
   {t^3} + \\ && \nonumber \frac{11\,Q^4\,{S_u} + 34\,Q^2\,s\,{S_u} - 26\,s^2\,{S_u}}{Q^2\,s\,{S_t}} - 
  \frac{2\,\left( 6\,Q^4\,{S_u} - Q^2\,s\,{S_u} - 13\,s^2\,{S_u} \right) }
   {Q^2\,s\,t} +\\ && \nonumber \frac{27\,Q^4\,{S_u} - 66\,Q^2\,s\,{S_u} + 26\,s^2\,{S_u}}
   {3\,s\,t^2} + \frac{2\,\left( 11\,Q^2\,{{S_u}}^2 - 13\,s\,{{S_u}}^2 \right) }
   {3\,Q^2\,s\,{S_t}} \\ && - \frac{2\,\left( 27\,Q^2\,{{S_u}}^2 - 8\,s\,{{S_u}}^2 \right) }
   {3\,s\,t^2} + \frac{9\,Q^2\,{{S_u}}^2 + 26\,s\,{{S_u}}^2}{3\,Q^2\,s\,t} + 
  \frac{16\,\left( Q^4\,{{S_u}}^2 - Q^2\,s\,{{S_u}}^2 \right) }{s\,t^3}, \nonumber \\
\eeqa
and 
\beq
C_1= \frac{N_1}{Q^2 \, s \, t^4 \, u \, (M^2-t)},
\eeq
where the numerator $N_1$ is
\beqa
N_1 &=& \nonumber
\left( {\left( M^2 - Q^2 \right) }^4 + s^4 \right) \,{{S_t}}^3\,
   \left( s + {S_u} \right) 
   - Q^{10}\,
   \left( 6\,s^3 - {S_t}\,\left( -2\,s^2 + 2\,s\,{S_t} + {{S_t}}^2 \right) 
\right. \\&& \nonumber \left. \hspace{1in}
 + 
     {\left( 4\,s + {S_t} \right) }^2\,{S_u} + 
     3\,\left( 4\,s + {S_t} \right) \,{{S_u}}^2 + 3\,{{S_u}}^3 \right)   \\&& \nonumber
 + Q^2\,{{S_t}}^2\,\left( 6\,s^5 + 2\,s^4\,\left( 7\,{S_t} + 3\,{S_u} \right) 
\right. \\&& \nonumber \left.\hspace{1in} + 
     s^3\,\left( 34\,{{S_t}}^2 + 34\,{S_t}\,{S_u} - 6\,{{S_u}}^2 \right) 
\right. \\&& \nonumber \left. \hspace{1in} + 
     s\,{\left( {S_t} + {S_u} \right) }^2\, 
      \left( 11\,{{S_t}}^2 + 22\,{S_t}\,{S_u} - 5\,{{S_u}}^2 \right) 
\right. \\&& \nonumber \left. \hspace{1in}  + 
     10\,s^2\,\left( {S_t} + {S_u} \right) \,
      \left( 3\,{{S_t}}^2 + 4\,{S_t}\,{S_u} - {{S_u}}^2 \right)   \right. \\&& \nonumber \left. \hspace{1in}+
     {\left( {S_t} + {S_u} \right) }^3\,
      \left( {{S_t}}^2 + 5\,{S_t}\,{S_u} - {{S_u}}^2 \right)  \right) 
 \\&& \nonumber  + 
  Q^6\,\left( 2\,s^5 + 6\,{{S_t}}^5 + {{S_t}}^4\,\left( 31\,s + 13\,{S_u} \right)   \right. \\&& \nonumber \left. \hspace{1in}
     +{{S_t}}^3\,\left( 40\,s^2 + 20\,s\,{S_u} - 2\,{{S_u}}^2 \right) 
\right. \\&& \nonumber \left. \hspace{1in}  - 
     {S_u}\,\left( 2\,s + {S_u} \right) \,
      \left( 4\,s^3 + 14\,s^2\,{S_u} + 11\,s\,{{S_u}}^2 + 3\,{{S_u}}^3 \right) \right. \\&& \nonumber \left. \hspace{1in} - 
     2\,{S_t}\,\left( s + {S_u} \right) \,
      \left( 7\,s^3 + 27\,s^2\,{S_u} + 23\,s\,{{S_u}}^2 + 6\,{{S_u}}^3 \right)   \right. \\&& \nonumber \left. \hspace{1in}
+     {{S_t}}^2\,\left( 16\,s^3 - 2\,{S_u}\,
         \left( 11\,s^2 + 25\,s\,{S_u} + 9\,{{S_u}}^2 \right)  \right)  \right)   \\&& \nonumber +
  Q^4\,{S_t}\,\left( 4\,{{S_t}}^5 + {{S_t}}^4\,\left( 29\,s + 17\,{S_u} \right)  + 
     2\,{{S_t}}^2\,\left( 2\,s + {S_u} \right) \,
      \left( 12\,s^2 + 9\,s\,{S_u} - {{S_u}}^2 \right)   \right. \\&& \nonumber \left. \hspace{1in}
+     2\,{{S_t}}^3\,\left( 29\,s^2 + 32\,s\,{S_u} + 9\,{{S_u}}^2 \right) 
\right. \\&& \nonumber \left. \hspace{1in}  + 
     2\,{S_t}\,\left( s + {S_u} \right) \,
      \left( 3\,s^3 - 5\,{S_u}\,\left( s + {S_u} \right) \,
         \left( 2\,s + {S_u} \right)  \right)  + \right. \\&& \nonumber \left. \hspace{1in}
     \left( s + {S_u} \right) \,\left( 6\,s^4 - 
        {S_u}\,\left( 12\,s^3 + 28\,s^2\,{S_u} + 16\,s\,{{S_u}}^2 + 3\,{{S_u}}^3
           \right)  \right)  \right)  \\&& \nonumber  - 
  2\,Q^8\,\left( -2\,{{S_t}}^4 - {{S_t}}^3\,\left( 7\,s + {S_u} \right)  + 
     {{S_t}}^2\,\left( -3\,s^2 + 8\,s\,{S_u} + 5\,{{S_u}}^2 \right)  + \right. \\&& \nonumber  \left. \hspace{1in}
     {S_t}\,\left( 5\,s + 3\,{S_u} \right) \,
      \left( s^2 + 2\,{S_u}\,\left( 2\,s + {S_u} \right)  \right) 
\right. \\&& \nonumber \left. \hspace{1in}  + 
     \left( s + {S_u} \right) \,\left( 6\,s^3 + 
        {S_u}\,\left( 16\,s^2 + 3\,{S_u}\,\left( 4\,s + {S_u} \right)  \right)  \right)
 \right). \\
\eeqa
Finally, The coefficient $C_\eta$ in equation (\ref{omega:3412}) is
\beqa
{\hspace{-1cm} C_\eta \  =  } \ \frac{10\,\left( Q^2 - 2\,s \right) \,s}{t\,u} & + &
  \frac{\left( {\left( M^2 - Q^2 \right) }^4 + s^4 \right) \,
     \left( 1 - \frac{2\,s}{{S_h}} \right) }{Q^2\,s\,t\,u} + \nonumber\\   
  \frac{4\,s\,\left( Q^4 - 3\,Q^2\,s + 3\,s^2 \right) }{t\,u\,{S_h}} & + & 
  \frac{2\,{S_h}\,\left( -4\,\left( Q^2 - 3\,s \right) \,s - 6\,s\,{S_h} + 
       {{S_h}}^2 \right) }{s\,t\,u}.
\eeqa

\subsection{$m(1^-,3^+, 2^+, 4^+)$}

Recalling $S_H=M^2+s-Q^2$, we write the result of the integration over the angular phase space as
\beqa
\Omega_{0}(1,3,2,4)&=& \nonumber
\frac{2\,M^2\,t^2}{s\,u^2} - \frac{16\,M^4\,Q^2\,{{p_t}}^2}{u^4}
 + 
 \frac{4\,M^4}{S_H^2 }\, 
   \frac{\log{\left(\frac{1+ {\eta}}{1- {\eta}} \right)}}{  {\eta} }   
\\&& \nonumber
+ \log \left(\frac{s\,u^2}{M^2\,{{S_u}}^2}\right)\,C_1
 + \log \left(\frac{s\,t^2}{M^2\,{{S_t}}^2}\right)\,C_2
\\&& \nonumber
 +   \frac{2\,t^3} 
   {\left( Q^2 - t \right) \,\left( M^2 - u \right) \,{S_H}}
\, \log \small{\left(\frac{{\left( s\, Q^2-(M^2-Q^2)\,S_t \right) }^2}{S_t^2\,M^2\,s}\right)}
\\&& +  \nonumber
  \frac{2\,\left( M^8 + {\left( M^2 - t \right) }^4 \right) \,
     }{\left( M^2 - t \right) \,
     \left( Q^2 - u \right) \,u\,{S_H}} 
\, \log \left(\frac{{\left( s\, Q^2-(M^2-Q^2)\,S_u \right) }^2}{S_u^2\,M^2\,s}\right).
\\&&
 \eeqa
The coefficients are
\beqa \nonumber
C_1&\hspace{-.15in}=&\hspace{-.15in}
 \frac{1}{  S_H  \,s\,\left( M^2 - u \right) \,u^4\, S_u  }\times 
\left[ -2\,\left( M^2 - s \right) \,t^3\,u^4 + 2\,M^2\,t^3\,u^3\,\left( 2\,Q^2 + t + 2\,u \right) 
\right. \\&& \left. \hspace{.5in} \nonumber +
      4\,M^{10}\,\left( 2\,Q^6 - 2\,Q^4\,u + 2\,Q^2\,u^2 - u^3 \right) 
\right. \\&& \left. \hspace{.5in} \nonumber - 
      4\,M^8\,\left( Q^4\,\left( t - 5\,u \right) \,u + Q^6\,\left( t + 3\,u \right) 
\right. \right. \\&& \left. \left. \hspace{1in} \nonumber
 - u^3\,\left( t + 3\,u \right)  + 
         Q^2\,u^2\,\left( t + 6\,u \right)  \right)
\right. \\&& \left. \hspace{.5in} \nonumber
  - 
      4\,M^4\,u^2\,\left( Q^4 \,\left( t^2 - u^2 \right)  + 
\right. \right. \\&& \left. \left. \hspace{1in}\nonumber
         Q^2\,\left( t + 2\,u \right) \,\left( 2\,t^2 + u^2 \right) 
   - u^2\,\left( 2\,t^2 + t\,u + u^2 \right)  \right) 
\right. \\&& \left. \hspace{.5in}\nonumber + 
      4\,M^6\,u\,\left( Q^6\,\left( t+u \right)  + Q^4\,\left( t^2 + t\,u - 4\,u^2 \right)   
\right.\right. \\&& \left.\left. \hspace{1in} + 
         2\,Q^2\,u\,\left( 2\,t^2 + t\,u + 3\,u^2 \right)  - u^2\,\left( 2\,t^2 + 2\,t\,u + 3\,u^2 \right)  \right) \right]
\eeqa
and
\beqa
C_2&=&
\frac{2\,t^3\,\left( M^2\,\left( 2\,Q^2 - 2\,t - u \right)  + t\,\left( -Q^2 + t + u \right)  \right) }
  {s\,\left( M^2 - t \right) \,\left( Q^2 - t \right) \,\left( 2\,M^2 - t - u \right) \,u}.
\eeqa

\subsection{$m(3^-, 2^+,4^+,1^+)$}

The last independent helicity amplitude is $m(3^-, 2^+,4^+,1^+)$.  We
write the finite remainder of the the azimuthal integration as
\beqa \nonumber
\Omega_{0}(3,2,4,1)&=& \frac{1}{2} C_0 +C_1 \log{\frac{s u^2}{M^2 S_u^2}} \\&&
+C_2 \log (\frac{{\left( Q^2\,s - \left( M^2 - Q^2 \right) \,{S_u} \right) }^2}{M^2\,s\,{{S_u}}^2}) +(u \leftrightarrow t),
\eeqa
where
\beqa
C_1&=& \nonumber
-\left( \frac{{\left( M^2 - t \right) }^3}{s\,t\,\left( -M^2 + u \right) } \right)  - 
  \frac{\left( 2\,t + u \right) \,\left( t^2 + t\,u + u^2 \right) }{s\,t\,u} 
 - 
  \frac{2\,M^6\,\left( Q^4\,t + \left( 7\,t - u \right) \,u^2 \right) }{s\,t\,u^4}
\\&&\nonumber + 
  \frac{M^2\,\left( -4\,t^3 + 3\,t\,u^2 + 3\,u^3 \right) }{s\,t\,u^2} + 
  \frac{M^4\,\left( 2\,Q^2\,t\,\left( t + u \right)  + 
       u\,\left( 14\,t^2 + 4\,t\,u - 3\,u^2 \right)  \right) }{s\,t\,u^3} 
\\&& -
  \frac{{\left( -2\,M^2 + t + u \right) }^3}{s\,u\,{S_u}},
\eeqa
\beq
C_2=-
\left( \frac{{ S_H }^3}{\left( M^2 - t \right) \,u \, S_u } \right),
\eeq
and finally, 
the coefficient $C_0$, which  is symmetric under the interchange $u \leftrightarrow t$:
\beqa
C_0&=&
5 + \frac{8\,M^4 - 4\,M^2\,t}{t^2} - \frac{2\,M^2\,Q^2\,\left( 4\,M^4 - M^2\,t - 3\,t^2 \right) }
   {s\,t^3} - \frac{8\,M^4\,Q^2\,t}{s\,u^3} - \frac{7\,M^4\,Q^2}{s\,u^2}  
\nonumber \\&&
+  \frac{M^2\,\left( M^4\,{\left( t - u \right) }^2 + 
       t^2\,\left( t + u \right) \,\left( t + 3\,u \right)  + 
       M^2\,t\,\left( -t^2 + t\,u + 4\,u^2 \right)  \right) }{s\,t^2\,u^2} 
\nonumber \\&&
+  \frac{M^2\,Q^4\,\left( -8\,M^2\,t\,u^4 + t^2\,u^3\,\left( -4\,t + u \right)  + 
       8\,M^4\,\left( t^4 + u^4 \right)  \right) }{s\,t^4\,u^4} + \frac{6\,M^4\,t^2}{{{S_t}}^4}
\nonumber \\&&
-  \frac{2\,\left( 3\,M^4\,Q^2 - 2\,M^2\,Q^2\,t \right) }{s\,t\,u}
-  \frac{M^2\,Q^8\,u}{t^2\,{{S_t}}^4} + \frac{2\,M^2\,Q^6\,u}{t\,{{S_t}}^4}
\nonumber \\&&  + 
  \frac{-10\,M^2\,Q^4\,u - 3\,M^2\,t^2\,u + Q^4\,u^2 + 4\,Q^2\,t\,u^2 + t^2\,u^2}{{{S_t}}^4}
\nonumber \\&& - 
  \frac{6\,M^2\,t\,\left( 2\,M^2 + t \right) }{{{S_t}}^3} + 
  \frac{2\,t\,\left( Q^2 + 2\,t \right) \,u}{{{S_t}}^3} + 
  \frac{10\,M^4 + 4\,M^2\,t + t^2}{{{S_t}}^2} 
\nonumber \\&& +
  \frac{2\,M^4\,t\,\left( M^2 - u \right) }{s\,u\,{{S_t}}^2} - 
  \frac{4\,M^2\,\left( M^2 + t \right) }{t\,{S_t}} - \frac{2\,M^4\,t}{s\,u\,{S_t}} + 
  \frac{M^2\,\left( -8\,M^2 + t \right) \,{S_t}}{t^3} \nonumber \\&& + 
  \frac{6\,Q^4\,{\left( M^2 - t \right) }^2}{{{S_u}}^4} - 
  \frac{6\,Q^2\,\left( M^2 - t \right) \,\left( Q^2 + t \right) }{{{S_u}}^3} + 
  \frac{4\,M^4 - 8\,M^2\,Q^2 + Q^4}{{{S_u}}^2} 
\nonumber \\&&
+ \frac{2\,M^4\,Q^2}{t\,{{S_u}}^2} - 
  \frac{2\,\left( 2\,M^2 - 5\,Q^2 \right) \,t}{{{S_u}}^2} + \frac{t^2}{{{S_u}}^2}
\nonumber \\&& + 
  \frac{2\,\left( M^2\,t\,\left( t - 3\,u \right)  + t\,\left( Q^2 + 2\,t \right) \,u + 
       M^4\,\left( -2\,t + u \right)  \right) }{t\,u\,{S_u}}.
\eeqa

\chapter{Formulas for Cross Sections}
Here we provide standard formulas for the evaluation of $2 \to 2$ and
$2 \to 3$ processes for 2 partons in the initial state, 1 massive
particle in the final state, and 1 or 2 massless particles in the final
state.  We also list convenient variable transformations for the
convolutions, which will allow us to directly evaluate the ``$+$'' functions.
\section{Expression for a $2 \to 2$ Cross Section}. 

The $2 \to 2$ contribution to the $p_\perp$ may be written
\beq
\frac{d \sigma_{\mathrm LO}}{d p_\perp \hspace{.05in} d y_H}
=\frac{1}{S}\int_{0}^{1} d\xi_a\, f(\xi_a) \int_{0}^{1} d\xi_b\, f(\xi_b)\, s \frac{d \sigma}{d t} \delta(Q^2),
\eeq
where
\beq
Q^2=s+t+u-M^2
\eeq
and
\beq
 \frac{d \sigma}{d t}=\frac{1}{4 \pi s^2} |\bar{M}|^2,
\eeq
where we have suppressed the parton indices for clarity.

The expression for the NLO cross section may be defined likewise: 
\beq
\frac{d \sigma_{\mathrm NLO}}{d p_\perp \hspace{.05in} d y_H}
=\frac{1}{S} \int_{0}^{1} d\xi_a \int_{0}^{1} d\xi_b \, s \frac{d
\sigma_{\Omega}}{d t \, du}\theta(Q^2), 
\eeq
where
\beq
\frac{d \sigma_{\Omega}}{ d t \, du} = \frac{1}{4 \pi s^2} \int \frac{d
\Omega}{2 \pi}   |\bar{M}|^2.
\eeq

\section{Evaluating the Convolutions}

First, we must evaluate the expression
\beq
\int^{1}_{0} d \xi_a \int^{1}_{0} d \xi_b \hspace{+.05in} \theta(Q^2) = 
\int^{1}_{0} d \xi_a \int^{1}_{0} d \xi_b  \hspace{+.05in}\theta(
\xi_a   \xi_b -x_T \xi_a -x_U \xi_b  +\tau),
\eeq
where, as before, $x_T=\frac{m_H^2-T}{S}$, $\hspace{+.02in}
x_U=\frac{m_H^2-U}{S}$, and  $\tau=\frac{m_H^2}{S}$.
Evaluating these in the hadron-hadron COM frame, we find that these expressions
become
\beqa
x_T &=& \sqrt{\frac{m_\perp^2}{S}} e^{-y_H} \nonumber, \\
x_U &=& \sqrt{\frac{m_\perp^2}{S}} e^{+y_H}.
\label{xeqn}
\eeqa

Substituting expressions (\ref{xeqn}) for $Q^2$, we find
\beq
\xi_a   \xi_b -x_T \xi_a -x_U \xi_b  +\tau
 = (\xi_a e^{-y_H} - \frac{m_\perp}{\sqrt{S}} )(\xi_b
e^{+y_H}-\frac{m_\perp}{\sqrt{S}} )- \frac{p_\perp^2}{S},
\eeq
which has the implication that 
\beq
 (\xi_a e^{+y_H} - \frac{m_\perp}{\sqrt{S}} )(\xi_b
e^{-y_H}-\frac{m_\perp}{\sqrt{S}} ) > \frac{p_\perp^2}{S}.
\eeq
We may break this up into four different phase space regions, each
defined by $ (\xi_{a,b} e^{\pm y_H} - m_\perp / \sqrt{S} ) $
greater than or less than $ p_\perp / \sqrt{S}$. 
One case of the four is automatically forbidden because it 
implies  $p_\perp <0$.  After 
evaluating the $\theta$-function for each term, we may recombine two
of the regions into one region, leaving us with 
\beq
\int^{1}_{0} d \xi_a \int^{1}_{0} d \xi_b \hspace{+.05in} \theta(Q^2)
=
\int^{1}_{x_{+}} d \xi_a \int^{1}_{\xi_{b}(\xi_{a})}
 d \xi_b +
\int^{1}_{x_{-}} d \xi_b \int^{x_{+}}_{\xi_{a}(\xi_{b})}
 d \xi_a.
\eeq
The lower limits of the internal integrals in this expression  are
solutions to the equation $Q^2=0$ for each of the variables.  Explicitly,
these are
\beqa
\xi_{b}(\xi_{a})&=&\frac{(x_T-\tau) \xi_a-\tau
(1-\xi_a)}{x_U-\xi_a},\nonumber \\
\xi_{a}(\xi_{b})&=&\frac{(x_U-\tau) \xi_b-\tau (1-\xi_b)}{x_T-\xi_b},
\eeqa
whereas the outer limits $x_{\pm}$ are
\beq
x_{\pm}=\exp{(\pm y_H)} \lp \frac{m_\perp+p_\perp}{\sqrt{S}} \rp.
\eeq
It is quite trivial to convert these terms back into the $Q^2$
variable.  If this is done, the expression for the phase
space in
reference \cite{DrellYan, Ellis:phase} is recovered identically:
\beq
\int^{1}_{0} d \xi_a \int^{1}_{0} d \xi_b \hspace{+.05in} \theta(Q^2)
\to
\int^{1}_{x_{+}}\frac{ d \xi_a}{\xi_a-x_U} \int^{A_1}_{0} 
d \left( \frac{Q^2}{S}\right) +
\int^{1}_{x_{-}}\frac{ d \xi_b}{\xi_b-x_T} \int^{A_2}_{0} 
d \left( \frac{Q^2}{S}\right).
\label{eqn:phase_space}
\eeq
In this formula, the limits of integration $A_i$ are
\beqa
A_1&=&    \xi_a (1-x_T)-x_U  +\tau,  \\
A_2&=&  x_+ \lp  \xi_b -x_T \rp  -x_U \xi_b  +\tau.
\eeqa
Note that it is quite trivial to evaluate the $\d(Q^2)$ of the
 $2 \to 2$ phase space using this form of the convolution.

\subsection{Evaluating Terms With ``+'' Functions}

In the regularization of the soft singularities, terms of the form
$f(z_{t,u})_{+}$ arise, and it is beneficial to change variables from
$d(Q^2)$ to
$dz_{t,u}$ in order to easily evaluate these ``+'' functions.  Recall
the  definition of $z_{t,u}$:
\beq
z_{t,u}=\frac{-t,u}{-t,u+Q^2}.
\eeq
Now, consider $z_t$ (a completely analogous analysis will follow for
$z_u$).  It is beneficial to first convert to the phase space
(\ref{eqn:phase_space})   as an intermediate step.
We wish to evaluate the Jacobian for the change of variables.  This is
summarily expressed as
\beq
\int^{A}_{0} d \left( Q^2 \right) 
\to \int^{1}_{z_t(A)} \frac{d z_t}{z_t^2} \frac{(-t)^2}{-t_0}
\to \int^{1}_{z_u(A)} \frac{d z_u}{z_u^2} \frac{(-u)^2}{-u_0},
\eeq
where $t_0=t|_{Q^2=0}$ and $u_0=u|_{Q^2=0}$.
Thus
\beqa
\int^{1}_{0} d \xi_a \int^{1}_{0} d \xi_b \hspace{+.05in} \theta(Q^2)
\to \nonumber\hspace{+.5in} && \\ 
\frac{1}{S}
\int^{1}_{x_{+}}\frac{ d \xi_a }{\xi_a-x_U} 
\int^{1}_{z_t(A1)}  \frac{d z_t}{z_t^2} \frac{(-t)^2}{-t_0}&+&
\frac{1}{S}
\int^{1}_{x_{-}}\frac{ d \xi_b}{\xi_b-x_T}
\int^{1}_{z_t(A_2)} \frac{d z_t}{z_t^2}  \frac{(-t)^2}{-t_0}
\eeqa
The evaluation of ``+'' functions in this basis is quite
straightforward. Converting the phase space into $z_u$ proceeds along
similar lines.

\addcontentsline{toc}{chapter}{\bibname}
\bibliographystyle{utcaps}
\bibliography{thesis}

\end{document}